\documentclass[12pt]{article}
\usepackage[utf8]{inputenc}
\usepackage[left]{lineno}
\usepackage{setspace}
\usepackage{comment}

\newcommand{\blind}{0} 

\usepackage{latexsym,amsmath,amssymb,amsfonts,amsthm}
\usepackage{bm} 
\usepackage{amsmath}
\usepackage{comment}
\newcommand{\bs}[1]{\boldsymbol{#1}}
\usepackage{graphicx}
\usepackage{booktabs}
\usepackage{array}
\usepackage{multirow}
\usepackage{adjustbox}
\usepackage{subcaption}
\usepackage{mathrsfs}
\usepackage{mathabx}

\usepackage{natbib}
\setcitestyle{authoryear,round,citesep={;},aysep={}}
\usepackage{xr-hyper}
\usepackage[colorlinks,citecolor=blue,urlcolor=blue]{hyperref}
\makeatletter
\newcommand*{\addFileDependency}[1]{\typeout{(#1)}\@addtofilelist{#1}\IfFileExists{#1}{}{\typeout{No file #1.}}}
\makeatother

\usepackage{setspace}
\usepackage{appendix}
\usepackage{enumitem}
\usepackage{authblk}
\usepackage{tikz}
\usepackage{float}
\usepackage{placeins}
\usepackage{algorithm}
\usepackage{algpseudocode}

\usepackage{makecell}
\usepackage{bbm} 
\usepackage{lscape} 
\usepackage{longtable}
\usepackage{etoolbox}
\usepackage{listofitems}
\usepackage{pifont}

\usepackage{xcolor}
\usepackage{url}
\usepackage{bbm}
\usepackage{lmodern} 
\usepackage[breakable,skins]{tcolorbox} 
\usepackage{textcomp} 
\usepackage{wrapfig} 
\usepackage[american]{babel}
\theoremstyle{definition}

\DeclareSymbolFont{AMSb}{U}{msb}{m}{n}
\DeclareMathSymbol{\Natural}{\mathbin}{AMSb}{"4E}
\DeclareMathSymbol{\Integer}{\mathbin}{AMSb}{"5A}
\DeclareMathSymbol{\Real}{\mathbin}{AMSb}{"52}
\DeclareMathSymbol{\Rational}{\mathbin}{AMSb}{"51}
\DeclareMathSymbol{\Imaginary}{\mathbin}{AMSb}{"49}
\DeclareMathSymbol{\Complex}{\mathbin}{AMSb}{"43} 
\DeclareMathSymbol{\Disk}{\mathbin}{AMSb}{"44} 
\def\bi{\begin{itemize}}
\def\ei{\end{itemize}}
\def\bd{\begin{description}}
\def\ed{\end{description}}
\def\ben{\begin{enumerate}}
\def\een{\end{enumerate}}
\def\bv{\begin{verbatim}}
\def\ev{\end{verbatim}}

\def\hat#1{{\widehat{#1}}}

\def\pr{{\rm Pr}}
\def\Pr{\pr}

\def\2to{{\ {\buildrel 2\over \longrightarrow}\ }}

\def\I1ton{{$I_1,\ldots,I_n$}}
\def\X1ton{{$X_1,\ldots,X_n$}}
\def\Y1ton{{$Y_1,\ldots,Y_n$}}
\def\Z1ton{{$Z_1,\ldots,Z_n$}}
\def\R1ton{{$R_1,\ldots,R_n$}}
\def\e1ton{{$e_1,\ldots,e_n$}}
\def\t1ton{{$t_1,\ldots,t_n$}}
\def\x1ton{{$x_1,\ldots,x_n$}}
\def\y1ton{{$y_1,\ldots,y_n$}}
\def\z1ton{{$z_1,\ldots,z_n$}}

\newtheorem{defn}{Definition}

\newtheorem{lemma}{Lemma}

\newtheorem{cor}[defn]{Corollary}

\newcommand{\figorpending}[2][\textwidth]{\IfFileExists{#2}{\includegraphics[width=#1]{#2}}{\figplaceholder[#1]{#2}}}
\newcommand{\chk}[1]{#1} 
\newcommand{\figplaceholder}[2][0.3\textwidth]{%
  \fbox{\parbox[c][\dimexpr #1*3/4\relax][c]{\dimexpr #1-2\fboxsep-2\fboxrule\relax}{%
    \centering\footnotesize Figure placeholder\\[2pt]\adjustbox{max width=\linewidth}{\texttt{\detokenize{#2}}}}}}
\newcommand{\bfs}{\mathbf{s}}

\newcommand{\bfW}{\mathbf{W}}
\newcommand{\bfw}{\mathbf{w}}

\newcommand{\bfu}{\mathbf{u}}
\newcommand{\bfZ}{\mathbf{Z}}

\newcommand{\bfpsi}{\boldsymbol{\psi}}

\newcommand{\bftheta}{\boldsymbol{\theta}}
\newtheorem{proposition}{Proposition}
\newtheorem{remark}{Remark}

\renewcommand{\arraystretch}{2}

\title{Bayesian Neural Factor Copula Model for Spatial Data}

\if0\blind
\author[1]{Pratik Nag}
\author[2]{David Gunawan}
\author[3]{Nadja Klein}

\affil[1]{Department of Statistics, University of Wisconsin-Madison, USA}
\affil[2]{School of Mathematics and Physics, University of Wollongong, Australia}
\affil[3]{Scientific Computing Center, Karlsruhe Institute of Technology, Karlsruhe, Germany}
\else
\author{}
\fi

\begin{document}
\date{}
\maketitle

\begin{abstract}
Factor copulas extend Gaussian process models for replicated spatial data by adding a common latent factor to a Gaussian process. The resulting copulas can be asymmetric and tail dependent, with densities that require only one-dimensional integrals. Their tail behavior is, however, fixed by a parametric factor family that must be chosen in advance, and the factor affects every location in the same way. We propose the Bayesian neural factor copula (BNFC), in which the factor is a neural network of a single uniform random variable and of the location, so that its effect can change in strength, shape and sign over space; a stationary version without the location input is a special case. The construction keeps the one-dimensional integrals, can approximate any factor distribution and, in its location-dependent form, induces factor covariances that can be any positive semidefinite matrix, so that long-range dependence may change in strength and sign over space. We characterize the tail properties of the resulting copulas and develop an approximate Bayesian inference procedure using variational Bayes for the model parameters and stochastic gradient Markov chain Monte Carlo (SG-MCMC) for the network weights. In simulations, BNFC outperforms common factor copulas, the Student-$t$ copula, Gaussian and gamma copulas with the same spatially varying loading, and the nonparametric Bayesian transport map when the factor effect varies over space and is competitive otherwise, and in an application to daily ERA5 precipitation over central Europe the location-dependent BNFC gives the best held-out log scores and predictions of all copula models.

\end{abstract}


\section{Introduction}\label{sec:intro}

Gaussian process (GP) regression and kriging are the standard tools for spatial data because of their interpretability and closed-form conditional distributions \citep{cressie1993statistics,stein1999interpolation,rasmussen2006gaussian}. Many environmental data sets, such as precipitation, however, have heavy tails, asymmetric dependence between low and high values, and stronger dependence among extremes than a Gaussian model allows \citep{davison2012statistical,davison2015statistics}. A Gaussian copula is reflection symmetric and tail independent, so it may underestimate the probability of simultaneous extremes at several sites.

Factor copulas \citep{krupskii2018factor} add a single random factor, common to all locations, to a GP. The copula keeps a parameterization by a spatial correlation function, its density requires only one-dimensional integrals, and for suitable factors it is asymmetric and tail dependent. Two features limit these models. First, the factor distribution is chosen from a parametric family in advance, and a misspecified family distorts the joint behavior of extremes; in our simulations it markedly reduces the held-out log score, the log predictive density of test data. Second, the factor acts in the same way at every location: it adds its variance to the covariance of every pair of locations, so that, with the nonnegative correlation functions used in practice, it cannot represent a factor whose strength, shape or sign changes over space, such as a weather system that brings heavy precipitation to one region and dry conditions to another. Gaussian scale mixtures \citep{huser2017bridging} keep the correlation of the GP and treat both tails symmetrically, and share this second limitation.


We propose the Bayesian neural factor copula (BNFC), which removes both limitations while keeping the one-dimensional integrals that make factor copulas tractable. The factor is a neural network $g(V,\bfs)$ of a single uniform random variable $V$ and of the location $\bfs\in\mathcal{D}\subset\mathbb{R}^d$, with a stationary version $g(V)$ as a special case. Since applying the quantile function of a univariate distribution to a
standard uniform random variable produces a draw from that
distribution \citep[Chapter~2]{devroye1986nonuniform}, this motivates
representing the factor as a neural transformation of a uniform random
variable. We estimate the transformation from the data without
prespecifying a parametric family for the factor distribution. In the
location-dependent model, the transformation can vary across space,
allowing the same realization of the uniform variable to raise the
latent process in one region and lower it in another, so that the strength, shape and sign of the large-scale dependence may all change over space. 
By allowing the factor's effect to vary across space, BNFC can
represent a weather system that brings heavy precipitation to one
region and dry conditions to another. Conditional on the factor, the
latent process is Gaussian, so standard Gaussian kriging formulas
apply. Our contributions are as follows.
\begin{enumerate}[label=(\roman*),leftmargin=*]
\item \emph{A spatially varying, learned factor.} BNFC can approximate any factor distribution with interval support, and any location-specific family of such distributions, and its location-dependent factor can induce any positive semidefinite covariance matrix. It therefore admits long-range dependence of either sign whose strength varies over space, which no common factor copula and no Gaussian scale mixture with a nonnegative correlation function can produce.
\item \emph{Tail behavior.} We characterize the tail dependence of the resulting copulas. A factor acting in opposite directions at two locations links the upper tail at one with the lower tail at the other. Every finite network gives asymptotic independence, but it can match the finite-level tail dependence of any target factor arbitrarily closely, and these finite levels are what a finite number of replicates can inform.
\item \emph{Scalable  Bayesian inference.} The network factor is overparameterized and not identifiable, so a single plug-in fit ignores which of many equally good networks is correct. We combine a Monte Carlo likelihood that is itself a valid copula density, variational Bayes for the factor and GP parameters, and stochastic gradient MCMC for the network weights. The result is posterior uncertainty for interpretable dependence summaries, predictive distributions that average over the parameters, and the posterior of the factor in each replicate. With the Vecchia approximation \citep{vecchia1988estimation,katzfuss2021general}, the cost per epoch is linear in the number of replicates and locations; a fit to more than $10^7$ precipitation values takes under an hour on one GPU.
\item \emph{Empirical evidence.} In simulations, BNFC matches the data-generating model under a true common gamma factor. When the effect of the factor varies over space, it has the best held-out log score,  CRPS and energy score, and near-nominal coverage. BNFC also outperforms the Gaussian and gamma factor copulas in the  simulations with spatially varying factor effects. Here, the
location-dependent BNFC closely approximates the posterior distribution
of the centered factor contribution under the corresponding
data-generating model. Across all scenarios, it
achieves a better mean held-out copula log score at the observed
locations than the nonparametric Bayesian transport map
\citep{katzfuss2024scalable}. For daily ERA5 precipitation over central Europe, BNFC has the best held-out log score, CRPS, energy score, and MAE of all copula models on $200$ test days in 2024, including the extreme rainfall of September 2024. Its factor reveals an interpretable structure: the share of the factor-induced variance ranges from $14\%$ to $61\%$ and follows the relief, and the skewness of the factor contribution changes its sign between the North European Plain and the Pannonian Basin.
\end{enumerate}


Section~\ref{sec:background} reviews factor copulas, Section~\ref{sec:methodology} introduces BNFC and its properties, Section~\ref{sec:estimation} describes estimation and prediction, Section~\ref{sec:empirical} reports the simulations and the application, and Section~\ref{sec:discussion} concludes. The Supplementary Material contains additional discussions, proofs, and algorithms, and results related to the simulation studies and data application. 

\section{Factor Copulas for Spatial Data}\label{sec:background}

We use the following notation throughout. $\Phi(\cdot)$ and $\phi(\cdot)$ denote the distribution function and density of the univariate standard normal distribution, and $\Phi_{\bs{\Sigma}}(\cdot)$ and $\phi_{\bs{\Sigma}}(\cdot)$ denote the distribution function and density of a multivariate normal distribution with mean zero and covariance matrix $\bs{\Sigma}$. For a distribution function $H$ on $\mathbb{R}$, its generalized inverse is $H^{-1}(p)=\inf\{x\in\mathbb{R}:H(x)\ge p\}$, $p\in(0,1)$.

\subsection{Copulas for spatial data}\label{sec:copulas_spatial}

Copulas separate the univariate marginal distributions from the dependence structure \citep{nelsen2006introduction,joe2014dependence}. By Sklar's theorem, a continuous joint distribution function $F$ with continuous margins $F_1,\ldots,F_n$ can be written as $F(y_1,\ldots,y_n)=C\{F_1(y_1),\ldots,F_n(y_n)\}$ for a unique copula $C$, and inference is typically carried out in two steps: the margins are estimated first, and a copula is then fitted to the data transformed to the uniform scale. Exchangeable, extreme-value and vine copulas are ill suited to spatial data, because they ignore the decay of dependence with distance or are hard to evaluate, interpret or fit in high dimensions, whereas implicit copulas, the copulas of latent processes with a tractable structure, avoid these difficulties \citep{smith2023implicitcopulas}; factor copulas belong to this class.

\subsection{The common factor copula}\label{sec:classicalfactorcopula}

Consider a real-valued spatial random field $\{W(\bfs),\bfs\in D\}$, $D\subseteq\mathbb{R}^d$, observed at locations $\bfs_1,\ldots,\bfs_n\in D$. Write $W_j=W(\bfs_j)$ and $Z_j=Z(\bfs_j)$, $j=1,\ldots,n$, and let $\bfW=(W_1,\ldots,W_n)^\top$ and $\bfZ=(Z_1,\ldots,Z_n)^\top$. The finite-dimensional representation of the common factor model of \citet{krupskii2018factor} is
\begin{equation}
W_j = Z_j + V_0, \qquad j=1,\ldots,n, \qquad \bfZ \sim \mathcal{N}_n(\bs{0}_n,\bs{\Sigma}_Z),
\label{eq:classical-factor-copula}
\end{equation}
where $\bs{\Sigma}_Z$ is a correlation matrix and $V_0\sim F_{V_0}$ is a common factor independent of $\bfZ$. The matrix $\bs{\Sigma}_Z$ is parameterized by a spatial correlation function, such as the exponential, squared exponential or Mat\'ern correlation function \citep{matern1986spatial,cressie1993statistics,stein1999interpolation}, so that $Z(\bfs)$ captures spatial correlation that decays with distance, while $V_0$ induces additional dependence shared by all locations. If $\sigma_0^2=\operatorname{var}(V_0)<\infty$, then $\operatorname{cor}(W_j,W_k)=\{(\bs{\Sigma}_Z)_{jk}+\sigma_0^2\}/(1+\sigma_0^2)$, so that, whatever its distribution, the factor adds the same nonnegative covariance $\sigma_0^2$ to every pair of locations.

Conditional on $V_0=v_0$, $\bfW$ is Gaussian with mean $v_0\mathbf{1}_n$, where $\mathbf{1}_n$ is an $n$-dimensional vector of ones. Hence the joint distribution function and density of $\bfW$ are
\begin{equation}
F_n^W(\bfw) = \int_{-\infty}^{\infty} \Phi_{\bs{\Sigma}_Z}(\bfw-v_0\mathbf{1}_n)\,dF_{V_0}(v_0),
\qquad
f_n^W(\bfw) = \int_{-\infty}^{\infty} \phi_{\bs{\Sigma}_Z}(\bfw-v_0\mathbf{1}_n)\,dF_{V_0}(v_0),
\label{eq:classical-factor-cdf}
\end{equation}
and the common marginal distribution function is $F_1^W(w)=\int\Phi(w-v_0)\,dF_{V_0}(v_0)$. The implied copula and its density are
\begin{equation}
\begin{gathered}
C_n^W(\bfu)=F_n^W\{(F_1^W)^{-1}(u_1),\ldots,(F_1^W)^{-1}(u_n)\},
\\
c_n^W(\bfu)=\frac{f_n^W\{(F_1^W)^{-1}(u_1),\ldots,(F_1^W)^{-1}(u_n)\}}{\prod_{j=1}^n f_1^W\{(F_1^W)^{-1}(u_j)\}},
\end{gathered}
\label{eq:classical-factor-copula-cdf}
\end{equation}
where $\bfu=(u_1,\ldots,u_n)^\top\in(0,1)^n$. The distribution $F_1^W$ is only used to construct the copula, which can then be combined with arbitrary univariate margins for the data.

The distribution of $V_0$ controls the tail behavior of $C_n^W$. If $V_0$ is Gaussian, the resulting copula remains Gaussian. For a pair of locations with Gaussian correlation $\rho<1$, a Weibull
common factor with shape parameter greater than one gives upper tail
independence ($\lambda_U=0$). An exponential factor gives intermediate
upper tail dependence ($0<\lambda_U<1$), and a Pareto factor gives
perfect upper tail dependence ($\lambda_U=1$). The model can also
produce tail asymmetry. For example, taking the common factor to be
the difference of two independent exponential variables with different
rates gives $\lambda_L\ne\lambda_U$ \citep{krupskii2018factor}.
The model has three limitations that motivate our construction. First, the family of $V_0$ must be chosen a priori, and this choice determines the tail behavior of the copula. Second, $F_1^W$ is available in closed form only for a few families, such as the exponential factor. Third, $V_0$ enters every location with the same loading, so that the factor effect cannot vary over space. A more flexible factor family, such as a finite mixture, addresses only the first limitation.

\subsection{Related approaches}\label{sec:related}

In a Gaussian scale mixture, such as the Student-$t$ copula \citep{demarta2005t} or the family of \citet{huser2017bridging}, a common positive random variable $R$ multiplies the Gaussian process, $W(\bfs)=R\,Z(\bfs)$. Its copula is reflection symmetric, and with a nonnegative correlation function its long-range dependence is nonnegative and the same wherever the locations are. Neural copulas \citep{tagasovska2019copulas,ling2020deep} and neural density models \citep{rezende2015variational,durkan2019neural} are flexible but not designed for spatial data: they neither respect the decay of dependence with distance nor provide predictions at unobserved locations. BNFC uses a neural network only for the one-dimensional factor and keeps the Gaussian process for the spatial structure. Section~\ref{S-app:related} of the Supplementary Material discusses these approaches in more detail, and Table~\ref{S-tab:positioning} there summarizes the dependence properties of BNFC and of the other models of this paper.

\section{The Bayesian Neural Factor Copula Model}\label{sec:methodology}

Section~\ref{subsec:neural_factors} introduces the neural factor construction and its approximation properties, and Section~\ref{subsec:neuralfactorcopula} defines the resulting copula for spatial data. Section~\ref{sec:tail_properties} studies its tail properties.

\subsection{Neural factors}\label{subsec:neural_factors}

We replace the parametric common factor $V_0$ by a neural transformation of a univariate uniform random variable. For $j=1,\ldots,n$, let
\begin{equation}
W_j = Z_j + g(V), \qquad \bfZ\sim \mathcal{N}_n(\bs{0}_n,\bs{\Sigma}_{\bs{\vartheta}}), \qquad V\sim U(\theta_1,\theta_2), \qquad 0<\theta_1<\theta_2,
\label{eq:model_definition}
\end{equation}
where $V$ is independent of $\bfZ$, $g:(\theta_1,\theta_2)\to\mathbb{R}$ is a feedforward neural network with weights $\bs{\varphi}$, and $\bftheta=(\theta_1,\theta_2)^\top$. The uniform variable $V$ is a simple base random variable, and $g$ transforms it into a flexible factor $g(V)$. In contrast to the common factor model, whose factor family must be specified a priori, the distribution of $g(V)$ is learned from the data. The following result shows that the construction can produce, and a neural network can approximate, any factor distribution whose support is an interval.

\begin{proposition}[Factor generation and approximation]
\label{prop:factor_generation}
Let $H$ be a distribution function on $\mathbb{R}$, let $V\sim U(\theta_1,\theta_2)$, and let $T_{\bftheta}(v)=H^{-1}\{(v-\theta_1)/(\theta_2-\theta_1)\}$ for $v\in(\theta_1,\theta_2)$.
\begin{itemize}
\item[(a)] $T_{\bftheta}$ is nondecreasing and $T_{\bftheta}(V)\sim H$.
\item[(b)] If $T_{\bftheta}$ is continuous on $(\theta_1,\theta_2)$, there exist feedforward neural networks $g_m$, $m\ge1$, with one hidden layer with $m$ units and a continuous nonpolynomial activation function, such that $\sup_{v\in K}|g_m(v)-T_{\bftheta}(v)|\to0$ as $m\to\infty$ for every compact set $K\subset(\theta_1,\theta_2)$. If, in addition, $H$ has bounded support, the convergence is uniform on $[\theta_1,\theta_2]$.
\item[(c)] If measurable functions $g_m$ converge to $T_{\bftheta}$ uniformly on every compact subset of $(\theta_1,\theta_2)$,  $g_m(V)\to T_{\bftheta}(V)$ almost surely, so that the law $H_m$ of $g_m(V)$ converges weakly to $H$.
\end{itemize}
\end{proposition}
Part (a) shows that the quantile transformation $T_{\bftheta}(V)$ has
distribution $H$. Thus, taking $g=T_{\bftheta}$ in
\eqref{eq:model_definition} gives the common factor model
\eqref{eq:classical-factor-copula} with $V_0\sim H$.
Parts (b) and (c) show that, when $T_{\bftheta}$ is continuous,
neural networks $g_m$ of increasing width can approximate
$T_{\bftheta}$ and that the distributions of $g_m(V)$ converge weakly
to $H$. If $H$ has unbounded support, $T_{\bftheta}$ diverges at one
or both endpoints of $(\theta_1,\theta_2)$. Uniform approximation
on compact subsets of this interval is sufficient for weak
convergence. Since a distribution with interval support has a
continuous quantile function on $(0,1)$, this result covers any
such factor distribution, including heavy-tailed and asymmetric
distributions.
Proofs are given in Sections~\ref{S-app:proofs_neural_factors}
and~\ref{S-app:tail_proofs} of the Supplementary Material.


Model \eqref{eq:model_definition}  assumes that the factor affects all locations in the same way. To allow the factor effect to vary over space, we generalize it to
\begin{equation}
W(\bfs) = Z(\bfs) + g(V,\bfs), \qquad V\sim U(\theta_1,\theta_2),
\label{eq:model_definition_varying_locs}
\end{equation}
where $g$ is a neural network that takes the factor value and the location as inputs. All locations share the same  variable $V$, but its effect is transformed differently at each location. The following lemma extends Proposition~\ref{prop:factor_generation}(a) to location-specific factor distributions.

\begin{lemma}[Location-specific factor generation]
\label{lemma:location_specific_factor}
For each $\bfs\in D$, let $H(\cdot,\bfs)$ be a distribution function on $\mathbb{R}$, let $V\sim U(\theta_1,\theta_2)$, and let $T_{\bftheta}(v,\bfs)=H^{-1}\{(v-\theta_1)/(\theta_2-\theta_1),\bfs\}$, where $H^{-1}(\cdot,\bfs)$ is the generalized inverse of $H(\cdot,\bfs)$. Then $T_{\bftheta}(V,\bfs)\sim H(\cdot,\bfs)$ for every $\bfs\in D$, and, for any $\bfs_1,\ldots,\bfs_n\in D$, the random vector $\{T_{\bftheta}(V,\bfs_1),\ldots,T_{\bftheta}(V,\bfs_n)\}^\top$ is comonotonic.
\end{lemma}

By Lemma~\ref{lemma:location_specific_factor}, the factor contribution in \eqref{eq:model_definition_varying_locs} can have a different distribution at each location, and the approximation results of Proposition~\ref{prop:factor_generation} apply at each location. Since $g$ needs not be monotone in $v$, the factor can also raise the process at some locations and lower it at others. Since $V$ is independent of the GP,
$
\operatorname{cov}(W_j,W_k)
=
(\bs{\Sigma}_{\bs{\vartheta}})_{jk}
+
\operatorname{cov}\{g(V,\bfs_j),g(V,\bfs_k)\}.
$
The factor contribution can therefore change the covariance between
locations in either direction. As the next proposition shows, at any
finite set of distinct locations, the covariance matrix of the factor
contribution can approximate any positive semidefinite matrix.

\begin{proposition}[Attainable covariance structures]
\label{thm:covariance}
Let $\bs{\Sigma}_{\bs{\vartheta}}$ be an $n\times n$ correlation matrix, let $\bs{C}$ be any $n\times n$ positive semidefinite matrix, and let $V\sim U(\theta_1,\theta_2)$. Then there exist continuous functions $g(\cdot,\bfs_1),\ldots,g(\cdot,\bfs_n)$ on $[\theta_1,\theta_2]$ such that $\bs{G}(V)=\{g(V,\bfs_1),\ldots,g(V,\bfs_n)\}^\top$ satisfies $\operatorname{cov}\{\bs{G}(V)\}=\bs{C}$. Hence, $\bfW=\bfZ+\bs{G}(V)$, with $\bfZ\sim\mathcal{N}_n(\bs{0}_n,\bs{\Sigma}_{\bs{\vartheta}})$ independent of $V$, has $\operatorname{cov}(\bfW)=\bs{\Sigma}_{\bs{\vartheta}}+\bs{C}$. In particular, if $C_{jj}=C_{kk}=\sigma^2$, then $\operatorname{cor}(W_j,W_k)$ can take any value in
\[
\left[\frac{(\bs{\Sigma}_{\bs{\vartheta}})_{jk}-\sigma^2}{1+\sigma^2},\ \frac{(\bs{\Sigma}_{\bs{\vartheta}})_{jk}+\sigma^2}{1+\sigma^2}\right],
\]
which tends to $[-1,1]$ as $\sigma^2\to\infty$. The functions can be approximated uniformly on $[\theta_1,\theta_2]$ by feedforward neural networks, so that the covariance matrix of the factor of a BNFC model can be made arbitrarily close to $\bs{C}$.
\end{proposition}

{In the common factor model, the factor adds the same nonnegative covariance $\sigma_0^2$ to every pair of locations, and in a Gaussian scale mixture with finite second moments, the latent
process $W$ inherits the correlation function of the Gaussian process
$Z$. If this correlation function is nonnegative and isotropic, then the correlation is nonnegative and depends only on the distance between
the two locations. In the location-dependent BNFC model, the factor can induce long-range dependence of either sign whose strength varies over space, such as that of a weather system that brings heavy rainfall to one region and dry conditions to another. Figure~\ref{S-fig:covariance_plot} in the Supplementary Material
illustrates how the factor can induce positive dependence between
some pairs of locations and negative dependence between others.
The sign-change simulation in Section~\ref{sec:simulation_studies}
shows that BNFC can recover this pattern from data. Section~\ref{S-app:three_properties} of the Supplementary Material discusses these properties in more detail.


Model \eqref{eq:model_definition} is the special case in which $g(v,\bfs)$ does not depend on $\bfs$. We refer to the copula model based on \eqref{eq:model_definition} or \eqref{eq:model_definition_varying_locs}, fitted with the Bayesian procedure of Section~\ref{sec:estimation}, as the Bayesian Neural Factor Copula (BNFC) model, and we call \eqref{eq:model_definition} and \eqref{eq:model_definition_varying_locs} the stationary and location-dependent BNFC, written BNFC (stat) and BNFC (nonstat). Here, stationarity refers to the factor: in BNFC (stat), the law of the factor contribution is the same at every location, and the copula is stationary when the correlation function of the Gaussian component is stationary, as for the squared exponential function used here. The margins of the data are modeled separately (Section~\ref{subsec:marginals}) and may differ between locations in both models.}

\begin{remark}[Identifiability]
\label{rem:identifiability}
The parameters $\bftheta$, the network $g$ and the location of the factor are not identifiable: an affine map of the interval $(\theta_1,\theta_2)$, or any other map that preserves the uniform distribution on it, changes $(\bftheta,g)$ but not the process, and adding a function $c(\bfs)$ to $g(v,\bfs)$ leaves the copula unchanged. At each location, we center the factor contribution by subtracting
its mean:
\[
\tilde g(V,\bfs)=g(V,\bfs)-\mathbb{E}\{g(V,\bfs)\}.
\]
The transformations above preserve the joint distribution of the
centered factor contributions across locations. Together with the
Gaussian covariance matrix $\bs{\Sigma}_{\bs{\vartheta}}$, this joint
distribution determines the copula.

The transformations also preserve the conditional distribution of
the centered factor contributions given the observed copula data for
a replicate. This allows us to compare the conditional distributions
from different fitted models with the one under the data-generating
model in Section~\ref{sec:simulation_studies}. We therefore interpret the distribution of the centered factor contribution and derived dependence summaries, such as $\lambda(q)$ in \eqref{eq:lambda_q} and the skewness measure $\zeta_1$ of Proposition~\ref{prop:subasymptotic_convergence}, but not $g$, $\bftheta$ or the location of the factor. Section~\ref{S-app:identifiability} of the Supplementary Material gives the details.
\end{remark}

\begin{remark}[Implementation]
\label{rem:implementation}
The factor network  is a multilayer perceptron, whose architecture is given in Section~\ref{sec:empirical}, and its output layer is chosen to suit the application. A linear output layer leaves the factor unrestricted, whereas a positive output activation gives $g\ge0$. By Remark~\ref{rem:identifiability}, such a restriction does not affect the copula through the location of the factor and only matters through its lower tail, see Section~\ref{sec:tail_properties}.

The network $g$ is not constrained to be monotone in $v$; Section~\ref{S-app:monotone} of the Supplementary Material explains why a monotonicity constraint would not change the stationary model and would exclude the negative dependence of Proposition~\ref{thm:covariance} in the location-dependent model.
\end{remark}

\subsection{The neural factor copula for spatial data}\label{subsec:neuralfactorcopula}

The finite-dimensional restriction of the BNFC model to $\bfs_1,\ldots,\bfs_n$ is
\begin{equation}
\bfW = \bfZ+\bs{G}(V), \qquad \bfZ\sim \mathcal{N}_n(\bs{0}_n,\bs{\Sigma}_{\bs{\vartheta}}), \qquad V\sim U(\theta_1,\theta_2),
\label{eq:bnfc_finite_model}
\end{equation}
where $\bs{G}(v)=\{g(v,\bfs_1),\ldots,g(v,\bfs_n)\}^\top$ in the location-dependent model, and $\bs{G}(v)=g(v)\mathbf{1}_n$ in the stationary model. The matrix $\bs{\Sigma}_{\bs{\vartheta}}$ is a correlation matrix indexed by $\bs{\vartheta}$. In this paper we use the squared exponential correlation function for the simulation studies and exponential correlation in the application. Here $\ell>0$ is the range and $\tau^2\in[0,1)$ the nugget, and $\bs{\vartheta}=(\ell,\tau^2)^\top$.

Conditional on $V=v$, $\bfW$ is Gaussian with mean $\bs{G}(v)$ and covariance matrix $\bs{\Sigma}_{\bs{\vartheta}}$. Integrating out $V$ gives the joint distribution function and density
\begin{equation}
F_n(\bfw) = \frac{1}{\theta_2-\theta_1}\int_{\theta_1}^{\theta_2} \Phi_{\bs{\Sigma}_{\bs{\vartheta}}}\{\bfw-\bs{G}(v)\}\,dv,
\label{eq:bnfc_marginal_cdf}
\end{equation}
\begin{equation}
f_n(\bfw) = \frac{1}{\theta_2-\theta_1}\int_{\theta_1}^{\theta_2} \phi_{\bs{\Sigma}_{\bs{\vartheta}}}\{\bfw-\bs{G}(v)\}\,dv,
\label{eq:bnfc_marginal_pdf}
\end{equation}
and the marginal distribution function and density of $W_j$,
\begin{equation}
F_{1,j}(w) = \frac{1}{\theta_2-\theta_1}\int_{\theta_1}^{\theta_2} \Phi\{w-g(v,\bfs_j)\}\,dv,
\qquad
f_{1,j}(w) = \frac{1}{\theta_2-\theta_1}\int_{\theta_1}^{\theta_2} \phi\{w-g(v,\bfs_j)\}\,dv,
\label{eq:bnfc_univariate_margin}
\end{equation}
where we have used the fact that $(\bs{\Sigma}_{\bs{\vartheta}})_{jj}=1$. In the stationary model, $F_{1,j}=F_1$ does not depend on $j$. Each $F_{1,j}$ is a mixture of normal distribution functions, hence continuous and strictly increasing. The BNFC copula and its density are
\begin{equation}
C_n(\bfu) = F_n\left\{F_{1,1}^{-1}(u_1),\ldots,F_{1,n}^{-1}(u_n)\right\},
\label{eq:bnfc_copula_cdf}
\end{equation}
\begin{equation}
c_n(\bfu) = \frac{ f_n\left\{F_{1,1}^{-1}(u_1),\ldots,F_{1,n}^{-1}(u_n)\right\}}{\prod_{j=1}^n f_{1,j}\left\{F_{1,j}^{-1}(u_j)\right\}},
\label{eq:bnfc_copula_density}
\end{equation}
for $\bfu\in(0,1)^n$. When we need to emphasize the dependence on the parameters, we write $c_n(\bfu\mid\bftheta,\bs{\varphi},\chk{\bs{\vartheta}})$, and similarly for $F_{1,j}$ and $f_{1,j}$\chk{, which do not depend on $\bs{\vartheta}$ because the Gaussian component has unit variance}. As in the common factor model, the copula density \eqref{eq:bnfc_copula_density} only requires one-dimensional integrals over $v$, whatever the number of locations. These integrals and the inverses $F_{1,j}^{-1}$ have no closed form, and Section~\ref{subsec:likelihood} describes how we evaluate them.

\subsection{Tail properties}\label{sec:tail_properties}

We now study the tail dependence of the BNFC copula. Detailed proofs are provided in Section~\ref{S-app:tail_proofs} of the Supplementary Material. For two locations $\bfs_i\neq\bfs_j$, write $W_i=Z_i+g(V,\bfs_i)$ and $W_j=Z_j+g(V,\bfs_j)$, let $\rho=\operatorname{cor}(Z_i,Z_j)<1$, and let $U_i=F_{1,i}(W_i)$ and $U_j=F_{1,j}(W_j)$. The lower and upper tail dependence coefficients of the bivariate copula $C_2$ of $(W_i,W_j)$ are
\[
\lambda_L^{(i,j)}=\lim_{q\to0}\frac{C_2(q,q)}{q}=\lim_{q\to 0}\Pr(U_i\le q\mid U_j\le q),
\qquad
\lambda_U^{(i,j)}=\lim_{q\to 0}\Pr(U_i>1-q\mid U_j>1-q),
\]
whenever these limits exist. The copula has lower or upper tail dependence if $\lambda_L^{(i,j)}>0$ or $\lambda_U^{(i,j)}>0$, respectively. For the Gaussian copula with $\rho<1$, $\lambda_L^{(i,j)}=\lambda_U^{(i,j)}=0$. Because these coefficients are limits, it is also useful to consider their values at a fixed level $q\in(0,1/2]$,
\begin{equation}
\lambda_L^{(i,j)}(q)=\frac{C_2(q,q)}{q},
\qquad
\lambda_U^{(i,j)}(q)=\frac{1-2(1-q)+C_2(1-q,1-q)}{q},
\label{eq:lambda_q}
\end{equation}
which describe the dependence at finite but extreme levels.

The tail behavior of the BNFC copula is determined by the factor distribution. Recall that a random variable $X$ is (right) regularly varying with index $\xi>0$ if $\Pr(X>x)=x^{-\xi}L(x)$ with $L$ slowly varying, that is, $L(tx)/L(x)\to1$ as $x\to\infty$ for all $t>0$. Adding an independent Gaussian variable does not change the leading-order tail of a regularly varying variable (Lemma~\ref{S-lemma:gaussian_invariance} in Section~\ref{S-app:gaussian_perturbation} of the Supplementary Material), so that a heavy-tailed factor dominates the Gaussian component in the joint tail.

\begin{proposition}[Heavy-tailed factors]
\label{thm:positive_tail_dependence}
Suppose that there exist a random variable $X=X(V)$ and constants $a_i,a_j>0$ and $b_i,b_j\in\mathbb{R}$ such that $g(V,\bfs_i)=a_iX+b_i$ and $g(V,\bfs_j)=a_jX+b_j$, and that $X$ is regularly varying with index $\xi>0$.
\begin{itemize}
\item[(a)] The copula is upper tail dependent with $\lambda_U^{(i,j)}=1$.
\item[(b)] If, in addition, $X$ is bounded below,  $\lambda_L^{(i,j)}=0$, so that the copula is tail asymmetric.
\end{itemize}
\end{proposition}

Proposition~\ref{thm:positive_tail_dependence} covers the stationary model \eqref{eq:model_definition} with a regularly varying factor, and part (a) agrees with the known result for power-law common factors. Intermediate and asymmetric values, $0<\lambda_L^{(i,j)}\neq\lambda_U^{(i,j)}<1$, arise when the two tails of the factor decay exponentially at different rates (Section~\ref{S-app:tail_discussion} of the Supplementary Material). When the factor acts in opposite directions at two locations, the joint tail is different.

\begin{cor}[Opposite loadings]
\label{cor:opposite_loadings}
Suppose that $g(V,\bfs_i)=a_iX+b_i$ and $g(V,\bfs_j)=-a_jX+b_j$ with $a_i,a_j>0$ and $b_i,b_j\in\mathbb{R}$, where $X=X(V)$ is regularly varying with index $\xi>0$ and bounded below, and that $\rho>-1$. Then
\begin{itemize}
\item[(a)] $\lim_{q\to0}\Pr(U_j\le q\mid U_i>1-q)=1$, so that an extreme high value at $\bfs_i$ occurs together with an extreme low value at $\bfs_j$;
\item[(b)] $\lambda_U^{(i,j)}=0$, although the factor is heavy tailed.
\end{itemize}
\end{cor}

This joint behavior of extremes is not available in the common factor copulas or in Gaussian scale mixtures with nonnegative correlation, in which a large value of the common variable raises the process at every location.

\begin{proposition}[Bounded factors and finite networks]
\label{thm:asymptotic_independence}
\begin{itemize}
\item[(a)] If $|g(V,\bfs_i)|$ and $|g(V,\bfs_j)|$ are bounded almost surely by a constant $c<\infty$, then $\lambda_L^{(i,j)}=\lambda_U^{(i,j)}=0$.
\item[(b)] In particular, for any feedforward neural network $g$ with continuous activation functions and any $0<\theta_1<\theta_2$, the bivariate copulas of models \eqref{eq:model_definition} and \eqref{eq:model_definition_varying_locs} satisfy $\lambda_L^{(i,j)}=\lambda_U^{(i,j)}=0$ for all $\bfs_i\neq\bfs_j$ with $\rho<1$.
\end{itemize}
\end{proposition}

Part (b) holds because a network with continuous activations is bounded on the bounded interval $(\theta_1,\theta_2)$. Every fitted BNFC copula is therefore asymptotically independent, but a finite network can reproduce the dependence of a tail-dependent copula at any fixed level, which is what matters for a data set of finite size. Let $T_{\bftheta}(v,\bfs)$ be a target factor map, and let $C_n^{(m)}$ and $C_n$ denote the copulas of $\{W^{(m)}(\bfs_1),\ldots,W^{(m)}(\bfs_n)\}$ and $\{W(\bfs_1),\ldots,W(\bfs_n)\}$, where $W^{(m)}(\bfs)=Z(\bfs)+g_m(V,\bfs)$ and $W(\bfs)=Z(\bfs)+T_{\bftheta}(V,\bfs)$.

\begin{proposition}[Convergence of the copula and of finite-level tail dependence]
\label{prop:subasymptotic_convergence}
Suppose that $g_m(\cdot,\bfs_k)\to T_{\bftheta}(\cdot,\bfs_k)$ uniformly on compact subsets of $(\theta_1,\theta_2)$ for $k=1,\ldots,n$. Then $C_n^{(m)}\to C_n$ uniformly on $[0,1]^n$. In particular, for every fixed $q\in(0,1/2]$, $\lambda_L^{(i,j)}(q)$ and $\lambda_U^{(i,j)}(q)$ in \eqref{eq:lambda_q} converge to their values under $C_n$, and so does the skewness measure $\zeta_1=\mathbb{E}\{(U_i+U_j-1)^3\}$ \citep{rosco2013measures}.
\end{proposition}
By Proposition~\ref{prop:factor_generation}(b), Proposition~\ref{prop:subasymptotic_convergence} applies whenever $T_{\bftheta}$ is continuous in $v$.

{\paragraph{Summary of the tail properties.} (i) An unbounded, regularly varying factor that acts in the same direction at two locations gives $\lambda_U=1$ and, if it is bounded below, $\lambda_L=0$ (Proposition~\ref{thm:positive_tail_dependence}). (ii) If it acts in opposite directions, it links the upper tail at one location with the lower tail at the other (Corollary~\ref{cor:opposite_loadings}). (iii) A bounded factor, and hence every finite network, gives $\lambda_L=\lambda_U=0$ (Proposition~\ref{thm:asymptotic_independence}). (iv) For any fixed tail probability $q\in(0,1/2]$, a neural network with
enough hidden units can approximate the target copula's lower and
upper tail dependence measures $\lambda_L^{(i,j)}(q)$ and
$\lambda_U^{(i,j)}(q)$ arbitrarily closely
(Proposition~\ref{prop:subasymptotic_convergence}), and these are the coefficients that $N$ replicates can inform. The output layer of $g$ can be chosen accordingly: by Proposition~\ref{thm:positive_tail_dependence}(b), a positive output activation suits data with stronger dependence in the upper than in the lower tail, whereas a linear output layer leaves both tails unrestricted.}


Section~\ref{S-app:zeta} of the Supplementary Material examines
$100$ factor networks $g(v,\bfs)$ with randomly drawn weights.
For the same pair of sites, the estimated asymmetry measure $\zeta_1$
ranges from $-0.007$ to $0.018$, illustrating asymmetry in both
directions. For the network with the largest $\zeta_1$, the estimated
upper and lower tail dependence measures at $q=0.05$ are $0.60$ and
$0.29$, respectively. Figure~\ref{S-fig:covariance_plot} shows that
selected networks can induce positive correlation between some
distant sites and negative correlation between others. The resulting
patterns depend on the locations of the sites as well as their
separation distance. In the additive common-factor copula with an
isotropic Gaussian correlation function, pairs at the same distance
have the same copula-scale correlation.

\section{Parameter Estimation and Spatial Interpolation}\label{sec:estimation}

We observe $N$ independent replicates $\bs{y}_i=(y_{i1},\ldots,y_{in})^\top$, $i=1,\ldots,N$, of a spatial process $Y(\bfs)$ at locations $\bfs_1,\ldots,\bfs_n$, with arbitrary marginal distributions. We assume that the dependence structure of $\bs{y}_i$ is the BNFC copula \eqref{eq:bnfc_copula_density}, and that $Y(\bfs)$ has marginal distribution function $G(\cdot\,;\bfs)$ at location $\bfs$. We estimate the model in two steps. In the first step, we transform the data to the copula scale (Section~\ref{subsec:marginals}). In the second step, we fit the copula by variational Bayes and explore the network weights by stochastic gradient MCMC (Sections~\ref{subsec:likelihood} and~\ref{subsec:vb_inference}). Section~\ref{subsec:cond_copula} derives the conditional distributions used for spatial interpolation.
The marginal distribution $G(\cdot\,;\bfs)$ of the data links the data and copula scales, whereas the marginal distribution $F_{1,j}$ of the latent variable in \eqref{eq:bnfc_univariate_margin} is part of the copula, depends on $(\bftheta,\bs{\varphi})$, and links the copula and latent scales.

\subsection{Marginal distributions}\label{subsec:marginals}

{In the simulations, the probability integral transforms $u_{ij}=G(y_{ij};\bfs_j)$ under the data-generating model are known and are passed directly to the copula fit. In the application, the copula data are rank-based scores $\hat u_{ij}$, computed among $y_{1j},\ldots,y_{Nj}$ at each location (Section~\ref{sec:application}). Spatial prediction on the data scale also requires the marginal quantile function at unobserved locations. We use a neural network
$T_{\bfpsi}(u,\bfs)$ to approximate the marginal quantile function
$G^{-1}(u;\bfs)$. The network is strictly increasing in $u$ by
construction and is fitted by least squares using
$(\hat u_{ij},\bfs_j)$ as inputs and $y_{ij}$ as the response.
We fit this network separately from the copula and use it to map
predictive draws to the data scale
(Section~\ref{S-app:T_psi} of the Supplementary Material).


\subsection{The copula likelihood}\label{subsec:likelihood}

Given the copula data $\bfu_i=(u_{i1},\ldots,u_{in})^\top$ for $i=1,...,N$, the copula log-likelihood is
\begin{equation}
\ell(\bftheta,\bs{\varphi},\bs{\vartheta})=\sum_{i=1}^N\log c_n(\bfu_i\mid\bftheta,\bs{\varphi},\bs{\vartheta})
=\sum_{i=1}^N\left[\log f_n(\bfw_i)-\sum_{j=1}^n\log f_{1,j}(w_{ij})\right],
\qquad w_{ij}=F_{1,j}^{-1}(u_{ij}),
\label{eq:copula_loglik}
\end{equation}
where $\bfw_i=(w_{i1},\ldots,w_{in})^\top$, and $\bs{\vartheta}$ enters only through the joint density $f_n$. Evaluating \eqref{eq:copula_loglik} requires $f_n$, $f_{1,j}$ and $F_{1,j}^{-1}$, all of which involve the one-dimensional integrals in \eqref{eq:bnfc_marginal_pdf} and \eqref{eq:bnfc_univariate_margin}. For an exponential factor, these integrals are available in closed form, but for a neural factor they are not, so we approximate them by Monte Carlo integration.

Write $V=\theta_1+(\theta_2-\theta_1)\varepsilon$ with $\varepsilon\sim U(0,1)$, draw $\varepsilon^{(1)},\ldots,\varepsilon^{(M_1)}\sim U(0,1)$ independently, and let $v^{(m)}=\theta_1+(\theta_2-\theta_1)\varepsilon^{(m)}$. We use the same draws for all three quantities,
\begin{equation}
\begin{gathered}
\hat{F}_{1,j}(w)=\frac{1}{M_1}\sum_{m=1}^{M_1}\Phi\{w-g(v^{(m)},\bfs_j)\},
\qquad
\hat{f}_{1,j}(w)=\frac{1}{M_1}\sum_{m=1}^{M_1}\phi\{w-g(v^{(m)},\bfs_j)\},
\\
\hat{f}_{n}(\bfw)=\frac{1}{M_1}\sum_{m=1}^{M_1}\phi_{\bs{\Sigma}_{\bs{\vartheta}}}\{\bfw-\bs{G}(v^{(m)})\}.
\end{gathered}
\label{eq:mc_estimators}
\end{equation}
Fresh draws are generated at every evaluation of the likelihood. For given draws, $\hat{F}_{1,j}$, $\hat{f}_{1,j}$ and $\hat{f}_n$ are exactly the margins and the joint density of model \eqref{eq:bnfc_finite_model} with $V$ uniform on $\{v^{(1)},\ldots,v^{(M_1)}\}$, so that the approximation $\hat{c}_n$ is itself a valid copula density, which converges to $c_n$ as $M_1\to\infty$. Since $\log\hat{c}_n$ involves the logarithm of an average, it is a consistent estimator of $\log c_n$ with bias of order $1/M_1$ (Section~\ref{S-app:supp_latent} of the Supplementary Material).

Since $\hat{F}_{1,j}$ is continuous and strictly increasing, $\hat{w}_{ij}=\hat{F}_{1,j}^{-1}(u_{ij})$ is the unique root of $\hat{F}_{1,j}(w_{ij})=u_{ij}$, which we compute by bisection. The root depends on any parameter $\bs{v}$ of $(\bftheta,\bs{\varphi})$ through $\hat{F}_{1,j}$, and the implicit function theorem gives its exact gradient
\begin{equation}
\frac{\partial\hat{w}_{ij}}{\partial\bs{v}}
=-\frac{\partial\hat{F}_{1,j}(w;\bs{v})/\partial\bs{v}\,\big|_{w=\hat{w}_{ij}}}{\hat{f}_{1,j}(\hat{w}_{ij};\bs{v})}.
\label{eq:implicit_gradient}
\end{equation}
The copula log-likelihood is invariant to additive shifts in $g$
(Remark~\ref{rem:identifiability}). The implicit gradient accounts
for the corresponding shift in the latent quantiles, so the
derivative in the shift direction is zero
(Section~\ref{S-app:supp_latent} of the Supplementary Material). When optimizing the ELBO in Section \ref{subsec:vb_inference}, we use \eqref{eq:implicit_gradient} to
account for how the latent quantiles change with the factor parameters
and network weights in the copula log-likelihood.}

\subsection{Variational Bayes inference}\label{subsec:vb_inference}

We collect the factor parameters $\bftheta=(\theta_1,\theta_2)^\top$
and the range and nugget parameters of the latent Gaussian process,
$\bs{\vartheta}=(\ell,\tau^2)^\top$, in the vector
$\bs{\eta}=(\theta_1,\theta_2,\ell,\tau^2)^\top$.
We assign a prior to $\bs{\eta}$ and approximate its posterior with
a variational distribution. We use independent priors, half-normal priors $\mathcal{HN}(\sigma_0)$ with scale $\sigma_0=4$ for the positive parameters $\theta_1$, $\theta_2$ and $\ell$, and a uniform prior on $(0,1)$ for the nugget $\tau^2$. We first obtain a point estimate of the network weights
$\bs{\varphi}$ by optimizing the ELBO. We then use stochastic gradient
Markov chain Monte Carlo (MCMC) to obtain approximate posterior draws of the weights, as
described at the end of this section.



For replicate $i=1,\ldots,N$, let $\bfu_i=(u_{i1},\ldots,u_{in})^\top$
denote the copula-scale observations at the $n$ spatial locations, where $u_{ij}$
is the copula-scale observation for replicate $i$ at spatial location $j$.
Let $\mathcal{U}=(\bfu_1^\top,\ldots,\bfu_N^\top)^\top$ denote the
complete copula data. For given weights $\bs{\varphi}$, the posterior distribution of $\bs{\eta}$ is
\[
\pi(\bs{\eta}\mid\mathcal{U},\bs{\varphi})=\frac{p(\bs{\eta})\prod_{i=1}^N c_n(\bfu_i\mid\bs{\eta},\bs{\varphi})}{p(\mathcal{U}\mid\bs{\varphi})},
\qquad
p(\mathcal{U}\mid\bs{\varphi})=\int p(\bs{\eta})\prod_{i=1}^N c_n(\bfu_i\mid\bs{\eta},\bs{\varphi})\,d\bs{\eta},
\]
where $p(\bs{\eta})$ is the prior density and $p(\mathcal{U}\mid\bs{\varphi})$ is the marginal likelihood of the
copula data $\mathcal{U}$ given the network weights $\bs{\varphi}$. 
Each evaluation of the copula likelihood requires numerical
integration over the factor and inversion of the latent marginal CDFs
$F_{1,j}$. With the network weights $\bs{\varphi}$ fixed, we use
variational inference to approximate
$\pi(\bs{\eta}\mid\mathcal{U},\bs{\varphi})$ by a mean-field approximation
$q_{\bs{\lambda}}(\bs{\eta})$ with parameter $\bs{\lambda}$. We use independent log-normal
distributions for $\theta_1$, $\theta_2$, and $\ell$, and a
logit-normal distribution for $\tau^2$. We generate draws from
$q_{\bs{\lambda}}(\bs{\eta})$ using the reparameterization \citep{kingma2013auto}
\begin{equation}
\eta_k=h_k(\mu_k+\sigma_k\kappa_k),\qquad
\kappa_k\stackrel{\mathrm{i.i.d.}}{\sim}\mathcal{N}(0,1),
\qquad k=1,\ldots,4,
\label{eq:variational_reparam}
\end{equation}
where $h_k(x)=\exp(x)$ for $k=1,2,3$ and
$h_4(x)=(1+\exp(-x))^{-1}$. The variational parameters are
$\bs{\lambda}=(\mu_1,\ldots,\mu_4,\log\sigma_1,\ldots,\log\sigma_4)^\top$.
For fixed $\bs{\varphi}$, the optimal approximation $q_{\bs{\lambda}}(\bs{\eta})$ minimizes the reverse Kullback--Leibler divergence
\begin{equation}
\label{eq:kl_divergence}
D_{\mathrm{KL}}\{q_{\bs{\lambda}}(\bs{\eta})\,\|\,\pi(\bs{\eta}\mid\mathcal{U},\bs{\varphi})\}
=\mathbb{E}_{q_{\bs{\lambda}}}\left[\log\frac{q_{\bs{\lambda}}(\bs{\eta})}{\pi(\bs{\eta}\mid\mathcal{U},\bs{\varphi})}\right].
\end{equation}
Substituting the expression for the posterior gives, for every $\bs{\varphi}$, the decomposition
\begin{equation}
\label{eq:elbo_decomposition}
\log p(\mathcal{U}\mid\bs{\varphi})=D_{\mathrm{KL}}\{q_{\bs{\lambda}}(\bs{\eta})\,\|\,\pi(\bs{\eta}\mid\mathcal{U},\bs{\varphi})\}+\mathcal{L}(\bs{\lambda},\bs{\varphi}),
\end{equation}
where
\begin{equation}
\label{eq:elbo}
\mathcal{L}(\bs{\lambda},\bs{\varphi})
=\sum_{i=1}^N\mathbb{E}_{q_{\bs{\lambda}}}\left\{\log c_n(\bfu_i\mid\bs{\eta},\bs{\varphi})\right\}
+\mathbb{E}_{q_{\bs{\lambda}}}\left\{\log p(\bs{\eta})-\log q_{\bs{\lambda}}(\bs{\eta})\right\}
\end{equation}
is the evidence lower bound (ELBO), $\mathcal{L}(\bs{\lambda},\bs{\varphi})\le\log p(\mathcal{U}\mid\bs{\varphi})$ \citep{blei2017variational}. 
Equation \eqref{eq:elbo} 
allows the use
of  stochastic gradient ascent (SGA)  methods
to maximize the ELBO $\mathcal{L}\left(\bs{\lambda}\right)$ \citep{kingma2013auto}.
For fixed $\bs{\varphi}$, maximizing the ELBO over $\bs{\lambda}$
is equivalent to minimizing the KL divergence in
\eqref{eq:kl_divergence}. We estimate the ELBO gradient by Monte Carlo
using the reparameterized draws in \eqref{eq:variational_reparam}
\citep{kingma2013auto}. We then take joint
stochastic gradient ascent steps in $(\bs{\lambda},\bs{\varphi})$,
yielding a point estimate $\hat{\bs{\varphi}}$ of the network weights
and a fitted variational distribution $q_{\hat{\bs{\lambda}}}(\bs\eta)$.
The mean-field variational approximation omits posterior dependence between the components of $\bs{\eta}$ and tends to understate posterior variance \citep{blei2017variational}, and the two-step procedure ignores the uncertainty about the margins; Section~\ref{S-app:supp_psgld} of the Supplementary Material discusses both, and we assess their effect through the calibration of the predictions in Section~\ref{sec:empirical}.


We estimate $\mathcal{L}(\bs{\lambda},\bs{\varphi})$ with a mini-batch $\mathcal{B}\subset\{1,\ldots,N\}$ of replicates and two levels of draws. The outer draws $\bs{\eta}^{(r)}$, $r=1,\ldots,M_2$, are drawn from $q_{\bs{\lambda}}(\bs\eta)$ by \eqref{eq:variational_reparam} and describe the uncertainty about the factor and latent Gaussian correlation parameters. For each $\bs{\eta}^{(r)}$, the inner draws $v^{(m,r)}=\theta_1^{(r)}+|\theta_2^{(r)}-\theta_1^{(r)}|\varepsilon^{(m,r)}$, $\varepsilon^{(m,r)}\sim U(0,1)$, $m=1,\ldots,M_1$, approximate the integrals over the factor and are shared by all quantities in \eqref{eq:mc_estimators}. Writing $\bs{G}(v;\bs{\varphi})$ and $\hat{f}_{1,j}(\cdot\,;\bs{\eta},\bs{\varphi})$ to show which quantities change when the weights are updated, the estimator is
\begin{equation}
\label{eq:elbo_mc}
\begin{aligned}
\hat{\mathcal{L}}(\bs{\lambda},\bs{\varphi})
=\frac{1}{M_2}\sum_{r=1}^{M_2}\Bigg[\frac{N}{|\mathcal{B}|}\sum_{i\in\mathcal{B}}\Bigg\{&\log\Bigg(\frac{1}{M_1}\sum_{m=1}^{M_1}\phi_{\bs{\Sigma}_{\bs{\vartheta}^{(r)}}}\big[\tilde{\bfw}_i^{(r)}-\bs{G}\{v^{(m,r)};\bs{\varphi}\}\big]\Bigg)\\
&-\sum_{j=1}^n\log\hat{f}_{1,j}\big(\tilde{w}_{ij}^{(r)};\bs{\eta}^{(r)},\bs{\varphi}\big)\Bigg\}\\
&+\log p\big(\bs{\eta}^{(r)}\big)-\log q_{\bs{\lambda}}\big(\bs{\eta}^{(r)}\big)\Bigg],
\end{aligned}
\end{equation}
where $\tilde{w}_{ij}^{(r)}=\hat{F}_{1,j}^{-1}(u_{ij})$ is computed under $\bs{\eta}^{(r)}$ and $\bs{\varphi}$ by bisection and carries the gradient \eqref{eq:implicit_gradient}. Using $M_1$ factor draws
introduces an $O(M_1^{-1})$ bias in \eqref{eq:elbo_mc}.
We leave $\theta_1$ and $\theta_2$ unordered under $q_{\bs{\lambda}}$
and set $V=\theta_1+|\theta_2-\theta_1|\varepsilon$, where
$\varepsilon\sim U(0,1)$. This gives $V\sim U(\theta_1,\theta_2)$
when $\theta_1<\theta_2$. Shifting the interval can be offset by
shifting the input to $g$ in the opposite direction, preserving
the copula (Remark~\ref{rem:identifiability}).


We maximize $\hat{\mathcal{L}}$ jointly over $(\bs{\lambda}^\top,\bs{\varphi}^\top)^\top$ by stochastic gradient ascent using Adam with a step-wise decreasing learning rate. Training stops early when the estimate \eqref{eq:elbo_mc}, averaged over the mini-batches of an epoch and divided by $N$, has stopped increasing, and the parameters of the epoch with the largest value are retained. Algorithm~\ref{S-alg:estimation} in Section~\ref{S-app:supp_alg_estimation} of the Supplementary Material summarizes the complete procedure, and the settings we use are given in Section~\ref{sec:empirical}.

The cost of a gradient step of \eqref{eq:elbo_mc} does not depend on $N$, and all parts of it except the Gaussian density $\phi_{\bs{\Sigma}_{\bs{\vartheta}}}$ are linear in $n$. With the Vecchia approximation used in the real data application \citep{vecchia1988estimation,katzfuss2021general}, in which each location is conditioned only on $m$ nearest previous neighbors, the Gaussian density costs $O(nm)$ instead of $O(n^2)$ per residual vector, so that, for fixed $m$, Monte Carlo sizes, mini-batch size and network, the cost of an epoch is linear in both $N$ and $n$ (Sections~\ref{S-app:vecchia} and~\ref{S-app:supp_cost} of the Supplementary Material).

The point estimate $\hat{\bs{\varphi}}$ ignores the uncertainty about the shape of the factor distribution, which is typically the dominant source of uncertainty about the copula, and can make prediction intervals too narrow. After fitting the model, we hold the variational density
$q_{\hat{\bs{\lambda}}}(\bs\eta)$ fixed and use stochastic gradient Markov chain
Monte Carlo (SG-MCMC) to account for uncertainty in the network weights
$\bs{\varphi}$. We obtain the sampling target by replacing the log
marginal likelihood $\log p(\mathcal{U}\mid\bs{\varphi})$ with the ELBO
$\mathcal{L}(\hat{\bs{\lambda}},\bs{\varphi})$ in the marginal posterior
of the weights, yielding the approximate posterior
\begin{equation}
\label{eq:weight_posterior}
\tilde{\pi}(\bs{\varphi}\mid\mathcal{U})\propto\exp\{\mathcal{L}(\hat{\bs{\lambda}},\bs{\varphi})\}\,\mathcal{N}\{\bs{\varphi};\mathbf{0},(N\omega)^{-1}I\},
\end{equation}
where the Gaussian prior corresponds to the weight decay $\omega=10^{-4}$ used in the optimization. We sample it by preconditioned stochastic gradient Langevin dynamics \citep{welling2011bayesian,li2016preconditioned}, started at $\hat{\bs{\varphi}}$, and retain $S$ thinned samples $\bs{\varphi}^{(1)},\ldots,\bs{\varphi}^{(S)}$ after burn-in; Section~\ref{S-app:supp_psgld} of the Supplementary Material gives the update and discusses how $\tilde{\pi}$ differs from the exact posterior. 


For all BNFC predictions reported below, we average over the $S$
retained network weight draws and integrate over the variational
distribution of the parameters:
\begin{equation}
\label{eq:weight_average}
\hat{p}(\cdot\mid\mathcal{U})
=
\frac{1}{S}\sum_{s=1}^{S}
\int p(\cdot\mid\bs{\eta},\bs{\varphi}^{(s)})
\,q_{\hat{\bs{\lambda}}}(\bs{\eta})\,d\bs{\eta}.
\end{equation}
This approach uses a single model fit, avoiding the repeated fits
required by a deep ensemble \citep{lakshminarayanan2017simple}.
We assess the calibration of the resulting predictive
distributions in Section~\ref{sec:empirical}.

\subsection{Conditional copula and spatial interpolation}\label{subsec:cond_copula}

We now derive the conditional distribution needed to predict the process at an unobserved location $\bfs_0$, given the observations of one replicate at $\bfs_1,\ldots,\bfs_n$. The derivation is given in Section~\ref{S-app:conditional_copula_derivation} of the Supplementary Material. For ease of notation, we leave the dependence on the model parameters
$\bs{\eta}$ and $\bs{\varphi}$ implicit. Let $\bfu=(u_1,\ldots,u_n)^\top$ be the copula data of the replicate, let $\bfw=(w_1,\ldots,w_n)^\top$ with $w_j=F_{1,j}^{-1}(u_j)$, for $j=1,...,n$, and let $F_{1,0}$ be the marginal distribution function \eqref{eq:bnfc_univariate_margin} at $\bfs_0$. Conditional on $V=v$, the vector $(W_0,\bfW^\top)^\top$ is Gaussian, so $W_0$ given $\bfW=\bfw$ and $V=v$ is normal with kriging mean and variance
\begin{equation}
m_0(v)=g(v,\bfs_0)+\bs{\sigma}_{0}^\top\bs{\Sigma}_{\bs{\vartheta}}^{-1}\{\bfw-\bs{G}(v)\},
\qquad
\sigma_{K,0}^2=1-\bs{\sigma}_{0}^\top\bs{\Sigma}_{\bs{\vartheta}}^{-1}\bs{\sigma}_{0},
\label{eq:kriging}
\end{equation}
where $\bs{\sigma}_0$ is the vector of correlations between $Z(\bfs_0)$ and $\bfZ$\chk{. Both $\bs{\Sigma}_{\bs{\vartheta}}$ and $\bs{\sigma}_0=\bs{\sigma}_0(\bs{\vartheta})$ are determined by the correlation parameters $\bs{\vartheta}$, and hence so are $m_0$ and $\sigma_{K,0}^2$}; we write $\sigma_{K,0}$ for the kriging standard deviation to distinguish it from the location $\bfs_0$. Integrating over the conditional distribution of the factor gives
\begin{equation}
\begin{gathered}
C_{0\mid n}(u_0\mid\bfu)
=\int_{\theta_1}^{\theta_2}\Phi\left\{\frac{F_{1,0}^{-1}(u_0)-m_0(v)}{\sigma_{K,0}}\right\}p(v\mid\bfw)\,dv,
\\
p(v\mid\bfw)=\frac{\phi_{\bs{\Sigma}_{\bs{\vartheta}}}\{\bfw-\bs{G}(v)\}}{\int_{\theta_1}^{\theta_2}\phi_{\bs{\Sigma}_{\bs{\vartheta}}}\{\bfw-\bs{G}(v')\}\,dv'}.
\end{gathered}
\label{eq:conditional_copula}
\end{equation}
The conditional distribution is a mixture of Gaussian kriging
distributions, weighted by the conditional density $p(v\mid\bfw)$. When a factor with a regularly varying right
tail enters the observed sites and the prediction site with positive
loadings, simultaneous upper extremes at the observed sites favor
large factor draws, which also raise the latent process at $\bfs_0$.
In the location-dependent model, the effect at $\bfs_0$ is determined
by the local factor map $g(v,\bfs_0)$.


We draw from the approximate posterior predictive distribution \eqref{eq:weight_average} of $Y(\bfs_0)$ with Algorithm~\ref{S-alg:prediction} in Section~\ref{S-app:supp_alg_prediction} of the Supplementary Material. Each predictive draw uses one weight sample and one draw of $\bs{\eta}$ from $q_{\hat{\bs{\lambda}}}(\bs\eta)$, selects a factor value among $L$ factor candidates with probabilities proportional to their conditional density given the observations, draws $W_0$ from the corresponding kriging distribution \eqref{eq:kriging}, and maps it to the copula and data scales by $\hat{F}_{1,0}$ and $T_{\hat{\bfpsi}}$ defined in \ref{subsec:marginals}. The draws therefore come from \eqref{eq:conditional_copula}, averaged over the parameter uncertainty, including that of the latent Gaussian process component. Several locations can be predicted jointly with the multivariate kriging mean and covariance.

\section{Empirical studies}\label{sec:empirical}


We evaluate the stationary and location-dependent BNFC models in
simulation studies and an application to precipitation data.
Section~\ref{sec:simulation_studies} compares BNFC with competing
copula models in three scenarios: a common factor, a factor whose
effect changes sign across space, and a factor whose distribution
varies across regions. Section~\ref{sec:application} applies the
models to daily ERA5 precipitation over central Europe and examines
how the shared factor affects different regions. In both studies, we
compare held-out copula log scores and assess the accuracy and
calibration of spatial predictions.

For the simulation studies, the factor network $g(v)$ of BNFC (stat) has two hidden layers of width $32$ with ReLU activations and $g(v,\bfs)$ of BNFC (nonstat) three hidden layers of width $64$ with GELU activations; both end in a single output unit with a softplus activation. The ELBO is maximized by Adam with $M_2=50$. The learning rate, the number of epochs, early stopping, $M_1$ and the mini-batch size of BNFC (stat) in scenario (i) and of BNFC (nonstat) in scenarios (ii) and (iii) are tuned on validation replicates, and the other BNFC model in each scenario uses default settings (Section~\ref{S-app:supp_tuning} of the Supplementary Material). The stochastic gradient MCMC uses step size $\epsilon=10^{-3}$ and retains $S=20$ samples, one every $50$ iterations. All computations use one NVIDIA GeForce RTX 4090 GPU.

\subsection{Simulation studies}\label{sec:simulation_studies}

We consider three data-generating processes on a regular $30\times30$ grid on $[0,1]^2$. In all three, $Z(\bfs)$ has squared exponential correlation with range $0.1$, which all models use without estimating it, the copula data are $U(\bfs)=F_1\{W(\bfs);\bfs\}$ under the data-generating model.
\begin{itemize}
\item[(i)] \mbox{\emph{Gamma-factor:} $W(\bfs)=Z(\bfs)+V_0$ with $V_0\sim\mathrm{Ga}(6,\text{scale }3)$, the common factor model \eqref{eq:classical-factor-copula}.}
\item[(ii)] \emph{Sign-change:} $W(\bfs)=Z(\bfs)+1.5\,b(\bfs)H^{-1}(U)$, where $U\sim U(0,1)$, $H$ is the $\mathrm{Ga}(2,1)$ distribution and $b(\bfs)=\tanh\{(s_2-0.5)/0.15\}$. The factor raises the field in the north and lowers it in the south, a structure that BNFC (nonstat) can represent (Proposition~\ref{thm:covariance}) but the common factor copulas cannot.
\item[(iii)] \emph{Regional mixture:} $W(\bfs)=Z(\bfs)+\{1-r(\bfs)\}H_A^{-1}(U)+r(\bfs)H_B^{-1}(U)$, where $r(\bfs)=[1+\exp\{-(s_1-0.5)/0.1\}]^{-1}$, $H_A$ is the bimodal gamma mixture $0.6\,\mathrm{Ga}(2,1)+0.4\,\mathrm{Ga}(20,0.5)$ and $H_B$ the lognormal $\mathrm{LN}(1.2,0.6^2)$. The factor family changes from west to east, and every model in the comparison, including BNFC, is misspecified.
\end{itemize}
For each scenario, we simulate $2{,}000$ training and $100$ test replicates,  at the same $n=100$ random locations.  For each test replicate we condition on these locations and predict at all $900$ grid locations; see Section~\ref{S-app:supp_design} of the Supplementary Material for further details.

We compare all models on the copula scale. For both the stationary
and location-dependent BNFC models, predictions average over the
$S=20$ network weight draws in \eqref{eq:weight_average}.
We fit four common-factor copulas \eqref{eq:classical-factor-copula}.
The factor distributions are exponential with rate $\delta$, gamma
with shape $6$ and rate $\delta$, Pareto with shape $2$ and scale
$\delta+1$, and the gamma mixture
$\pi\,\mathrm{Ga}(2,\text{rate }\delta_1)
+(1-\pi)\,\mathrm{Ga}(20,\text{rate }\delta_2)$.
The mixture uses the same component shapes as $H_A$ in scenario~(iii).
We also fit a Student-$t$ copula \citep{demarta2005t} with correlation
matrix $(1-\beta)\bs{\Sigma}+\beta\mathbf{1}\mathbf{1}^\top$.
Finally, we fit two copulas with a spatially varying factor loading.
Both use $W(\bfs)=Z(\bfs)+b(\bfs)X$, where
$b(\bfs)=\sum_{k=1}^{25}\beta_k\psi_k(\bfs)$ is a linear combination
of Gaussian basis functions with coefficients of either sign.
The Gaussian loading copula uses a Gaussian factor $X$, and the gamma
loading copula uses $X\sim\mathrm{Ga}(\kappa,1)$. These benchmarks
allow the strength and sign of the factor's effect to vary across
space and provide a basis for assessing the additional flexibility
of the location-dependent BNFC. The parametric factor copula models are fitted by Hamiltonian Monte Carlo with the No-U-Turn sampler \citep{hoffman2014no} in Pyro \citep{bingham2019pyro}, with factor integrals discretized by quantile nodes, and the Student-$t$ posterior is computed on a grid (Section~\ref{S-app:supp_design} of the Supplementary Material for further details). The main-text tables report results for the best-performing
common-factor copula in each scenario. Results for all four common-factor copulas
are given in Section~\ref{S-app:supp_all_models} of the Supplementary Material.


The main measure is the held-out copula log score, the log posterior predictive density of the copula data of a test replicate \citep{gneiting2007strictly}, at the $100$ observed and at all $900$ grid locations. We assess predictions from Algorithm~\ref{S-alg:prediction} of the
Supplementary Material using the continuous ranked probability score
(CRPS), the energy score, and mean absolute error (MAE) over all
$900$ grid locations. At the $800$ unobserved locations, we report
the prediction interval coverage probability (PICP) to assess
calibration and the mean prediction interval width (MPIW). We also compare each model's posterior distribution of the
centered factor contribution with the corresponding posterior under
the data-generating model (Remark~\ref{rem:identifiability}).
We average the performance measures over the $100$ test replicates.
For pairwise model comparisons, we compute the difference in each
measure on the same replicate and report the mean paired difference
with its standard error in parentheses. Section~\ref{S-app:supp_measures}
of the Supplementary Material gives further details.

\begin{table}[t]
\centering
\caption{Held-out copula log score and predictive performance on the copula scale, averaged over $100$ test replicates, with standard errors in parentheses. Log scores are at the $100$ observed and at all $900$ grid locations (higher is better); CRPS, energy score and MAE are those of the predictions given the $100$ observed locations, averaged over all $900$ grid locations, including the $100$ conditioning locations (lower is better); the nominal $90$\% PICP (closest to $0.90$ is better) and MPIW (lower is better) are computed at the $800$ unobserved locations. ``Best common factor'' is the common factor copula with the highest log score over all locations: the gamma factor in (i), the exponential factor in (ii) and the gamma mixture in (iii); all four are in Table~\ref{S-tab:supp_sim_all} of the Supplementary Material. The best value of each measure is in bold. The settings of BNFC (stat) in (i) and of BNFC (nonstat) in (ii) and (iii) are tuned on validation replicates. Fit times are wall-clock times on one GPU, including the stochastic gradient MCMC for BNFC.}
\label{tab:simulation_comparison}
\renewcommand{\arraystretch}{1.15}
\resizebox{\textwidth}{!}{%
\begin{tabular}{lcccccc}
\toprule
 & \multicolumn{2}{c}{BNFC} & \multicolumn{2}{c}{Common factor / scale mixture} & \multicolumn{2}{c}{Varying loading} \\
\cmidrule(lr){2-3}\cmidrule(lr){4-5}\cmidrule(lr){6-7}
Measure & stat & nonstat & Best common factor & Student-$t$ & Gaussian & Gamma \\
\midrule
\multicolumn{7}{l}{\textbf{(i) Gamma-factor data} \quad (best common factor: gamma)} \\[2pt]
\quad CRPS & $0.0045$\,{\scriptsize$(0.0002)$} & $0.0045$\,{\scriptsize$(0.0002)$} & $0.0045$\,{\scriptsize$(0.0002)$} & $0.0045$\,{\scriptsize$(0.0002)$} & $0.0045$\,{\scriptsize$(0.0002)$} & $\mathbf{0.0045}$\,{\scriptsize$(0.0002)$} \\
\quad Energy score & $0.232$\,{\scriptsize$(0.009)$} & $0.233$\,{\scriptsize$(0.009)$} & $\mathbf{0.232}$\,{\scriptsize$(0.009)$} & $0.232$\,{\scriptsize$(0.009)$} & $0.233$\,{\scriptsize$(0.009)$} & $0.232$\,{\scriptsize$(0.009)$} \\
\quad MAE & $\mathbf{0.0063}$\,{\scriptsize$(0.0002)$} & $0.0063$\,{\scriptsize$(0.0002)$} & $0.0063$\,{\scriptsize$(0.0002)$} & $0.0063$\,{\scriptsize$(0.0002)$} & $0.0063$\,{\scriptsize$(0.0002)$} & $0.0063$\,{\scriptsize$(0.0002)$} \\
\quad Log score (observed) & $266.5$\,{\scriptsize$(4.9)$} & $265.1$\,{\scriptsize$(4.6)$} & $\mathbf{266.8}$\,{\scriptsize$(5.0)$} & $265.1$\,{\scriptsize$(4.9)$} & $260.0$\,{\scriptsize$(4.3)$} & $266.3$\,{\scriptsize$(4.9)$} \\
\quad Log score (all locations) & $4899$\,{\scriptsize$(44)$} & $4892$\,{\scriptsize$(41)$} & $\mathbf{4908}$\,{\scriptsize$(45)$} & $4893$\,{\scriptsize$(45)$} & $4839$\,{\scriptsize$(38)$} & $4902$\,{\scriptsize$(44)$} \\
\quad PICP $90\%$ & $0.916$\,{\scriptsize$(0.004)$} & $0.930$\,{\scriptsize$(0.004)$} & $0.914$\,{\scriptsize$(0.004)$} & $0.876$\,{\scriptsize$(0.005)$} & $\mathbf{0.902}$\,{\scriptsize$(0.008)$} & $0.893$\,{\scriptsize$(0.004)$} \\
\quad MPIW $90\%$ & $0.030$\,{\scriptsize$(0.001)$} & $0.032$\,{\scriptsize$(0.001)$} & $0.030$\,{\scriptsize$(0.001)$} & $\mathbf{0.028}$\,{\scriptsize$(0.001)$} & $0.031$\,{\scriptsize$(0.001)$} & $0.029$\,{\scriptsize$(0.001)$} \\
\quad Fit time (s) & $126$ & $726$ & $143$ & $19$ & $325$ & $3601$ \\
\midrule
\multicolumn{7}{l}{\textbf{(ii) Sign-change data} \quad (best common factor: exponential)} \\[2pt]
\quad CRPS & $0.0215$\,{\scriptsize$(0.0004)$} & $\mathbf{0.0163}$\,{\scriptsize$(0.0005)$} & $0.0219$\,{\scriptsize$(0.0004)$} & $0.0218$\,{\scriptsize$(0.0004)$} & $0.0177$\,{\scriptsize$(0.0004)$} & $0.0166$\,{\scriptsize$(0.0005)$} \\
\quad Energy score & $1.087$\,{\scriptsize$(0.014)$} & $\mathbf{0.831}$\,{\scriptsize$(0.023)$} & $1.106$\,{\scriptsize$(0.014)$} & $1.099$\,{\scriptsize$(0.016)$} & $0.904$\,{\scriptsize$(0.020)$} & $0.857$\,{\scriptsize$(0.021)$} \\
\quad MAE & $0.0297$\,{\scriptsize$(0.0005)$} & $\mathbf{0.0229}$\,{\scriptsize$(0.0007)$} & $0.0299$\,{\scriptsize$(0.0005)$} & $0.0303$\,{\scriptsize$(0.0005)$} & $0.0247$\,{\scriptsize$(0.0006)$} & $0.0233$\,{\scriptsize$(0.0006)$} \\
\quad Log score (observed) & $106.7$\,{\scriptsize$(2.2)$} & $\mathbf{144.2}$\,{\scriptsize$(5.7)$} & $105.6$\,{\scriptsize$(2.5)$} & $111.7$\,{\scriptsize$(3.8)$} & $133.8$\,{\scriptsize$(4.6)$} & $143.2$\,{\scriptsize$(5.7)$} \\
\quad Log score (all locations) & $2741$\,{\scriptsize$(176)$} & $\mathbf{3764}$\,{\scriptsize$(46)$} & $2767$\,{\scriptsize$(121)$} & $3072$\,{\scriptsize$(38)$} & $3336$\,{\scriptsize$(56)$} & $3688$\,{\scriptsize$(35)$} \\
\quad PICP $90\%$ & $0.880$\,{\scriptsize$(0.006)$} & $0.892$\,{\scriptsize$(0.004)$} & $\mathbf{0.893}$\,{\scriptsize$(0.005)$} & $0.873$\,{\scriptsize$(0.003)$} & $0.841$\,{\scriptsize$(0.007)$} & $0.871$\,{\scriptsize$(0.004)$} \\
\quad MPIW $90\%$ & $0.160$\,{\scriptsize$(0.003)$} & $\mathbf{0.104}$\,{\scriptsize$(0.003)$} & $0.172$\,{\scriptsize$(0.004)$} & $0.157$\,{\scriptsize$(0.002)$} & $0.109$\,{\scriptsize$(0.003)$} & $0.107$\,{\scriptsize$(0.003)$} \\
\quad Fit time (s) & $93$ & $960$ & $524$ & $20$ & $41$ & $4355$ \\
\midrule
\multicolumn{7}{l}{\textbf{(iii) Regional mixture data} \quad (best common factor: gamma mixture)} \\[2pt]
\quad CRPS & $0.0109$\,{\scriptsize$(0.0004)$} & $\mathbf{0.0105}$\,{\scriptsize$(0.0004)$} & $0.0109$\,{\scriptsize$(0.0005)$} & $0.0118$\,{\scriptsize$(0.0005)$} & $0.0119$\,{\scriptsize$(0.0005)$} & $0.0110$\,{\scriptsize$(0.0005)$} \\
\quad Energy score & $0.544$\,{\scriptsize$(0.023)$} & $\mathbf{0.536}$\,{\scriptsize$(0.023)$} & $0.545$\,{\scriptsize$(0.023)$} & $0.575$\,{\scriptsize$(0.025)$} & $0.586$\,{\scriptsize$(0.024)$} & $0.550$\,{\scriptsize$(0.023)$} \\
\quad MAE & $0.0151$\,{\scriptsize$(0.0006)$} & $\mathbf{0.0147}$\,{\scriptsize$(0.0006)$} & $0.0152$\,{\scriptsize$(0.0006)$} & $0.0162$\,{\scriptsize$(0.0007)$} & $0.0160$\,{\scriptsize$(0.0007)$} & $0.0152$\,{\scriptsize$(0.0006)$} \\
\quad Log score (observed) & $188.0$\,{\scriptsize$(6.9)$} & $\mathbf{197.4}$\,{\scriptsize$(7.0)$} & $188.9$\,{\scriptsize$(7.5)$} & $181.1$\,{\scriptsize$(7.8)$} & $172.7$\,{\scriptsize$(7.0)$} & $187.7$\,{\scriptsize$(6.9)$} \\
\quad Log score (all locations) & $4097$\,{\scriptsize$(65)$} & $\mathbf{4248}$\,{\scriptsize$(63)$} & $4093$\,{\scriptsize$(67)$} & $3989$\,{\scriptsize$(77)$} & $3870$\,{\scriptsize$(72)$} & $4034$\,{\scriptsize$(58)$} \\
\quad PICP $90\%$ & $\mathbf{0.902}$\,{\scriptsize$(0.005)$} & $0.908$\,{\scriptsize$(0.005)$} & $0.889$\,{\scriptsize$(0.005)$} & $0.875$\,{\scriptsize$(0.006)$} & $0.889$\,{\scriptsize$(0.011)$} & $0.874$\,{\scriptsize$(0.006)$} \\
\quad MPIW $90\%$ & $0.073$\,{\scriptsize$(0.003)$} & $\mathbf{0.069}$\,{\scriptsize$(0.003)$} & $0.070$\,{\scriptsize$(0.003)$} & $0.076$\,{\scriptsize$(0.003)$} & $0.082$\,{\scriptsize$(0.003)$} & $0.071$\,{\scriptsize$(0.003)$} \\
\quad Fit time (s) & $66$ & $681$ & $1465$ & $20$ & $32$ & $3327$ \\
\bottomrule
\end{tabular}%
}
\end{table}

\begin{figure}[t]
\centering
\includegraphics[width=\textwidth]{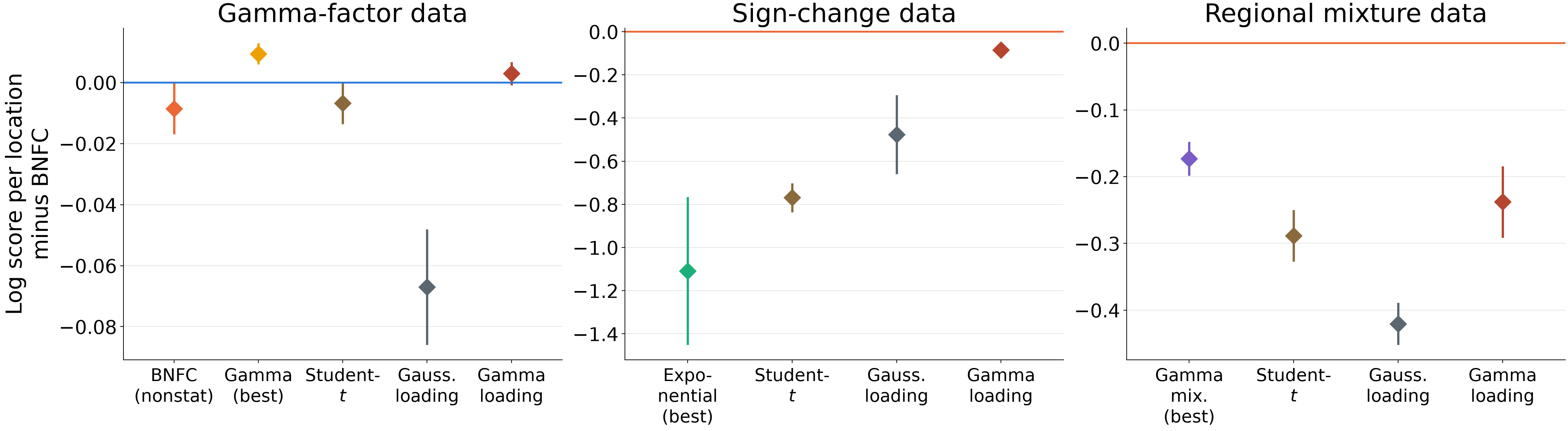}
\caption{Held-out copula log score at all $900$ grid locations of the models of Table~\ref{tab:simulation_comparison} minus that of the reference BNFC model, per location and averaged over the $100$ test replicates. The reference (horizontal line) is BNFC (stat) for the gamma-factor data and BNFC (nonstat) for the sign-change and regional mixture data. ``(best)'' marks the best common factor copula of each scenario. Bars are $\pm2$ standard errors of the paired differences. Values below the reference line mean that the reference BNFC model assigns a higher density to the test data.}
\label{fig:logscore}
\end{figure}

Table~\ref{tab:simulation_comparison} and Figure~\ref{fig:logscore}
summarize the simulation results. Section~\ref{S-app:additional_sim}
of the Supplementary Material reports results for all four
common-factor copulas, illustrates predictions for one test replicate,
and presents additional analyses of predictive calibration and
performance when conditioning on fewer locations. For the gamma-factor data, BNFC (stat) is close to the data-generating model without being told the factor family. In the two scenarios in which the effect of the factor varies over space, BNFC (nonstat) has the highest log scores, including for the regional mixture data, for which every model is misspecified, and its prediction intervals are close to the nominal level. In these scenarios it is also ahead of both loading copulas, so that its gain does not come from the spatially varying covariance alone, and, at the observed locations, of a nonparametric Bayesian transport map \citep{katzfuss2024scalable} (Section~\ref{S-app:supp_transport_map} of the Supplementary Material). The conclusions are unchanged with $25$ conditioning locations (Section~\ref{S-app:supp_sparse} of the Supplementary Material). For the gamma-factor data, the differences between BNFC (stat) and the data-generating gamma factor copula are within two standard errors on every measure, whereas in the other two scenarios the log scores of BNFC (nonstat) are significantly higher than those of every other model, with paired differences larger than two standard errors (Figure~\ref{fig:logscore}).

Section~\ref{S-app:supp_factor_posterior} of the Supplementary Material compares the posterior of the centered factor contribution given the $100$ observed locations, which is comparable across models (Remark~\ref{rem:identifiability}), with its true posterior under the data-generating model along a transect through the domain (Figure~\ref{S-fig:factor_posterior}). For the sign-change and regional mixture data, BNFC (nonstat) follows the true posterior closely, whereas the gamma loading copula departs from it in parts of the domain and the factor of a common factor copula shifts the whole field by the same amount and cannot follow it. For the gamma-factor data, the posterior from BNFC (stat) agrees closely with the reference posterior along the transect.

\FloatBarrier

\subsection{Application}\label{sec:application}

\begin{table}[t]
\centering
\caption{ERA5 precipitation: held-out copula log score and predictive performance on the copula scale, averaged over the $200$ test days from 15 June to 31 December 2024, with standard errors in parentheses. Log scores are at the $100$ observed and at all $1{,}024$ grid locations (higher is better); CRPS, energy score and MAE are those of the predictions given the $100$ observed locations, averaged over all $1{,}024$ grid locations, including the $100$ conditioning locations (lower is better); the nominal $90$\% PICP (closest to $0.90$ is better) and MPIW (lower is better) are computed at the $924$ unobserved locations. ``Best common factor'' is the exponential factor copula, which has the highest log score over the full grid of the three common factor copulas; all models are in Table~\ref{S-tab:supp_app_all} of the Supplementary Material. The best value of each measure is in bold. Fit times are wall-clock times on one GPU, including the stochastic gradient MCMC for BNFC.}
\label{tab:application_comparison}
\renewcommand{\arraystretch}{1.15}
\resizebox{\textwidth}{!}{%
\begin{tabular}{lcccccc}
\toprule
 & \multicolumn{2}{c}{BNFC} & \multicolumn{2}{c}{Common factor / scale mixture} & \multicolumn{2}{c}{Varying loading} \\
\cmidrule(lr){2-3}\cmidrule(lr){4-5}\cmidrule(lr){6-7}
Measure & stat & nonstat & Best common factor & Student-$t$ & Gaussian & Gamma \\
\midrule
\quad CRPS & $0.0373$\,{\scriptsize$(0.0009)$} & $\mathbf{0.0326}$\,{\scriptsize$(0.0007)$} & $0.0378$\,{\scriptsize$(0.0009)$} & $0.0374$\,{\scriptsize$(0.0009)$} & $0.0373$\,{\scriptsize$(0.0009)$} & $0.0373$\,{\scriptsize$(0.0009)$} \\
\quad Energy score & $1.717$\,{\scriptsize$(0.038)$} & $\mathbf{1.573}$\,{\scriptsize$(0.034)$} & $1.793$\,{\scriptsize$(0.041)$} & $1.777$\,{\scriptsize$(0.040)$} & $1.726$\,{\scriptsize$(0.037)$} & $1.726$\,{\scriptsize$(0.037)$} \\
\quad MAE & $0.0502$\,{\scriptsize$(0.0012)$} & $\mathbf{0.0443}$\,{\scriptsize$(0.0010)$} & $0.0506$\,{\scriptsize$(0.0012)$} & $0.0507$\,{\scriptsize$(0.0012)$} & $0.0503$\,{\scriptsize$(0.0012)$} & $0.0502$\,{\scriptsize$(0.0012)$} \\
\quad Log score (observed) & $112.9$\,{\scriptsize$(3.5)$} & $\mathbf{122.2}$\,{\scriptsize$(3.6)$} & $104.4$\,{\scriptsize$(3.7)$} & $113.5$\,{\scriptsize$(3.6)$} & $111.6$\,{\scriptsize$(3.7)$} & $111.7$\,{\scriptsize$(3.7)$} \\
\quad Log score (all locations) & $1684$\,{\scriptsize$(37)$} & $\mathbf{1995}$\,{\scriptsize$(32)$} & $1619$\,{\scriptsize$(46)$} & $1817$\,{\scriptsize$(36)$} & $1685$\,{\scriptsize$(38)$} & $1690$\,{\scriptsize$(38)$} \\
\quad PICP $90\%$ & $0.850$\,{\scriptsize$(0.006)$} & $\mathbf{0.887}$\,{\scriptsize$(0.005)$} & $0.826$\,{\scriptsize$(0.008)$} & $0.829$\,{\scriptsize$(0.003)$} & $0.858$\,{\scriptsize$(0.006)$} & $0.858$\,{\scriptsize$(0.007)$} \\
\quad MPIW $90\%$ & $0.194$\,{\scriptsize$(0.003)$} & $0.204$\,{\scriptsize$(0.003)$} & $0.191$\,{\scriptsize$(0.003)$} & $\mathbf{0.190}$\,{\scriptsize$(0.004)$} & $0.205$\,{\scriptsize$(0.003)$} & $0.204$\,{\scriptsize$(0.003)$} \\
\quad Fit time (s) & $274$ & $4745$ & $2314$ & $74$ & $136$ & $48114$ \\
\bottomrule
\end{tabular}%
}
\end{table}
We use daily total precipitation from the ERA5 reanalysis \citep{hersbach2020era5} over central Europe ($15^\circ$E--$25^\circ$E, $45^\circ$N--$55^\circ$N) from 1994 to 2024, averaged over two consecutive days on a regular $32\times32$ grid. Each day is a replicate. The $N=10{,}591$ days up to 2022 form the training set and the $200$ days from 15 June to 31 December 2024, which include the extreme rainfall of September 2024, the test set, all observed at the same $n=100$ random locations. The copula data are rank scores whose ties at the point mass at zero, $6.0\%$ of the values, are broken at random, and the marginal network $T_{\hat{\bfpsi}}$ is used only to map predictions to the data scale. We fit the models of Section~\ref{sec:simulation_studies} with expoential correlation function for the Gaussian component. The architecture of BNFC (nonstat), four hidden layers of width $128$, a linear output and an exponential correlation function, is chosen by the held-out log score on the days of 2022 after fitting to the days up to 2021, and the model is then refitted to all training days. Section~\ref{S-app:application_tail} of the Supplementary Material gives the details. For each test day, we condition on the $100$ observed locations and predict all $1{,}024$ grid locations.

\begin{figure}[htbp]
\centering
\includegraphics[width=\textwidth]{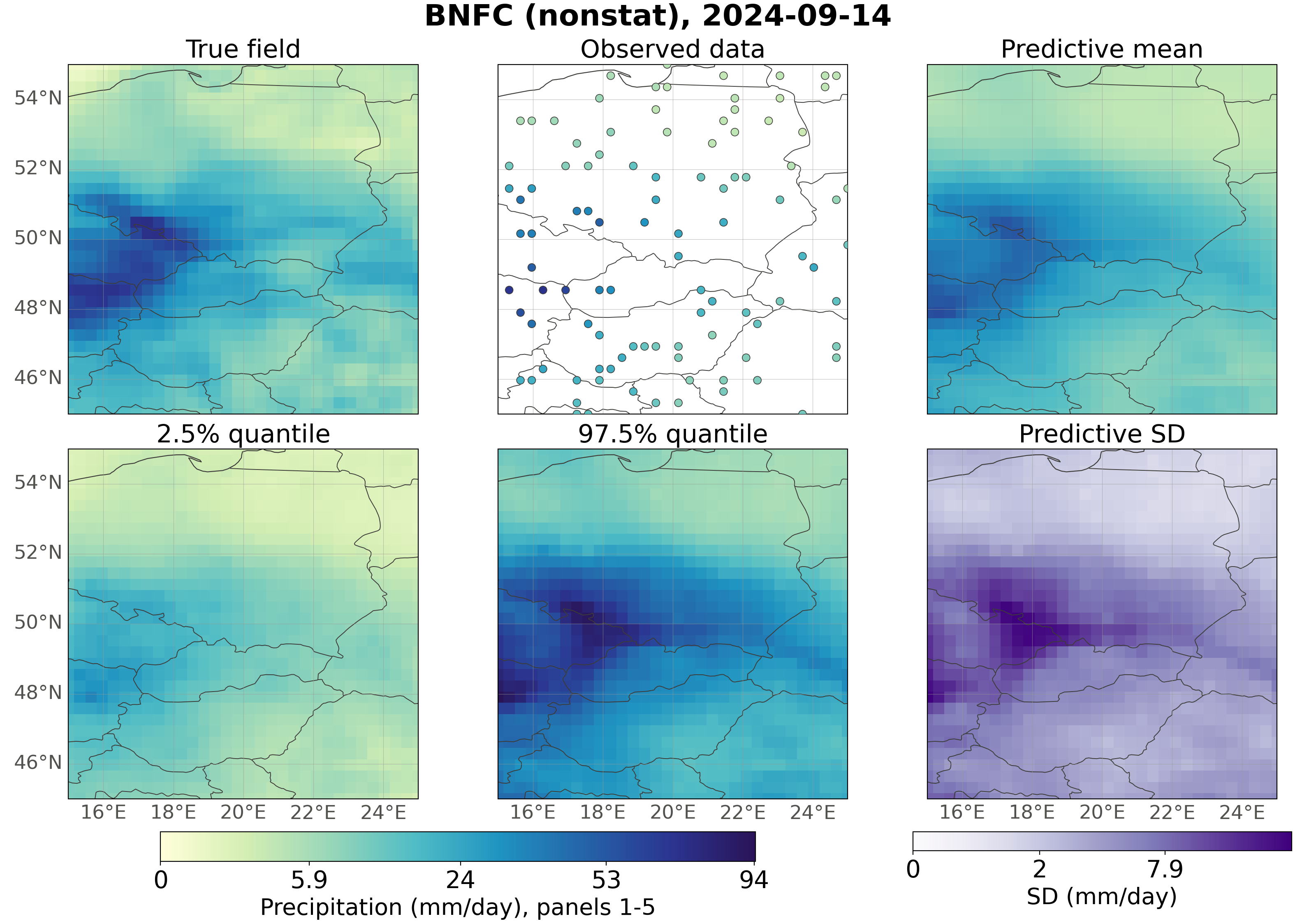}
\caption{ERA5 precipitation on 14 September 2024 (mm/day): predictions of BNFC (nonstat) given the $100$ observed locations. Top row: ERA5 field, data at the $100$ conditioning locations and predictive mean; bottom row: $2.5\%$ and $97.5\%$ predictive quantiles and predictive standard deviation. The first five panels share one color scale; all color scales are square-root scales. }
\label{fig:application_map}
\end{figure}

Table~\ref{tab:application_comparison} summarizes the results. BNFC (nonstat) is best on every measure except the MPIW: its $90\%$ intervals are slightly wider than those of the common factor and Student-$t$ copulas, which, however, cover the true values too rarely. Its improvements in CRPS, energy score, MAE and the log score over all locations exceed two standard errors of the difference against every other model. Most of its advantage over the other models comes from the learned factor rather than from the spatially varying loading, since it is also clearly ahead of the gamma loading copula. At the observed locations, a nonparametric Bayesian transport map \citep{katzfuss2024scalable} has a slightly higher log score, but it is defined only at the locations with training data and gives no predictions at unobserved locations (Section~\ref{S-app:supp_transport_map} of the Supplementary Material). Figure~\ref{fig:application_map} shows the predictions for 14 September 2024, the wettest day of the test period. With the Vecchia approximation, BNFC (nonstat) can also be fitted to all $1{,}024$ grid locations on all training days, more than $10^7$ observations, on one GPU (Section~\ref{S-app:vecchia} of the Supplementary Material).


\subsection{Inference on the dependence of precipitation}\label{sec:application_inference}

\begin{figure}[htbp]
\centering
\includegraphics[width=\textwidth]{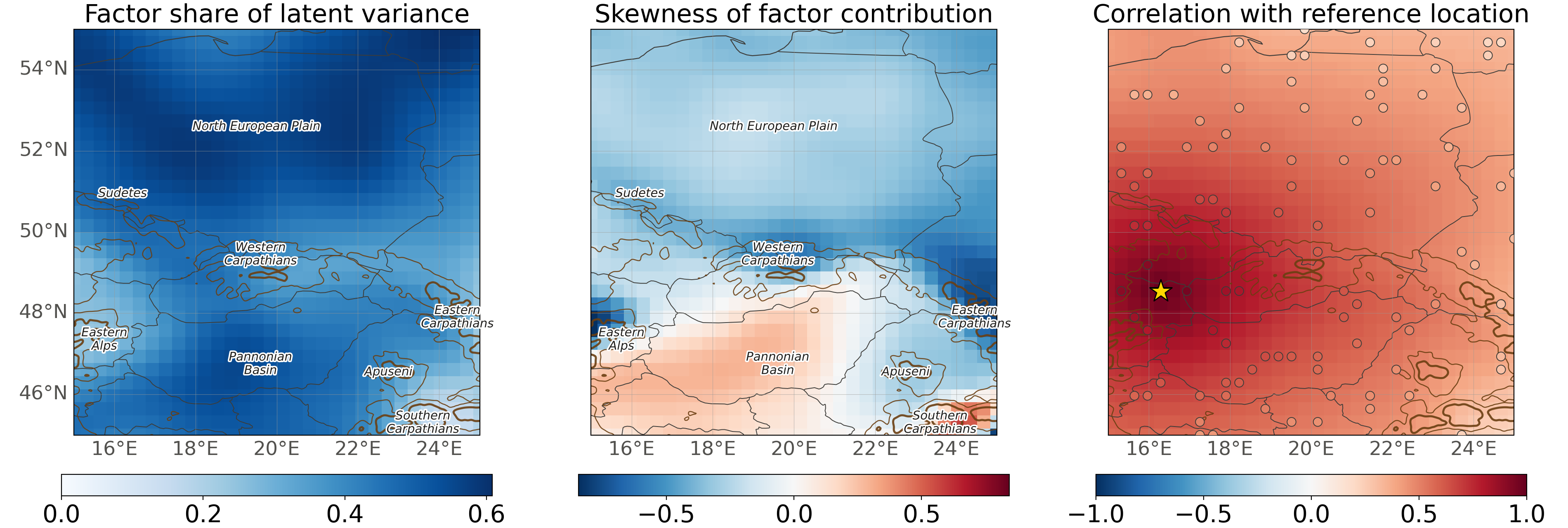}
\caption{ERA5 precipitation, BNFC (nonstat), averaged over the weight samples: share $\varrho(\bfs)$ of the latent variance due to the factor (left), skewness of the factor contribution (middle), and correlation of the latent field with the reference location near Vienna (star; right), with the empirical correlation of the normal scores of the training data at the observed locations (dots). Brown lines: elevation contours at $500$~m (thin) and $1{,}000$~m (thick) from the ETOPO1 relief model \citep{amante2009etopo1}, with the names of the main relief units.}
\label{fig:application_factor_maps}
\end{figure}

In BNFC (nonstat), $W(\bfs)=Z(\bfs)+g(V,\bfs)$, and the share
$
\varrho(\bfs)=\frac{\operatorname{var}\{g(V,\bfs)\}}{1+\operatorname{var}\{g(V,\bfs)\}}
$
of the latent variance at $\bfs$ due to the factor measures how strongly a location takes part in the large-scale dependence; it is defined on the latent scale and is invariant to the transformations of Remark~\ref{rem:identifiability}. The skewness of $g(V,\bfs)$ describes the asymmetry of the factor contribution. Figure~\ref{fig:application_factor_maps} shows both, together with the model-implied and empirical correlations with a location near Vienna. The Gaussian component has an exponential correlation and describes short-range residual dependence. The factor explains a little under half of the latent variance on average, most over the lowlands and least in the mountains, and its skewness changes sign over the domain, from negative in the north to positive over the Pannonian basin, which neither a common factor copula nor the gamma loading copula can represent. These patterns are stable across the weight samples, and the implied correlations closely match the empirical ones. Section~\ref{S-app:application_tail} of the Supplementary Material relates the maps to the relief, shows the factor over the test period and the tail dependence of the fitted model, and discusses what the posterior adds to a plug-in fit.

\section{Discussion}\label{sec:discussion}

The BNFC model removes two restrictions of factor copulas while keeping their one-dimensional integrals: the factor distribution is learned from the data, and in the location-dependent model its effect may change over space in strength, shape and sign. The simulations show that BNFC performs well when the factor family is unknown or misspecified, and that the larger gains come from allowing the effect of the factor to vary across locations rather than from learning its distribution: even a gamma mixture whose components match the true factor falls behind once the factor behaves differently in different regions. The loading copulas show that this gain is not only second-order. Averaging over the approximate weight samples gives prediction intervals with close to nominal coverage in the simulations and slightly below nominal coverage in the application (Table~\ref{S-tab:supp_app_all} of the Supplementary Material). In the application, BNFC (nonstat) has the best held-out scores of all copula models, and its factor changes the sign of its skewness over the domain.

The flexibility has a limit in the tails: every fitted network gives an asymptotically independent copula, and what the model learns is dependence at fixed tail levels, which is what a finite number of replicates can inform. A heavy-tailed base variable would restore asymptotic dependence, but with a slope extrapolated beyond the data. 
Additionally in this work a neural network is preferred over a spline in $(v,\bfs)$ because it scales with the input dimension and represents loadings of either sign without constraints; finite mixtures \citep{mclachlan2000finite} and Bayesian nonparametric priors such as Dirichlet process mixtures \citep{ferguson1973bayesian,escobar1995bayesian} are less convenient because the Monte Carlo likelihood needs a reparameterizable sampler. With the Vecchia approximation \citep{vecchia1988estimation,katzfuss2021general}, the per-epoch cost scales linearly with the numbers of replicates and locations for fixed approximation settings; accuracy at larger scales requires separate assessment.

Several promising directions for future research remain. In the precipitation application, the copula density assumes continuous marginal distributions, whereas precipitation exhibits a point mass at zero, which the randomized rank transformation addresses approximately. Additionally, the days are treated as independent replicates after removing the overall temporal trend, despite the presence of serial dependence. A natural extension would therefore be to develop a Bayesian factor copula model for spatio-temporal data, which would explicitly account for temporal dependence while retaining the flexibility of the proposed framework. Another promising direction is to consider multiple factors operating across different spatial regions and temporal intervals, allowing for strong tail dependence within regions while permitting weaker or vanishing dependence across regions. Although this extension may introduce additional computational challenges through multidimensional integrals, structured constructions such as nested or bi-factor copulas offer potential solutions \citep{krupskii2015structured}.

\if0\blind
\section*{Acknowledgements}
Nadja Klein acknowledges funding from the German Research Foundation (DFG, Deutsche
Forschungsgemeinschaft) through the Emmy Noether Grant KL3037/1-1 and project A07 of
the TRR391 (INST 212/537-1).
\fi

\section*{Supplementary Material}
\begin{description}
\item[Supplement:] Related approaches, proofs, algorithms, further simulation and application results, the Vecchia approximation, and implementation details. The Supplementary Material follows the References of this article.
\item[Code:] Python code to reproduce all results, tables and figures is available at \if0\blind\url{https://github.com/pratiknag/BNFC_code.git}\else\texttt{<github link>}\fi{}; instructions for running the code are given in \texttt{README.md}. The ERA5 data are publicly available from the Copernicus Climate Data Store; the code includes a script to download them, with instructions in \texttt{README.md}.
\end{description}

\begin{singlespace}
\bibliographystyle{dcu}
\bibliography{ref}
\end{singlespace}
\end{document}


\title{Supplementary Material for ``Bayesian Neural Factor Copula Model for Spatial Data''}	
\author{}
\date{}
\renewcommand\Authands{ and }
\maketitle

References to equations, sections, results, figures and tables of the main paper are marked explicitly, such as Equation~(1) or Section~1 of the main text, and those with the prefix S, such as (S1) or Section~S1, refer to this Supplementary Material. Acronyms that are not defined here are defined in the main paper.

This Supplementary Material follows the main text of the article ``Bayesian Neural Factor Copula Model for Spatial Data''. It contains a discussion of related approaches (Section~\ref{app:related}), the proofs of all results in Section 3 (Sections~\ref{app:proofs_neural_factors} and~\ref{app:tail_proofs}), the derivation of the conditional copula of Section 4 (Section~\ref{app:conditional_copula_derivation}), the details of the simulation design and competing models with additional results on calibration and sparse conditioning (Section~\ref{app:additional_sim}), additional results for the precipitation application, including the tail dependence of the fitted model (Section~\ref{app:application_tail}), the Vecchia approximation with timing benchmarks and the fit to the complete precipitation data (Section~\ref{app:vecchia}), and the construction of the marginal network and further implementation details of the estimation procedure (Section~\ref{app:implementation}). The prediction and estimation algorithms are given in Section~\ref{app:supp_algorithm}. Code to reproduce all results is available at \if0\blind\url{https://github.com/pratiknag/BNFC_code.git}\else\texttt{<github link>}\fi{}. The ERA5 data are publicly available from the Copernicus Climate Data Store.

\section{Related approaches}\label{app:related}

This section expands Section~\ref{M-sec:related} of the main text. In a Gaussian scale mixture, a common positive random variable $R$, independent of $Z$, multiplies the Gaussian process, $W(\bfs)=R\,Z(\bfs)$. The distribution of $R$ determines the tails. In the Student-$t$ copula \citep{demarta2005t}, $R=G^{-1/2}$ with $G\sim\mathrm{Ga}(\nu/2,\text{rate }\nu/2)$, and \citet{huser2017bridging} use a parametric family for $R$ that interpolates between asymptotic dependence and asymptotic independence. When second moments exist, the correlation of $W$ is that of $Z$, and a large value of $R$ makes positive values more positive and negative values more negative rather than raising the process at every location. The copula is reflection symmetric, so that the dependence of two high extremes equals that of two low extremes, and that of a high and a low extreme is the same in both orders; a Student-$t$ copula can link a high extreme at one location with a low extreme at another only through a negative Gaussian correlation \citep{demarta2005t}. With a nonnegative correlation function, as in the comparison models of this paper, the dependence between distant locations is therefore nonnegative and the same wherever the locations are. Neural networks have been used to build flexible distributions \citep{rezende2015variational,durkan2019neural} and copulas \citep{tagasovska2019copulas,ling2020deep}. These neural copulas are, however, not designed for spatial data: they neither respect the decay of dependence with distance nor provide predictions at unobserved locations. BNFC uses a neural network only for the one-dimensional factor and keeps the Gaussian process for the spatial structure. For inference, we use variational methods \citep{kingma2013auto,blei2017variational} (Section~\ref{M-subsec:vb_inference} of the main text), which are separate from the use of neural networks for modeling. Table~\ref{tab:positioning} summarizes how BNFC relates to Gaussian processes, Gaussian scale mixtures and the common factor copula. Related constructions within the class of factor copulas include structured and linear factor copulas \citep{krupskii2015structured,krupskii_genton_2018}, spatio-temporal factor copulas \citep{krupskii2017spatiotemporal} and local likelihood methods for complex tail dependence \citep{castrocamilo_huser_2020}.

\begin{table}[tbp]
\centering
\caption{Dependence properties of the models, for the nonnegative correlation functions of the Gaussian component used in this paper. BNFC (stat) and BNFC (nonstat) are models (4) and \eqref{M-eq:model_definition_varying_locs} of the main text, with factor map $g(v)$ and $g(v,\bfs)$ (Section~\ref{M-subsec:neural_factors}); the loading copulas $W(\bfs)=Z(\bfs)+b(\bfs)X$ have a Gaussian or gamma factor $X$ and a loading $b(\bfs)$ of either sign (Section~\ref{M-sec:simulation_studies}). $\lambda_L$, $\lambda_U$: lower and upper tail dependence coefficients, $\lambda=\lim_{q\to0}\lambda(q)$, with $\lambda(q)$ in Equation~\eqref{M-eq:lambda_q} of the main text; \emph{intermediate}: $0<\lambda<1$; \emph{tail asymmetry}: $\lambda_L\neq\lambda_U$ or $\lambda_L(q)\neq\lambda_U(q)$; \emph{spatial variation}: the effect of the factor can differ between locations; \emph{flexible factor}: its distribution is learned rather than chosen. Sections and propositions are those of the main text. \yes: yes; \no: no.}
\label{tab:positioning}
\renewcommand{\arraystretch}{1.3}
\footnotesize
\newcolumntype{L}[1]{>{\raggedright\arraybackslash}p{#1}}
\newcolumntype{C}[1]{>{\centering\arraybackslash}p{#1}}
\begin{tabular}{@{}L{2.5cm}L{3.0cm}L{2.5cm}C{1.5cm}L{2.4cm}C{1.3cm}@{}}
\toprule
Model & Limiting $\lambda_L$, $\lambda_U$ & Finite-level $\lambda(q)$ & Tail asymmetry & Spatial variation & Flexible factor \\
\midrule
\multicolumn{6}{@{}l}{\textit{Existing models}} \\
Gaussian process        & $0$ & Gaussian & \no & \na & \na \\
Gaussian scale mixture  & $\lambda_L=\lambda_U\geq0$, fixed by mixing family & fixed by mixing family & \no & \no & \no \\
Common factor copula    & $0$, intermediate or $1$, fixed by family & fixed by family & fixed by family & \no & \no \\
\addlinespace[2pt]
\multicolumn{6}{@{}l}{\textit{Spatially varying loading}} \\
Gaussian loading copula & $0$ & Gaussian & \no & scale, sign & \no \\
Gamma loading copula    & fixed by family and loadings & fixed by family and loadings & \yes & scale, sign & \no \\
\addlinespace[2pt]
\multicolumn{6}{@{}l}{\textit{Proposed}} \\
BNFC (stat)    & $0$ (Prop.~4) & any target (Prop.~5) & \yes & \no & \yes \\
BNFC (nonstat) & $0$ (Prop.~4) & any target (Prop.~5) & \yes & scale, sign, shape (Prop.~2) & \yes \\
\bottomrule
\end{tabular}
\end{table}

\newpage 

\section{Proofs for \texorpdfstring{Section~\ref{M-subsec:neural_factors}}{Section~3.1} of the main text}
\label{app:proofs_neural_factors}

This section provides the proofs of the results stated in Section~\ref{M-subsec:neural_factors} of the main text. Throughout, $V\sim U(\theta_1,\theta_2)$ with $\theta_1<\theta_2$, and, for a distribution function $H$ on $\mathbb{R}$, $H^{-1}(p)=\inf\{x\in\mathbb{R}:H(x)\ge p\}$ for $p\in(0,1)$. We use the following property of the generalized inverse, which follows from the right-continuity of $H$: for all $p\in(0,1)$ and $x\in\mathbb{R}$,
\begin{equation}
H^{-1}(p)\le x \iff p\le H(x).
\label{eq:supp_galois}
\end{equation}

\subsection{Proof of \texorpdfstring{Proposition~\ref{M-prop:factor_generation}}{Proposition~1} of the main text}
\label{app:proof_factor_generation}

\noindent\textbf{Proposition~\ref{M-prop:factor_generation}
(Factor generation and approximation).}
Let $H$ be a distribution function on $\mathbb{R}$ and let
$V\sim U(\theta_1,\theta_2)$. Define
\[
T_{\bftheta}(v)
=
H^{-1}\!\left\{\frac{v-\theta_1}{\theta_2-\theta_1}\right\},
\qquad v\in(\theta_1,\theta_2),
\]
where $H^{-1}(p)=\inf\{x\in\mathbb{R}:H(x)\ge p\}$ for $p\in(0,1)$.

\begin{enumerate}
\item[(a)] $T_{\bftheta}$ is nondecreasing and
$T_{\bftheta}(V)\sim H$.

\item[(b)] If $T_{\bftheta}$ is continuous on $(\theta_1,\theta_2)$,
there exist feedforward neural networks $g_m$, $m\ge1$, each with one
hidden layer of $m$ units and a continuous nonpolynomial activation
function, such that
\[
\sup_{v\in K}|g_m(v)-T_{\bftheta}(v)|\longrightarrow 0
\qquad\text{as }m\to\infty
\]
for every compact set $K\subset(\theta_1,\theta_2)$. If, in addition,
$H$ has bounded support, the convergence is uniform on
$[\theta_1,\theta_2]$, with $T_{\bftheta}$ extended continuously
to the endpoints.

\item[(c)] If measurable functions $g_m$ converge to $T_{\bftheta}$
uniformly on every compact subset of $(\theta_1,\theta_2)$, then
$g_m(V)\to T_{\bftheta}(V)$ almost surely. Consequently, the law $H_m$
of $g_m(V)$ converges weakly to $H$.
\end{enumerate}

\begin{proof}
\emph{Part (a).} Let $P=(V-\theta_1)/(\theta_2-\theta_1)$. Since $V\sim U(\theta_1,\theta_2)$, $P\sim U(0,1)$, and $P\in(0,1)$ almost surely. By \eqref{eq:supp_galois}, for every $x\in\mathbb{R}$,
\[
\Pr\{T_{\bftheta}(V)\le x\}=\Pr\{H^{-1}(P)\le x\}=\Pr\{P\le H(x)\}=H(x),
\]
so $T_{\bftheta}(V)\sim H$. Since $H^{-1}$ is nondecreasing and $v\mapsto(v-\theta_1)/(\theta_2-\theta_1)$ is increasing, $T_{\bftheta}$ is nondecreasing.

\emph{Part (b).} For $r\ge1$, let
\[
\delta_r=\frac{\theta_2-\theta_1}{2r+2},
\qquad
K_r=[\theta_1+\delta_r,\theta_2-\delta_r].
\]
Fix a continuous nonpolynomial hidden activation. By the universal
approximation theorem \citep[Theorem~1]{leshno1993multilayer}, there
exists a finite-width network $h_r$ with one hidden layer, hidden-unit
biases, and a linear output layer such that
\[
\sup_{v\in K_r}|h_r(v)-T_{\bftheta}(v)|<\frac{1}{r}.
\]

Choose strictly increasing integers $m_1,m_2,\ldots$ such that $m_r$
is at least the width of $h_r$. For $m_r\le m<m_{r+1}$, define $g_m$
by adding hidden units with zero output weights to $h_r$ until its
width is exactly $m$. The added units contribute zero, so $g_m=h_r$
as functions. For $m<m_1$, take a network of width $m$ representing
the zero function.

Let $K\subset(\theta_1,\theta_2)$ be compact. Since $K$ lies a
positive distance from the endpoints, $K\subseteq K_r$ for all
sufficiently large $r$. Also, the index $r$ satisfying
$m_r\le m<m_{r+1}$ tends to infinity as $m\to\infty$. Thus, for all
sufficiently large $m$,
\[
\sup_{v\in K}|g_m(v)-T_{\bftheta}(v)|
\le \sup_{v\in K_r}|h_r(v)-T_{\bftheta}(v)|
<\frac{1}{r}\longrightarrow0.
\]

If $H$ has bounded support, then $T_{\bftheta}$ is bounded and
nondecreasing, so it has finite one-sided limits at both endpoints.
Assigning these limits as endpoint values gives a continuous
extension to $[\theta_1,\theta_2]$. Applying the same construction
with $K_r=[\theta_1,\theta_2]$ for every $r$ gives uniform convergence
on the closed interval.



\emph{Part (c).} Let $\Omega_0=\{\omega:V(\omega)\in(\theta_1,\theta_2)\}$, so that $\Pr(\Omega_0)=1$. Fix $\omega\in\Omega_0$ and let $v=V(\omega)$. Choose $\delta>0$ such that $K=[v-\delta,v+\delta]\subset(\theta_1,\theta_2)$. Since $g_m\to T_{\bftheta}$ uniformly on the compact set $K$,
\[
|g_m(v)-T_{\bftheta}(v)|\le\sup_{v'\in K}|g_m(v')-T_{\bftheta}(v')|\longrightarrow0.
\]
Hence $g_m(V)\to T_{\bftheta}(V)$ on $\Omega_0$, that is, almost surely. Almost sure convergence implies convergence in distribution, which is equivalent to the weak convergence $H_m\Rightarrow H$ of the induced laws.
\end{proof}

Note that part (c) does not require uniform convergence on all of $(\theta_1,\theta_2)$, which is important when $T_{\bftheta}$ is unbounded, that is, when the target factor distribution has unbounded support.

\begin{remark}[Networks with a positive output activation]
Let $\sigma_{\mathrm{out}}\colon\mathbb{R}\to(0,\infty)$ be
continuous, strictly increasing, and surjective output activation. On each compact interval
$K_r$ from part~(b), shift the continuous target $T_{\bftheta}$ by
\[
c_r=1-\min_{v\in K_r}T_{\bftheta}(v).
\]
The shifted target $T_{\bftheta}(v)+c_r$ is at least $1$ throughout
$K_r$, so the inverse activation is defined at every target value.

By the universal approximation theorem, a network $\tilde g_r$
with a linear output layer can approximate the continuous function
\[
v\longmapsto
\sigma_{\mathrm{out}}^{-1}\{T_{\bftheta}(v)+c_r\}
\]
uniformly on $K_r$. Applying $\sigma_{\mathrm{out}}$ to its output
gives the positive-output network
$g_r=\sigma_{\mathrm{out}}\circ\tilde g_r$. The inverse-transformed
target has compact range, and $\sigma_{\mathrm{out}}$ is uniformly
continuous on a compact interval containing this range and the
nearby network outputs. We can therefore choose $\tilde g_r$ so that
\[
\sup_{v\in K_r}
\bigl|g_r(v)-\{T_{\bftheta}(v)+c_r\}\bigr|<\frac1r.
\]

Thus the recentered maps $g_r-c_r$ converge to $T_{\bftheta}$
uniformly on every compact subset of $(\theta_1,\theta_2)$.
Adding $c_r$ to the recentered factor shifts every latent coordinate
by the same constant and preserves the copula
(Remark~\ref{M-rem:identifiability} of the main text).
The positive-output network $g_r$ therefore induces the same copula
as $g_r-c_r$ at each stage. Proposition~\ref{M-prop:subasymptotic_convergence}
of the main text gives convergence of these copulas to the target copula.
\end{remark}


\subsection{Proof of \texorpdfstring{Lemma~\ref{M-lemma:location_specific_factor}}{Lemma~1} of the main text}
\label{app:proof_location_specific_factor}

\noindent\textbf{Lemma~\ref{M-lemma:location_specific_factor}
(Location-specific factor generation).}
For each $\bfs\in D$, let $H(\cdot,\bfs)$ be a distribution function
on $\mathbb{R}$. Let $V\sim U(\theta_1,\theta_2)$ and define
\[
T_{\bftheta}(v,\bfs)
=
H^{-1}\!\left\{
\frac{v-\theta_1}{\theta_2-\theta_1},\bfs
\right\},
\qquad v\in(\theta_1,\theta_2),
\]
where $H^{-1}(\cdot,\bfs)$ is the generalized inverse of
$H(\cdot,\bfs)$. Then $T_{\bftheta}(V,\bfs)\sim H(\cdot,\bfs)$
for every $\bfs\in D$. For any $\bfs_1,\ldots,\bfs_n\in D$, the
random vector
\[
\{T_{\bftheta}(V,\bfs_1),\ldots,T_{\bftheta}(V,\bfs_n)\}^{\top}
\]
is comonotonic.

\begin{proof}
\emph{Step 1. Verify the distribution at each location.}
Let $P=(V-\theta_1)/(\theta_2-\theta_1)$. As in the proof of
Proposition~\ref{M-prop:factor_generation}(a), $P\sim U(0,1)$.
At a fixed location $\bfs$, we evaluate the quantile function
$H^{-1}(\cdot,\bfs)$ at $P$. Using the generalized inverse property
in \eqref{eq:supp_galois}, we obtain, for every $x\in\mathbb{R}$,
\[
\begin{aligned}
\Pr\{T_{\bftheta}(V,\bfs)\le x\}
&=\Pr\{H^{-1}(P,\bfs)\le x\}\\
&=\Pr\{P\le H(x,\bfs)\}\\
&=H(x,\bfs).
\end{aligned}
\]
Thus the factor contribution at $\bfs$ has the 
distribution function $H(\cdot,\bfs)$.

\smallskip
\noindent\emph{Step 2. Show comonotonicity across locations.}
Each location uses its own quantile function
$H^{-1}(\cdot,\bfs_k)$, and all locations evaluate their quantile
functions at the same $P$. Since each quantile function is
nondecreasing, an increase in $P$ can only increase each
coordinate or leave it equal to its previous value. Hence all
coordinates are nondecreasing functions of a single random
variable, which gives comonotonicity.

\end{proof}

\subsection{Proof of \texorpdfstring{Proposition~\ref{M-thm:covariance}}{Proposition~2} of the main text}
\label{app:proof_covariance}

\noindent\textbf{Proposition~\ref{M-thm:covariance}
(Attainable covariance structures).}
Let $\bs{\Sigma}_{\bs{\vartheta}}$ be an $n\times n$ correlation
matrix, let $\bs{C}$ be any $n\times n$ positive semidefinite
matrix, and let $V\sim U(\theta_1,\theta_2)$. At any $n$ distinct
locations $\bfs_1,\ldots,\bfs_n$, there exist continuous functions
$g(\cdot,\bfs_1),\ldots,g(\cdot,\bfs_n)$ on $[\theta_1,\theta_2]$
such that
\[
\bs{G}(V)=\{g(V,\bfs_1),\ldots,g(V,\bfs_n)\}^{\top}
\qquad\text{and}\qquad
\operatorname{cov}\{\bs{G}(V)\}=\bs{C}.
\]
Hence $\bfW=\bfZ+\bs{G}(V)$, with
$\bfZ\sim\mathcal{N}_n(\bs{0}_n,\bs{\Sigma}_{\bs{\vartheta}})$
independent of $V$, has
$\operatorname{cov}(\bfW)=\bs{\Sigma}_{\bs{\vartheta}}+\bs{C}$.
In particular, for $j\ne k$, if $C_{jj}=C_{kk}=\sigma^2$, then
$\operatorname{cor}(W_j,W_k)$ can take any value in
\[
\left[
\frac{(\bs{\Sigma}_{\bs{\vartheta}})_{jk}-\sigma^2}{1+\sigma^2},
\;
\frac{(\bs{\Sigma}_{\bs{\vartheta}})_{jk}+\sigma^2}{1+\sigma^2}
\right],
\]
whose endpoints tend to $-1$ and $1$ as $\sigma^2\to\infty$.
The functions can be approximated uniformly on $[\theta_1,\theta_2]$
by a feedforward neural network, so that the factor covariance
matrix of a BNFC model can be made arbitrarily close to $\bs{C}$.

\begin{proof}
Let $h_1,\ldots,h_n$ be the shifted Legendre polynomials of degrees
$1,\ldots,n$ on $[\theta_1,\theta_2]$, normalized under the uniform
distribution of $V$. Their orthonormality gives
$\mathbb{E}\{h_\ell(V)h_m(V)\}=\delta_{\ell m}$, where
$\delta_{\ell m}$ is the Kronecker delta. Each polynomial also
has mean zero because it is orthogonal to the constant polynomial.

Since $\bs{C}$ is positive semidefinite, choose a matrix $\bs{A}$
such that $\bs{A}\bs{A}^{\top}=\bs{C}$. For instance, $\bs{A}$
can be the symmetric square root of $\bs{C}$. Define the continuous
factor function at location $\bfs_k$ by
\[
g(v,\bfs_k)=\sum_{\ell=1}^n A_{k\ell}h_\ell(v).
\]
These factor contributions have mean zero. Using orthonormality,
\[
\begin{aligned}
\operatorname{cov}\{g(V,\bfs_j),g(V,\bfs_k)\}
&=\sum_{\ell=1}^n\sum_{m=1}^n
 A_{j\ell}A_{km}\,\mathbb{E}\{h_\ell(V)h_m(V)\}\\
&=\sum_{\ell=1}^n A_{j\ell}A_{k\ell}\\
&=(\bs{A}\bs{A}^{\top})_{jk}=C_{jk}.
\end{aligned}
\]
Thus $\operatorname{cov}\{\bs{G}(V)\}=\bs{C}$. Independence of
$\bfZ$ and $V$ then gives
$\operatorname{cov}(\bfW)=\bs{\Sigma}_{\bs{\vartheta}}+\bs{C}$.

For the correlation claim, the Gaussian marginal variances are
one, so $C_{jj}=C_{kk}=\sigma^2$ gives
\[
\operatorname{cor}(W_j,W_k)
=
\frac{(\bs{\Sigma}_{\bs{\vartheta}})_{jk}+C_{jk}}{1+\sigma^2}.
\]
The $2\times2$ block of $\bs{C}$ for locations $j$ and $k$ has
eigenvalues $\sigma^2+C_{jk}$ and $\sigma^2-C_{jk}$. Both are
nonnegative exactly when $C_{jk}\in[-\sigma^2,\sigma^2]$.
Each value in this interval is attainable by placing the
corresponding block at coordinates $j,k$ and setting the other
entries of $\bs{C}$ to zero. Substituting the two extreme values
of $C_{jk}$ gives the stated correlation interval. Its lower
and upper endpoints tend to $-1$ and $1$, respectively.

Finally, the functions $g(\cdot,\bfs_k)$ together define a
continuous function of $(v,\bfs)$ on the compact set
$[\theta_1,\theta_2]\times\{\bfs_1,\ldots,\bfs_n\}$.
The universal approximation theorem \citep{hornik1989multilayer}
therefore gives a single feedforward network with inputs $v$
and $\bfs$ that approximates $g$ uniformly on this set. Uniform convergence on this compact set gives entrywise convergence
of the factor covariance matrix to $\bs{C}$.
\end{proof}


\subsection{Monotonicity of the factor network}\label{app:monotone}

The network $g$ is not constrained to be monotone in $v$. Monotonicity in Proposition~\ref{M-prop:factor_generation} of the main text is a property of the target $T_{\bftheta}$, a quantile function, and not a requirement on $g$: the approximating networks in (b) and (c) need not be monotone, and the copula density in Equation~\eqref{M-eq:bnfc_copula_density} of the main text integrates over $v$ and involves neither the inverse nor the derivative of $g$, so that it is defined for any measurable $g$. In the stationary model, the copula depends on $g$ only through the law of $g(V)$, which is also the law of its nondecreasing rearrangement $g^\ast(V)$, where $g^\ast$ is the quantile function of $g(V)$ composed with the affine map of $(\theta_1,\theta_2)$ onto $(0,1)$. A non-monotone $g$ therefore defines the same copula as the monotone $g^\ast$, which is a further instance of the non-identifiability in Remark~\ref{M-rem:identifiability} of the main text, and a monotonicity constraint would neither enlarge nor restrict the model. In the location-dependent model, the copula depends on the joint law of $\{g(V,\bfs_1),\ldots,g(V,\bfs_n)\}^\top$, and here monotonicity would be restrictive. If $g(\cdot,\bfs)$ were nondecreasing for every $\bfs$, the factor contributions would be comonotonic, as in Lemma~\ref{M-lemma:location_specific_factor} of the main text, and all covariances $\operatorname{cov}\{g(V,\bfs_j),g(V,\bfs_k)\}$ would be nonnegative, which excludes the negative dependence of Proposition~\ref{M-thm:covariance} of the main text. Leaving $g$ unconstrained allows the factor to act in opposite directions in different regions, as in the sign-change scenario of Section~\ref{M-sec:simulation_studies} of the main text.

\subsection{Properties of the location-dependent model}\label{app:three_properties}

This section expands the discussion of Proposition~\ref{M-thm:covariance} in Section~\ref{M-subsec:neural_factors} of the main text.
Three properties of the location-dependent model should be distinguished. First, the covariance structure is flexible. Writing $\sigma_k^2=\operatorname{var}\{g(V,\bfs_k)\}$ and $c_{jk}=\operatorname{cov}\{g(V,\bfs_j),g(V,\bfs_k)\}$, we have $\operatorname{cor}(W_j,W_k)=\{(\bs{\Sigma}_{\bs{\vartheta}})_{jk}+c_{jk}\}/\{(1+\sigma_j^2)(1+\sigma_k^2)\}^{1/2}$ with $|c_{jk}|\le\sigma_j\sigma_k$. For given distributions of the two factor contributions, $c_{jk}$ is largest when they are comonotone and smallest when they are countermonotone, that is, nondecreasing and nonincreasing functions of one another, by the Hoeffding--Fr\'echet bounds, and the lower bound of Proposition~\ref{M-thm:covariance} of the main text is attained with $g(\cdot,\bfs_k)=-g(\cdot,\bfs_j)$. Arbitrarily strong negative dependence between two regions is therefore obtained when the factor acts on them in opposite directions with a large effect. Second, a single random variable drives all locations. The factor contributions at all locations are functions of the same $V$, so that the model cannot produce, for example, two regions whose large-scale states vary independently of each other; this is a restriction on the higher-order structure, not on the covariances. Third, the factor links the tails of different locations in a way determined by the direction of its effect. If the factor raises the process at $\bfs_j$ and lowers it at $\bfs_k$, a large value of the factor produces a high extreme at $\bfs_j$ together with a low extreme at $\bfs_k$. In the sign-change scenario of Section~\ref{M-sec:simulation_studies} of the main text, for example, a factor value that produces heavy rainfall in the north produces dry conditions in the south. The factor then associates the upper tail at one location with the lower tail at the other, rather than creating upper--upper tail dependence between them; Corollary~\ref{M-cor:opposite_loadings} in Section~\ref{M-sec:tail_properties} of the main text makes this precise.

\subsection{Identifiability}\label{app:identifiability}

This section gives the details of Remark~\ref{M-rem:identifiability} of the main text. The parameters $\bftheta$, the network $g$ and the location of the factor are not identifiable. First, for any other interval $(\theta_1',\theta_2')$, the affine map $\tau(v)=\theta_1'+(\theta_2'-\theta_1')(v-\theta_1)/(\theta_2-\theta_1)$ maps $U(\theta_1,\theta_2)$ to $U(\theta_1',\theta_2')$, so $(\bftheta,g)$ and $(\bftheta',g\circ\tau^{-1})$ define the same process. More generally, any map $\rho$ that preserves the uniform distribution on $(\theta_1,\theta_2)$, such as the reflection $\rho(v)=\theta_1+\theta_2-v$, gives the same process with $g\circ\rho$ in place of $g$. Second, adding a function $c(\bfs)$ to $g(v,\bfs)$ shifts each $W(\bfs)$ by a constant and leaves the copula unchanged, since copulas are invariant to strictly increasing transformations of the margins. What these transformations leave unchanged is the law of the centered factor contribution
\[
\tilde g(V,\bfs)=g(V,\bfs)-\mathbb{E}\{g(V,\bfs)\},
\]
jointly over the locations, which, together with $\bs{\Sigma}_{\bs{\vartheta}}$, determines the copula, because the Gaussian component has unit variance. The same holds conditionally on the data: the conditional law of $\tilde g(V,\bfs)$ given the copula data of a replicate does not depend on the representation, so that it can be compared across models and with a data-generating model, as in Section~\ref{M-sec:simulation_studies} of the main text. In the stationary model, for a given $\bs{\Sigma}_{\bs{\vartheta}}$, the joint law of $\bfW-\mathbb{E}\bfW=\bfZ+\tilde g(V)\mathbf{1}_n$ also determines the law of $\tilde g(V)$, because the characteristic function of the Gaussian component has no zeros. We therefore interpret the law of the centered factor contribution, such as its shape, skewness and tail weight, and derived dependence summaries, such as $\lambda(q)$ in \eqref{M-eq:lambda_q} of the main text and the skewness measure $\zeta_1$ of Proposition~\ref{M-prop:subasymptotic_convergence} of the main text, but not $g$, $\bftheta$ or the location of the factor. In particular, $\bftheta$ parameterizes the input range of the network, and its posterior is informed by the prior and by the inductive bias of $g$. In the location-dependent model, we interpret only derived dependence summaries.

\section{Tail properties: proofs, illustration and discussion}
\label{app:tail_proofs}

Throughout this section, $\bfs_i\neq\bfs_j$ are fixed, $W_k=Z_k+g(V,\bfs_k)$ for $k\in\{i,j\}$, $(Z_i,Z_j)$ is bivariate normal with mean zero, common variance $\sigma^2$ and correlation $\rho<1$, and $(Z_i,Z_j)$ is independent of $V$. In the main text, $\sigma^2=1$. Each $W_k$ has the continuous density $\mathbb{E}[\sigma^{-1}\phi\{(w-g(V,\bfs_k))/\sigma\}]>0$, so its distribution function $F_k$ is continuous and strictly increasing, and $U_k=F_k(W_k)\sim U(0,1)$. For $q\in(0,1)$, let $t_k(q)=F_k^{-1}(1-q)$ and $r_k(q)=F_k^{-1}(q)$, so that
\[
\begin{gathered}
\Pr(U_i>1-q,U_j>1-q)=\Pr\{W_i>t_i(q),W_j>t_j(q)\},\\
C_2(q,q)=\Pr\{W_i\le r_i(q),W_j\le r_j(q)\}.
\end{gathered}
\]
Let $\bar{\Phi}=1-\Phi$. We use the following auxiliary lemmas.

\begin{lemma}[Gaussian joint tail bound]
\label{lemma:supp_gauss_joint}
Let $(Z_i,Z_j)$ be a centered bivariate Gaussian vector with common
variance $\sigma^2>0$ and correlation $\rho$. For $-1<\rho<1$, set
$\kappa=\{2/(1+\rho)\}^{1/2}>1$. Then, for every $a\in\mathbb{R}$,
\[
\Pr(Z_i>a,Z_j>a)\le\bar{\Phi}(\kappa a/\sigma),
\qquad
\Pr(Z_i\le a,Z_j\le a)\le\Phi(\kappa a/\sigma).
\]
If $\rho=-1$, the joint upper-tail probability is zero for $a\ge0$,
and the joint lower-tail probability is zero for $a\le0$.
\end{lemma}

\begin{proof}
If both $Z_i$ and $Z_j$ exceed $a$, their sum exceeds $2a$.
If both are at most $a$, their sum is at most $2a$. For
$-1<\rho<1$, the sum is Gaussian with mean zero and variance
\[
\operatorname{var}(Z_i+Z_j)
=\sigma^2+\sigma^2+2\rho\sigma^2
=2\sigma^2(1+\rho).
\]
Standardizing the threshold $2a$ gives
$2a/\{\sigma\sqrt{2(1+\rho)}\}=\kappa a/\sigma$. Consequently,
\[
\begin{aligned}
\Pr(Z_i>a,Z_j>a)
&\le \Pr(Z_i+Z_j>2a)
=\bar{\Phi}(\kappa a/\sigma),\\
\Pr(Z_i\le a,Z_j\le a)
&\le \Pr(Z_i+Z_j\le 2a)
=\Phi(\kappa a/\sigma).
\end{aligned}
\]
When $\rho=-1$, $Z_i+Z_j=0$ almost surely. The joint upper event
is then impossible for $a\ge0$, and the joint lower event is
impossible for $a<0$. At $a=0$, the joint lower event requires
$Z_i=Z_j=0$, which has probability zero.
\end{proof}

\begin{lemma}[Comparison of shifted Gaussian tails]
\label{lemma:supp_mills}
Let $\kappa>1$ and $c,d\in\mathbb{R}$ be fixed. Then
\[
\frac{\bar{\Phi}\{\kappa(x-c)\}}{\bar{\Phi}(x+d)}
\longrightarrow 0
\qquad\text{as }x\to\infty,
\]
and
\[
\frac{\Phi\{\kappa(x+c)\}}{\Phi(x-d)}
\longrightarrow 0
\qquad\text{as }x\to-\infty.
\]
\end{lemma}

\begin{proof}
As $x\to\infty$, both $\kappa(x-c)$ and $x+d$ tend to infinity.
Applying Mills' ratio, $\bar{\Phi}(y)\sim\phi(y)/y$, to the
numerator and denominator gives
\[
\frac{\bar{\Phi}\{\kappa(x-c)\}}{\bar{\Phi}(x+d)}
\sim
\frac{x+d}{\kappa(x-c)}
\exp\left\{-\frac{\kappa^2(x-c)^2-(x+d)^2}{2}\right\}.
\]
The ratio tends to $1/\kappa$. The difference of squares is
\[
\kappa^2(x-c)^2-(x+d)^2
=(\kappa^2-1)x^2-2(\kappa^2c+d)x+\kappa^2c^2-d^2.
\]
Since $\kappa>1$, the leading quadratic term is positive, and
this difference tends to infinity. The exponential term therefore
tends to zero, proving the upper-tail limit.

For the lower-tail limit, symmetry gives $\Phi(t)=\bar{\Phi}(-t)$
and $\phi(-t)=\phi(t)$. By Mills' ratio,
\[
\Phi(t)\sim\frac{\phi(t)}{-t}
\qquad\text{as }t\to-\infty.
\]
As $x\to-\infty$, both $\kappa(x+c)$ and $x-d$ tend to $-\infty$.
Applying this approximation to the numerator and denominator gives
\[
\frac{\Phi\{\kappa(x+c)\}}{\Phi(x-d)}
\sim
\frac{x-d}{\kappa(x+c)}
\exp\left\{-\frac{\kappa^2(x+c)^2-(x-d)^2}{2}\right\}.
\]
The ratio $(x-d)/\{\kappa(x+c)\}$ tends to $1/\kappa$. Also,
\[
\kappa^2(x+c)^2-(x-d)^2
=
(\kappa^2-1)x^2+2(\kappa^2c+d)x+\kappa^2c^2-d^2.
\]
Since $\kappa>1$, the positive quadratic term dominates as
$x\to-\infty$, so this difference tends to $+\infty$. The
exponential therefore tends to zero, and so does the ratio of
the two lower-tail probabilities.
\end{proof}

In the proofs of Proposition~\ref{M-thm:positive_tail_dependence}(b)
and Proposition~\ref{M-thm:asymptotic_independence},
Lemma~\ref{lemma:supp_gauss_joint} bounds the joint tail probability
by a univariate Gaussian tail probability.
Lemma~\ref{lemma:supp_mills} shows that this bound is negligible
relative to the corresponding marginal tail probability, establishing
lower-tail independence for factor contributions bounded below and
independence in both tails for bounded factor contributions.





\subsection{Gaussian perturbation of a regularly varying variable}\label{app:gaussian_perturbation}

\begin{lemma}[Gaussian perturbation invariance]
\label{lemma:gaussian_invariance}
Suppose that the right tail of $X$ satisfies
$\Pr(X>x)=x^{-\xi}L(x)$ for all sufficiently large $x$, where
$\xi>0$ and $L$ is positive and slowly varying. Let $Z$ be a
Gaussian random variable independent of $X$. Then
\[
\Pr(X+Z>x)\sim\Pr(X>x)
\qquad\text{as }x\to\infty.
\]
\end{lemma}

\begin{proof}
Fix $\epsilon\in(0,1)$ and take $x>0$. If $X>(1+\epsilon)x$
and $Z>-\epsilon x$, then $X+Z>x$. Independence therefore gives
the lower bound
\[
\Pr(X+Z>x)
\ge
\Pr\{X>(1+\epsilon)x\}\Pr(Z>-\epsilon x).
\]
For the upper bound, if $X\le(1-\epsilon)x$ and $Z\le\epsilon x$,
then $X+Z\le x$. Thus, for $X+Z>x$ to occur, at least one of
these two thresholds must be exceeded. By the union bound,
\[
\Pr(X+Z>x)
\le
\Pr\{X>(1-\epsilon)x\}+\Pr(Z>\epsilon x).
\]

To compare the probabilities involving $X$, note that for every
fixed $t>0$, slow variation of $L$ gives
\[
\frac{\Pr(X>tx)}{\Pr(X>x)}
=
t^{-\xi}\frac{L(tx)}{L(x)}
\longrightarrow t^{-\xi}.
\]
Taking $t=1+\epsilon$ and $t=1-\epsilon$ gives the limits
$(1+\epsilon)^{-\xi}$ and $(1-\epsilon)^{-\xi}$, respectively.

For the Gaussian probabilities, $\Pr(Z>-\epsilon x)\to1$
because $-\epsilon x\to-\infty$. We also need to show that
$\Pr(Z>\epsilon x)$ tends to zero faster than $\Pr(X>x)$.
By Potter's bound with exponent $1$, we can fix $x_0$
sufficiently large that
\[
\frac{L(x_0)}{L(x)}\le 2\frac{x}{x_0},
\qquad x\ge x_0.
\]
Rearranging gives $L(x)\ge c/x$, where $c=x_0L(x_0)/2>0$.
Consequently, $\Pr(X>x)\ge cx^{-\xi-1}$ for $x\ge x_0$.
Since $Z$ is Gaussian, $\mathbb{E}(e^Z)<\infty$. Markov's
inequality applied to $e^Z$ gives
$\Pr(Z>\epsilon x)\le\mathbb{E}(e^Z)e^{-\epsilon x}$.
Combining these bounds yields
\[
0\le
\frac{\Pr(Z>\epsilon x)}{\Pr(X>x)}
\le
\frac{\mathbb{E}(e^Z)}{c}\,x^{\xi+1}e^{-\epsilon x}
\longrightarrow 0,
\]
because exponential decay dominates polynomial growth for
fixed $\epsilon>0$.

Dividing the lower and upper bounds for $\Pr(X+Z>x)$ by
$\Pr(X>x)$ and taking limits gives
\[
(1+\epsilon)^{-\xi}
\le
\liminf_{x\to\infty}\frac{\Pr(X+Z>x)}{\Pr(X>x)}
\le
\limsup_{x\to\infty}\frac{\Pr(X+Z>x)}{\Pr(X>x)}
\le
(1-\epsilon)^{-\xi}.
\]
These inequalities hold for every fixed $\epsilon\in(0,1)$.
As $\epsilon\downarrow0$, both outer bounds tend to $1$.
Hence $\Pr(X+Z>x)/\Pr(X>x)\to1$.
\end{proof}
In the proof of Proposition~\ref{M-thm:positive_tail_dependence}(a),
we apply Lemma~\ref{lemma:gaussian_invariance} to
$W_k/a_k=X+(Z_k+b_k)/a_k$ for $k=i,j$. The resulting equivalence
$\Pr(W_k/a_k>x)\sim\Pr(X>x)$ as $x\to\infty$ lets us compare the
upper quantiles at the two locations on the scale of the common
factor $X$ when proving that $\lambda_U^{(i,j)}=1$.



\subsection{Proof of \texorpdfstring{Proposition~\ref{M-thm:positive_tail_dependence}}{Proposition~3} of the main text}

\noindent\textbf{Proposition~\ref{M-thm:positive_tail_dependence}
(Heavy-tailed factors).}
Suppose that there exist a random variable $X=X(V)$ and constants
$a_i,a_j>0$ and $b_i,b_j\in\mathbb{R}$ such that
$g(V,\bfs_i)=a_iX+b_i$ and $g(V,\bfs_j)=a_jX+b_j$, and that $X$
is regularly varying with index $\xi>0$.
\begin{enumerate}
\item[(a)] The copula is upper tail dependent with
$\lambda_U^{(i,j)}=1$.
\item[(b)] If, in addition, $X$ is bounded below, then
$\lambda_L^{(i,j)}=0$, so that the copula is tail asymmetric.
\end{enumerate}

\begin{proof}
For $k\in\{i,j\}$, write $W_k=a_kX+b_k+Z_k$, where $X=X(V)$ is
independent of $(Z_i,Z_j)$. Let $\sigma^2>0$ be the common variance
of $Z_i$ and $Z_j$, with $\sigma=1$ in the main-text model, and let
$\rho=\operatorname{cor}(Z_i,Z_j)<1$. Write
$U_k=F_{1,k}(W_k)$ and define the upper and lower marginal thresholds
by
\[
t_k(q)=F_{1,k}^{-1}(1-q),
\qquad
r_k(q)=F_{1,k}^{-1}(q).
\]
The Gaussian convolution gives continuous, strictly increasing
marginal distribution functions with support $\mathbb{R}$. Thus
$\Pr\{W_k>t_k(q)\}=\Pr\{W_k\le r_k(q)\}=q$, with
$t_k(q)\to\infty$ and $r_k(q)\to-\infty$ as $q\downarrow0$.

\emph{Part (a).}
Let $\bar{F}_X(x)=\Pr(X>x)$ and
$x_q=\inf\{x:\bar{F}_X(x)\le q\}$. The right tail of $X$ is
unbounded, so $x_q\to\infty$ as $q\downarrow0$. By the definition
of $x_q$ and the right continuity of $\bar{F}_X$,
\[
\bar{F}_X(x_q)\le q\le\bar{F}_X(x_q-),
\]
where $\bar{F}_X(x_q-)$ is the left limit at $x_q$.

To compare the two bounds, fix $\epsilon\in(0,1)$. By monotonicity
of $\bar{F}_X$ and regular variation, for $x>0$,
\[
1\le
\frac{\bar{F}_X(x-)}{\bar{F}_X(x)}
\le
\frac{\bar{F}_X\{(1-\epsilon)x\}}{\bar{F}_X(x)}
\longrightarrow (1-\epsilon)^{-\xi}
\qquad\text{as }x\to\infty.
\]
Letting $\epsilon\downarrow0$ shows that
$\bar{F}_X(x-)/\bar{F}_X(x)\to1$. Applying this at $x_q$ gives
\[
1\le
\frac{q}{\bar{F}_X(x_q)}
\le
\frac{\bar{F}_X(x_q-)}{\bar{F}_X(x_q)}
\longrightarrow 1.
\]
Hence $\bar{F}_X(x_q)\sim q$.

\emph{Step 1. Bound the marginal thresholds.}
Since $a_k>0$, we have $W_k/a_k=X+(Z_k+b_k)/a_k$. The second
term is Gaussian and independent of $X$, so
Lemma~\ref{lemma:gaussian_invariance} gives
$\Pr(W_k/a_k>x)\sim\bar{F}_X(x)$ as $x\to\infty$. Fix
$\epsilon\in(0,1)$. Using $\bar{F}_X(x_q)\sim q$ and regular
variation, we obtain
\[
\frac{\Pr\{W_k>(1+\epsilon)a_kx_q\}}{q}
\sim
\frac{\bar{F}_X\{(1+\epsilon)x_q\}}{\bar{F}_X(x_q)}
\longrightarrow (1+\epsilon)^{-\xi}<1.
\]
For sufficiently small $q$, the exceedance probability at
$(1+\epsilon)a_kx_q$ is therefore smaller than $q$. Since
$\Pr\{W_k>t_k(q)\}=q$ and the marginal tail probability is
strictly decreasing, we have
\[
t_k(q)<(1+\epsilon)a_kx_q,
\qquad k\in\{i,j\}.
\]

\emph{Step 2. Bound the joint exceedance probability.}
Consider the event
\[
E_q
=
\{X>(1+2\epsilon)x_q\}
\cap\{Z_i+b_i>-\epsilon a_ix_q\}
\cap\{Z_j+b_j>-\epsilon a_jx_q\}.
\]
On $E_q$, the large value of $X$ is sufficient for both
thresholds to be exceeded. Indeed, for $k\in\{i,j\}$,
\[
W_k
>
(1+2\epsilon)a_kx_q-\epsilon a_kx_q
=
(1+\epsilon)a_kx_q
>
t_k(q)
\]
for sufficiently small $q$. Independence of $X$ and
$(Z_i,Z_j)$ then gives
\[
\begin{aligned}
\Pr\{W_i>t_i(q),W_j>t_j(q)\}
&\ge \Pr(E_q)\\
&=
\bar{F}_X\{(1+2\epsilon)x_q\}
\Pr(Z_i+b_i>-\epsilon a_ix_q,\,
    Z_j+b_j>-\epsilon a_jx_q).
\end{aligned}
\]
The Gaussian probability tends to one because both thresholds
tend to $-\infty$. Regular variation and $\bar{F}_X(x_q)\sim q$
give $\bar{F}_X\{(1+2\epsilon)x_q\}/q\to(1+2\epsilon)^{-\xi}$.
Since $\Pr(U_j>1-q)=q$, it follows that
\[
\begin{aligned}
\liminf_{q\downarrow0}\Pr(U_i>1-q\mid U_j>1-q)
&=
\liminf_{q\downarrow0}
\frac{\Pr\{W_i>t_i(q),W_j>t_j(q)\}}{q}\\
&\ge (1+2\epsilon)^{-\xi}.
\end{aligned}
\]
Letting $\epsilon\downarrow0$ gives a lower limit of at least one.
The conditional probability is at most one, so
$\lambda_U^{(i,j)}=1$.

\emph{Part (b).}
By part (a), $\lambda_U^{(i,j)}=1$. To study the lower tail,
choose $x_0$ such that $X\ge x_0$ almost surely and set
$c_k=a_kx_0+b_k$. Then $W_k\ge Z_k+c_k$. Also choose a finite
$M>x_0$ with $\pi=\Pr(X\le M)>0$ and set $d_k=a_kM+b_k$.
On the event $X\le M$, we have $W_k\le Z_k+d_k$. Hence,
by independence, for all $t\in\mathbb{R}$,
\[
\Pr(W_k\le t)
\ge
\Pr(X\le M,\ Z_k\le t-d_k)
=
\pi\,\Phi\{(t-d_k)/\sigma\}.
\]

Write $r_k=r_k(q)$ and let
$r^\ast=\max\{r_i-c_i,r_j-c_j\}$. A joint lower-tail event
requires both Gaussian variables to lie below $r^\ast$, since
\[
\begin{aligned}
C_2(q,q)
&=\Pr(W_i\le r_i,W_j\le r_j)\\
&\le \Pr(Z_i\le r_i-c_i,\ Z_j\le r_j-c_j)\\
&\le \Pr(Z_i\le r^\ast,Z_j\le r^\ast).
\end{aligned}
\]
As $q\downarrow0$, $r^\ast\to-\infty$. If $\rho=-1$, then
$Z_i+Z_j=0$ almost surely, so the last event is impossible
once $r^\ast<0$. This gives $\lambda_L^{(i,j)}=0$ in that case.

For $-1<\rho<1$, Lemma~\ref{lemma:supp_gauss_joint} gives
$C_2(q,q)\le\Phi(\kappa r^\ast/\sigma)$, where
$\kappa=\{2/(1+\rho)\}^{1/2}>1$. To bound the denominator,
choose an index $k^\ast$ attaining the maximum in $r^\ast$ and
let $D=\max\{d_i-c_i,d_j-c_j\}$. Then
\[
r_{k^\ast}-d_{k^\ast}
=
r^\ast-(d_{k^\ast}-c_{k^\ast})
\ge r^\ast-D.
\]
Applying the marginal lower bound at $t=r_{k^\ast}$ yields
\[
q
=\Pr(W_{k^\ast}\le r_{k^\ast})
\ge
\pi\,\Phi\{(r^\ast-D)/\sigma\}.
\]
Consequently,
\[
0\le
\frac{C_2(q,q)}{q}
\le
\frac{\Phi(\kappa r^\ast/\sigma)}
     {\pi\,\Phi\{(r^\ast-D)/\sigma\}}
\longrightarrow0
\qquad\text{as }q\downarrow0.
\]
The limit follows from Lemma~\ref{lemma:supp_mills} with
$x=r^\ast/\sigma$ and the fixed shift $D/\sigma$ in the
denominator. Hence $\lambda_L^{(i,j)}=0$.
\end{proof}

The lower-tail argument applies whenever the factor contributions
at both locations are bounded below. Take $c_i,c_j$ to be their
lower bounds and choose finite $d_i,d_j$ such that both contributions
lie below their respective $d_k$ with positive probability.
This gives $\lambda_L^{(i,j)}=0$, including for a network with a
positive output activation.

\begin{remark}[Two regularly varying tails]
Suppose that $X$ and $-X$ both have regularly varying right tails.
Part (a) gives $\lambda_U^{(i,j)}=1$ for $(W_i,W_j)$. It also
applies to the reflected pair, since
$-W_k=a_k(-X)-b_k-Z_k$ still has a positive factor loading.
Reflecting both variables replaces $U_k$ by $1-U_k$, so upper
tail dependence of the reflected pair gives
$\lambda_L^{(i,j)}=1$ for the original pair. Thus both limiting
coefficients equal one for any two positive tail indices. The
finite-level coefficients $\lambda_U^{(i,j)}(q)$ and
$\lambda_L^{(i,j)}(q)$ can still differ.
\end{remark}

\subsection{Proof of \texorpdfstring{Corollary~\ref{M-cor:opposite_loadings}}{Corollary~1} of the main text}

\noindent\textbf{Corollary~\ref{M-cor:opposite_loadings}
(Opposite loadings).}
Suppose that $g(V,\bfs_i)=a_iX+b_i$ and
$g(V,\bfs_j)=-a_jX+b_j$ with $a_i,a_j>0$ and
$b_i,b_j\in\mathbb{R}$, where $X=X(V)$ is regularly varying
with index $\xi>0$ and bounded below, and that $\rho>-1$. Then
\begin{enumerate}
\item[(a)] $\lim_{q\downarrow0}\Pr(U_j\le q\mid U_i>1-q)=1$,
so that an extreme high value at $\bfs_i$ occurs together with
an extreme low value at $\bfs_j$.
\item[(b)] $\lambda_U^{(i,j)}=0$, although the factor is heavy tailed.
\end{enumerate}


\begin{proof}
\emph{Part (a).}
Under the stated loadings,
\[
W_i=a_iX+b_i+Z_i,
\qquad
-W_j=a_jX-b_j-Z_j.
\]
Reflecting $W_j$ makes both loadings of $X$ positive. The pair
$(Z_i,-Z_j)$ is centered bivariate Gaussian with common marginal
variance $\sigma^2$ and correlation $-\rho<1$, since $\rho>-1$.
It is independent of $X$. Therefore
Proposition~\ref{M-thm:positive_tail_dependence}(a) of the main
text applies to $(W_i,-W_j)$ and gives an upper tail dependence
coefficient of one for this pair.

To express this result in terms of $(U_i,U_j)$, let $F_{-W_j}$
be the marginal distribution function of $-W_j$. Since
$F_{1,j}$ is continuous, for every $w\in\mathbb{R}$,
\[
F_{-W_j}(-w)=\Pr(W_j\ge w)=1-F_{1,j}(w).
\]
Thus $F_{-W_j}(-W_j)=1-U_j$, and the event that $-W_j$ exceeds
its $(1-q)$-quantile is $\{U_j<q\}$. Both upper-quantile exceedance events for the
reflected pair have probability $q$. The result for that pair
therefore gives
\[
\Pr(U_j\le q\mid U_i>1-q)
=
\frac{\Pr\{U_i>1-q,\ F_{-W_j}(-W_j)>1-q\}}{q}
\longrightarrow 1
\qquad\text{as }q\downarrow0.
\]

\emph{Part (b).}
For $0<q\le 1/2$, the events $\{U_j\le q\}$ and
$\{U_j>1-q\}$ are disjoint. Conditioning on $\{U_i>1-q\}$
and using part~(a) gives
\[
\begin{aligned}
0\le
\frac{\Pr(U_i>1-q,\ U_j>1-q)}{q}
&=\Pr(U_j>1-q\mid U_i>1-q)\\
&\le 1-\Pr(U_j\le q\mid U_i>1-q)
\longrightarrow 0.
\end{aligned}
\]
Since $\Pr(U_j>1-q)=q$ as well, the joint probability divided
by $q$ is also $\Pr(U_i>1-q\mid U_j>1-q)$. Hence
$\lambda_U^{(i,j)}=0$.
\end{proof}

\subsection{Proof of \texorpdfstring{Proposition~\ref{M-thm:asymptotic_independence}}{Proposition~4} of the main text}

\noindent\textbf{Proposition~\ref{M-thm:asymptotic_independence}
(Bounded factors and finite networks).}
\begin{enumerate}
\item[(a)] If $|g(V,\bfs_i)|$ and $|g(V,\bfs_j)|$ are bounded almost surely
by a constant $c<\infty$, then
$\lambda_L^{(i,j)}=\lambda_U^{(i,j)}=0$.

\item[(b)] In particular, for any feedforward neural network $g$
with continuous activation functions and any $0<\theta_1<\theta_2$,
the bivariate copulas of the stationary and location-dependent
BNFC models satisfy $\lambda_L^{(i,j)}=\lambda_U^{(i,j)}=0$
for all $\bfs_i\ne\bfs_j$ with $\rho<1$.
\end{enumerate}

\begin{proof}
\emph{Part (a). Upper tail.}
For $k\in\{i,j\}$, the bound on the factor contribution gives
\[
Z_k-c\le W_k\le Z_k+c
\qquad\text{almost surely}.
\]
Thus $Z_k>w+c$ implies $W_k>w$, and $W_k>w$ implies $Z_k>w-c$.
Since $Z_k$ is centered Gaussian with variance $\sigma^2$
($\sigma=1$ in the main-text model), we obtain
\[
\bar{\Phi}\{(w+c)/\sigma\}
\le \Pr(W_k>w)
\le \bar{\Phi}\{(w-c)/\sigma\},
\qquad w\in\mathbb{R}.
\]

Let $t_k=t_k(q)=F_{1,k}^{-1}(1-q)$. The margins are continuous
and strictly increasing with support $\mathbb{R}$, so
$\Pr(W_k>t_k)=q$ and $t_k\to\infty$ as $q\downarrow0$.
Set $t=\min\{t_i,t_j\}$. At the location whose threshold equals
$t$, the marginal lower bound gives
\[
q\ge\bar{\Phi}\{(t+c)/\sigma\}.
\]
For the joint exceedance, $W_k>t_k$ implies
$Z_k>t_k-c\ge t-c$. Hence
\[
\Pr(W_i>t_i,W_j>t_j)
\le \Pr(Z_i>t-c,Z_j>t-c).
\]

For $-1<\rho<1$, Lemma~\ref{lemma:supp_gauss_joint} bounds the
right-hand side by $\bar{\Phi}\{\kappa(t-c)/\sigma\}$, where
$\kappa=\{2/(1+\rho)\}^{1/2}>1$. Combining the joint upper bound
with the marginal lower bound gives
\[
\Pr(U_i>1-q\mid U_j>1-q)
=
\frac{\Pr(W_i>t_i,W_j>t_j)}{q}
\le
\frac{\bar{\Phi}\{\kappa(t-c)/\sigma\}}
     {\bar{\Phi}\{(t+c)/\sigma\}}
\longrightarrow 0.
\]
The limit follows from Lemma~\ref{lemma:supp_mills} with
$x=t/\sigma$ and both shifts equal to $c/\sigma$, since
$t\to\infty$ as $q\downarrow0$. If $\rho=-1$, then
$Z_i+Z_j=0$ almost surely, so the joint Gaussian exceedance
is impossible once $t-c>0$. Thus $\lambda_U^{(i,j)}=0$
in both cases.

\emph{Lower tail.}
For the reflected pair $(-W_i,-W_j)$, the Gaussian pair
$(-Z_i,-Z_j)$ has correlation $\rho$, and the reflected factor
contributions satisfy $|-g(V,\bfs_k)|\le c$. The upper-tail
argument therefore applies to this pair. On the uniform scale,
reflection replaces $U_k$ by $1-U_k$, so an upper exceedance
of the reflected variable corresponds to $U_k<q$. Since
$\Pr(U_k=q)=0$, the upper tail dependence coefficient of the
reflected pair equals $\lambda_L^{(i,j)}$ of the original pair.
Hence $\lambda_L^{(i,j)}=0$.

\emph{Part (b).}
A feedforward neural network with continuous activation functions
is continuous as a composition of affine maps and continuous
activations. For each fixed location $\bfs_k$, the function
$v\mapsto g(v,\bfs_k)$ is therefore continuous on the compact
interval $[\theta_1,\theta_2]$. In the stationary model this
function is $g(v)$ at both locations. In either model,
\[
c=\max_{k\in\{i,j\}}\max_{v\in[\theta_1,\theta_2]}
  |g(v,\bfs_k)|
\]
is finite. Since $V\in[\theta_1,\theta_2]$ almost surely,
this $c$ bounds the factor contributions at both locations,
and part (a) applies.

For fixed parameter values, the same compactness argument applies
to the parameterization $V=\theta_1+|\theta_2-\theta_1|\varepsilon$
in Section~\ref{M-subsec:vb_inference} of the main text. Since
$\varepsilon\in[0,1]$ almost surely, $V$ lies in the compact
interval $[\theta_1,\theta_1+|\theta_2-\theta_1|]$.
\end{proof}

\subsection{Proof of \texorpdfstring{Proposition~\ref{M-prop:subasymptotic_convergence}}{Proposition~5} of the main text}

\noindent\textbf{Proposition~\ref{M-prop:subasymptotic_convergence}
(Convergence of the copula and of finite-level tail dependence).}
Let $C_n^{(m)}$ and $C_n$ be the copulas of
$\bfW^{(m)}=(W_1^{(m)},\ldots,W_n^{(m)})^\top$ and
$\bfW=(W_1,\ldots,W_n)^\top$, respectively, where
\[
W_k^{(m)}=Z_k+g_m(V,\bfs_k),
\qquad
W_k=Z_k+T_{\bftheta}(V,\bfs_k),
\qquad k=1,\ldots,n.
\]
Suppose that $g_m(\cdot,\bfs_k)\to T_{\bftheta}(\cdot,\bfs_k)$
uniformly on compact subsets of $(\theta_1,\theta_2)$ for
$k=1,\ldots,n$. Then $C_n^{(m)}\to C_n$ uniformly on $[0,1]^n$.
In particular, for every fixed $q\in(0,1/2]$, the finite-level
coefficients $\lambda_L^{(i,j)}(q)$ and $\lambda_U^{(i,j)}(q)$
in Equation~\eqref{M-eq:lambda_q}, evaluated under $C_n^{(m)}$,
converge to their values under $C_n$. The same holds for the
skewness measure $\zeta_1=\mathbb{E}\{(U_i+U_j-1)^3\}$.

\begin{proof}
Use the same $V$ and Gaussian vector $\bfZ$ for all $m$. The
assumed convergence on compact subsets implies pointwise
convergence at every $v\in(\theta_1,\theta_2)$. Since $V$ lies
in this interval almost surely, we have
$g_m(V,\bfs_k)\to T_{\bftheta}(V,\bfs_k)$ almost surely for each
$k$, as in the proof of
Proposition~\ref{M-prop:factor_generation}(c) in
Section~\ref{app:proof_factor_generation}. There are finitely
many locations, so $\bfW^{(m)}\to\bfW$ almost surely.

Let $F_{1,k}^{(m)}$ and $F_{1,k}$ be the marginal distribution
functions of $W_k^{(m)}$ and $W_k$. Almost-sure convergence implies convergence
in distribution, so $F_{1,k}^{(m)}(w)\to F_{1,k}(w)$ for every
$w\in\mathbb{R}$.

Fix $\bfu\in(0,1)^n$ and write
\[
w_k^{(m)}=(F_{1,k}^{(m)})^{-1}(u_k),
\qquad
w_k=F_{1,k}^{-1}(u_k).
\]
For every $\delta>0$, strict monotonicity gives
\[
F_{1,k}(w_k-\delta)<u_k<F_{1,k}(w_k+\delta).
\]
By convergence at $w_k-\delta$ and $w_k+\delta$, the same
inequalities hold with $F_{1,k}^{(m)}$ for sufficiently large
$m$. Consequently, $w_k-\delta<w_k^{(m)}<w_k+\delta$.
Since $\delta$ is arbitrary, $w_k^{(m)}\to w_k$.

The event $\{W_k=w_k\}$ has probability zero for each $k$.
Thus, for almost every draw, $W_k^{(m)}-w_k^{(m)}$ eventually
has the same sign as $W_k-w_k$ at every location. The indicator
of the joint event $\{W_k^{(m)}\le w_k^{(m)},\ k=1,\ldots,n\}$
therefore converges almost surely to the indicator of
$\{W_k\le w_k,\ k=1,\ldots,n\}$. Taking expectations by bounded
convergence gives
\[
C_n^{(m)}(\bfu)
=
\Pr\{W_k^{(m)}\le w_k^{(m)},\ k=1,\ldots,n\}
\longrightarrow
\Pr\{W_k\le w_k,\ k=1,\ldots,n\}
=
C_n(\bfu).
\]

To obtain uniform convergence, recall that changing one argument
of a copula from $u_k$ to $v_k$ changes its value by at most
$|u_k-v_k|$, because the $k$th margin is uniform. Changing the
arguments one at a time therefore bounds the total change by
$\sum_{k=1}^n|u_k-v_k|$. For any $\delta>0$, choose a finite
grid $\mathcal{G}_\delta\subset(0,1)^n$ such that for every
$\bfu\in[0,1]^n$ there is a grid point $\mathbf{v}$ with
$\sum_{k=1}^n|u_k-v_k|\le\delta$. Applying the copula bound to
both $C_n^{(m)}$ and $C_n$ yields
\[
\sup_{\bfu\in[0,1]^n}|C_n^{(m)}(\bfu)-C_n(\bfu)|
\le
2\delta+
\max_{\mathbf{v}\in\mathcal{G}_\delta}
|C_n^{(m)}(\mathbf{v})-C_n(\mathbf{v})|.
\]
The maximum tends to zero because the grid is finite and
pointwise convergence holds at each grid point. Letting
$\delta\downarrow0$ proves uniform convergence on $[0,1]^n$.

For the pair $(i,j)$, let $C_2^{(m)}$ and $C_2$ denote the
bivariate margins of $C_n^{(m)}$ and $C_n$, obtained by setting
the other arguments to one. Uniform convergence of the full
copulas implies uniform convergence of these bivariate margins.
For fixed $q>0$, the difference between the lower-tail
coefficients in Equation~\eqref{M-eq:lambda_q} is
$\{C_2^{(m)}(q,q)-C_2(q,q)\}/q$. The difference between the
upper-tail coefficients has the same form, evaluated at
$(1-q,1-q)$. Both differences tend to zero.

Finally, write
$U_k^{(m)}=F_{1,k}^{(m)}(W_k^{(m)})$ and $U_k=F_{1,k}(W_k)$.
Convergence of the bivariate distribution functions
$C_2^{(m)}$ to the continuous limit $C_2$ gives
$(U_i^{(m)},U_j^{(m)})\to(U_i,U_j)$ in distribution.
Since $(u,v)\mapsto(u+v-1)^3$ is bounded and continuous
on $[0,1]^2$, it follows that
\[
\mathbb{E}\{(U_i^{(m)}+U_j^{(m)}-1)^3\}
\longrightarrow
\mathbb{E}\{(U_i+U_j-1)^3\}.
\]
\end{proof}

\subsection{Tail flexibility of simulated factor networks}\label{app:zeta}

{This section gives the details of the illustration in Section~\ref{M-sec:tail_properties} of the main text.

To illustrate the tail flexibility of the BNFC copula, we consider the location-dependent model~\eqref{M-eq:model_definition_varying_locs} of the main text and the bivariate copula $C_2$ of $(W_1,W_2)$ at the two locations $\bfs_1=(0.25,0.25)^\top$ and $\bfs_2=(0.25,0.35)^\top$, which are the spatial inputs of the network, where $(Z_1,Z_2)$ is bivariate standard normal with correlation $0.5$. As a measure of asymmetry of $C_2$, we use
\[
\zeta_1 = \mathbb{E}\big\{(U_1 + U_2 - 1)^3\big\},\qquad (U_1,U_2)\sim C_2,
\]
which can be interpreted as a measure of skewness of $C_2$ \citep{rosco2013measures}. It is zero for radially symmetric copulas, such as the Gaussian copula, and values of order $10^{-2}$ already indicate markedly asymmetric dependence. A negative value of $\zeta_1$ indicates stronger dependence in the lower tail, whereas a positive value indicates stronger dependence in the upper tail. By Proposition~\ref{M-prop:subasymptotic_convergence} of the main text, $\zeta_1$ is a continuous functional of the factor map, so it is a natural measure of the flexibility of the copulas that the networks can represent.

We simulate 100 factor networks $g(v,\bfs)$ with input $(v,\bfs^\top)^\top$, three hidden layers of width $64$ with GELU activations, and a linear output layer with a single unit, that is, the architecture of BNFC (nonstat) in Section~\ref{M-sec:empirical} of the main text without the softplus output activation, with weights drawn from $\mathcal{N}(0,9/d_{\mathrm{in}})$, where $d_{\mathrm{in}}$ is the input dimension of the layer, biases drawn from $\mathcal{N}(0,1)$, and $\bftheta$ drawn from its prior distributions (Section~\ref{M-subsec:vb_inference} of the main text). For each simulated network, we compute $\zeta_1$ and the finite-level tail dependence coefficients in~\eqref{M-eq:lambda_q} of the main text at $q=0.05$ from $10^5$ draws of $(W_1,W_2)$, transformed to the copula scale by their empirical marginal distribution functions. Figure~\ref{fig:zeta_plot} shows the values of $\zeta_1$. They range from $-0.007$ to $0.018$, so the BNFC model can represent dependence that is stronger in the upper tail, stronger in the lower tail, or symmetric, without specifying the factor distribution in advance. For the network with the largest value, $\lambda_U(0.05)=0.60$ and $\lambda_L(0.05)=0.29$, whereas the Gaussian copula with correlation $0.5$ gives $\lambda_U(0.05)=\lambda_L(0.05)\approx0.24$.
\begin{figure}
    \centering
    \includegraphics[width=0.8\linewidth]{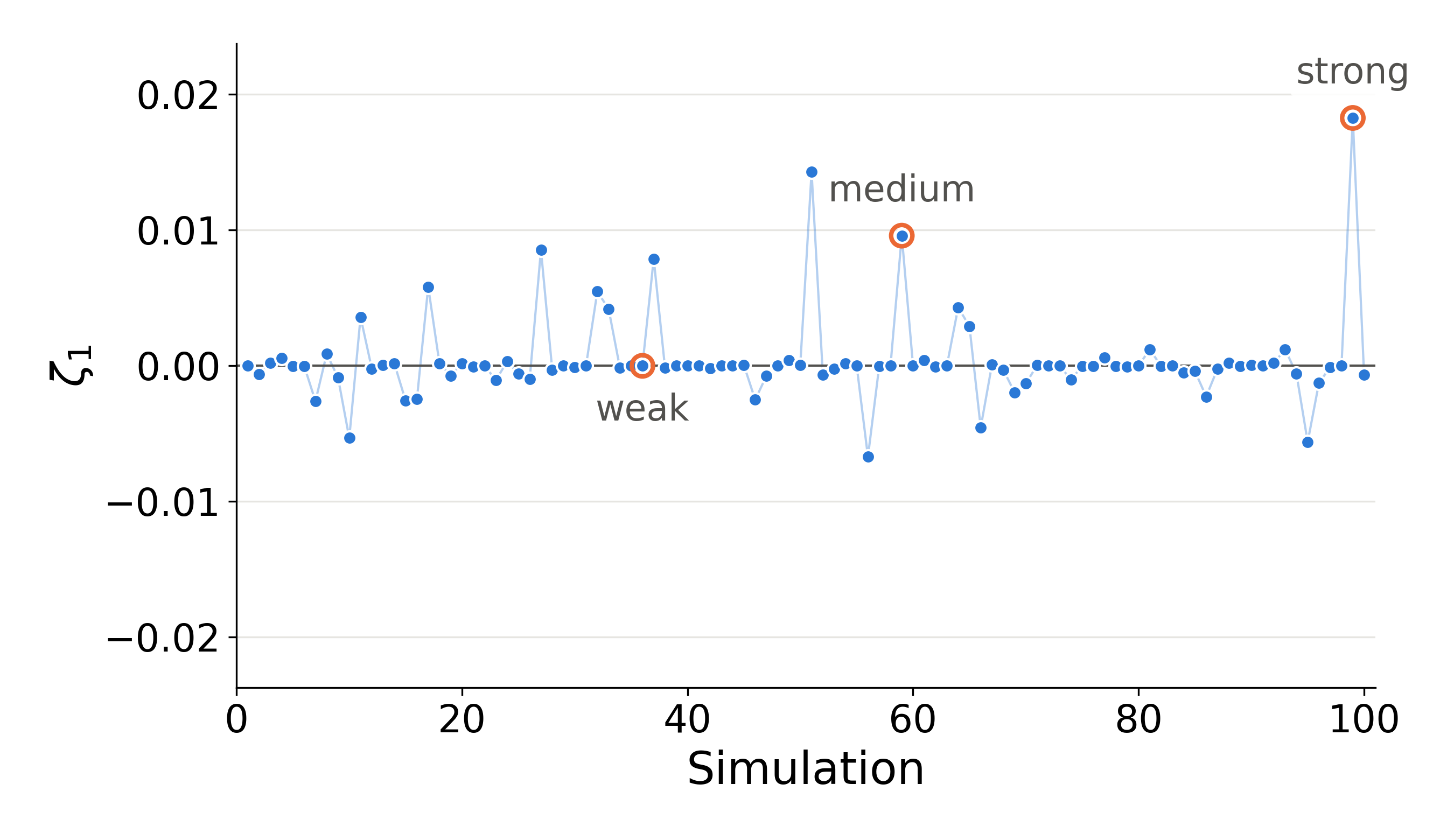}
    \caption{$\zeta_1$ for the bivariate BNFC copula of $\{W(\bfs_1),W(\bfs_2)\}$, $\bfs_1=(0.25,0.25)^\top$, $\bfs_2=(0.25,0.35)^\top$, with Gaussian correlation $0.5$, over 100 simulated factor networks. Circles mark the three networks used in Figure~\ref{fig:covariance_plot}.}
    \label{fig:zeta_plot}
\end{figure}
}

\begin{figure}[tbp]
    \centering
    \includegraphics[width=\textwidth]{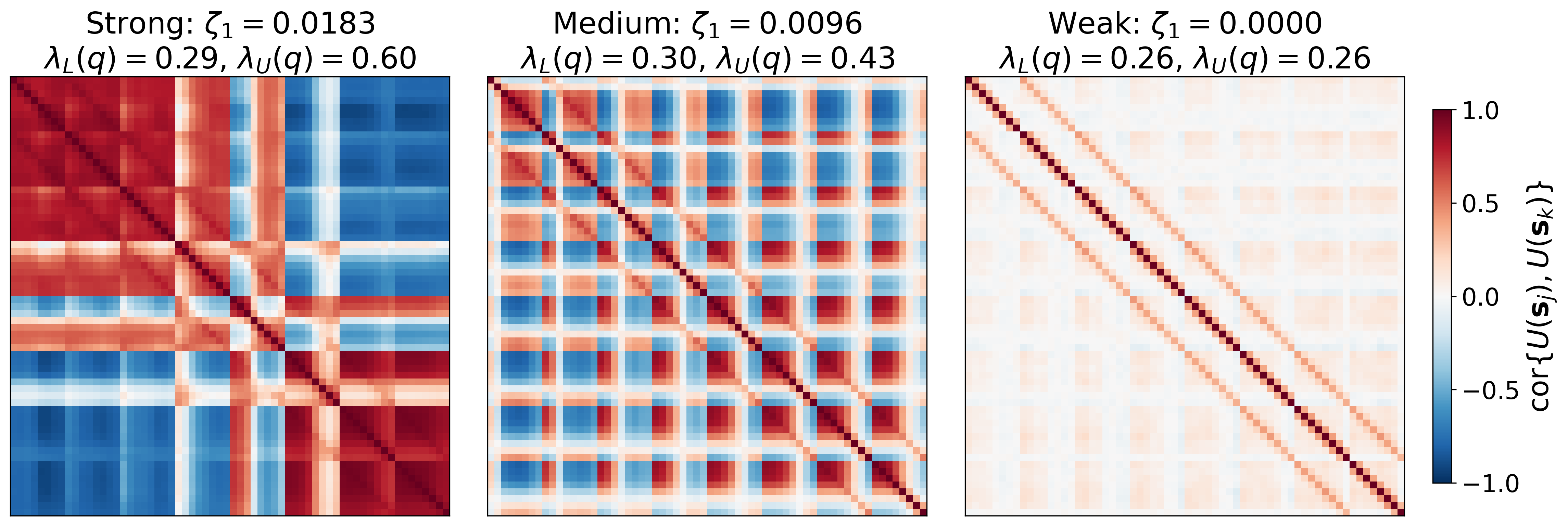}
    \caption{(Left to right) Correlation matrices of the BNFC copula field on an $8\times8$ grid for the three factor networks circled in Figure~\ref{fig:zeta_plot}, with strong ($\zeta_1=0.018$), medium ($\zeta_1=0.010$) and weak ($\zeta_1\approx0$) tail asymmetry. Panel titles give $\zeta_1$ and $\lambda_L(q)$, $\lambda_U(q)$, $q=0.05$, of the bivariate copula at $\bfs_1=(0.25,0.25)^\top$ and $\bfs_2=(0.25,0.35)^\top$.}
    \label{fig:covariance_plot}
\end{figure}

To show the spatial dependence that these networks induce, we simulate the copula field $\{U(\bfs_1),\ldots,U(\bfs_{64})\}$ on an $8\times8$ grid on $[0,1]^2$ for three of the networks above, with strong, medium and weak tail asymmetry, using the squared exponential correlation function with range $0.1$ for $Z(\bfs)$. Figure~\ref{fig:covariance_plot} shows the correlation matrices of the simulated fields. With weak asymmetry, the correlation is close to that of the Gaussian component and decays quickly with distance. With strong and medium asymmetry, the factor induces long-range dependence, which forms patterns that are not functions of distance alone. This dependence is positive between locations at which $g(\cdot,\bfs)$ moves in the same direction as $v$ increases and negative between locations at which it moves in opposite directions, as permitted by Proposition~\ref{M-thm:covariance} of the main text. The networks in Figure~\ref{fig:covariance_plot} are drawn from the distribution above and are not fitted to data. They illustrate the range of dependence structures that the model contains, whereas a fitted network contains negative dependence only if the data support it. No common factor copula can produce such a structure, since its factor adds the same nonnegative covariance to every pair of locations; with the squared exponential correlation function, its correlation matrix would show the decay of the Gaussian component plus a constant.

\subsection{Discussion of the tail properties}\label{app:tail_discussion}

This section expands the discussion of the results of Section~\ref{M-sec:tail_properties} of the main text.

Proposition~\ref{M-thm:positive_tail_dependence} of the main text covers the stationary model~\eqref{M-eq:model_definition} of the main text with a regularly varying factor $g(V)$, where $a_i=a_j=1$ and $b_i=b_j=0$, and part (a) agrees with the known result for power-law common factors: a single large value of the factor produces simultaneous extremes at all locations. More generally, any factor that is bounded below, such as the output of a network with a positive output activation, produces no lower tail dependence, whereas a factor with two regularly varying tails gives $\lambda_L^{(i,j)}=\lambda_U^{(i,j)}=1$ whatever the tail indices. Intermediate and asymmetric values, $0<\lambda_L^{(i,j)}\neq\lambda_U^{(i,j)}<1$, arise when the two tails of the factor decay exponentially at different rates, for example when the factor is the difference of two independent exponential random variables with rates $\delta_1\neq\delta_2$, for which $\lambda_U^{(i,j)}=2\Phi[-\delta_1\{(1-\rho)/2\}^{1/2}]$ and $\lambda_L^{(i,j)}=2\Phi[-\delta_2\{(1-\rho)/2\}^{1/2}]$ \citep[Example~2]{krupskii2018factor}. 


Corollary~\ref{M-cor:opposite_loadings} of the main text shows that a single heavy-tailed factor with loadings of opposite sign creates perfect dependence between the upper tail at one location and the lower tail at the other, and no upper tail dependence between the two. In the sign-change scenario of Section~\ref{M-sec:simulation_studies} of the main text, for example, a factor value that produces heavy rainfall in the north produces dry conditions in the south. This joint behavior of extremes is not available in the common factor copulas or in Gaussian scale mixtures with nonnegative correlation, in which a large value of the common variable raises the process at every location, with the nonnegative correlation functions used here.

Part (a) of Proposition~\ref{M-thm:asymptotic_independence} of the main text parallels the asymptotic independence of the common factor copula with a light-tailed factor, such as a Weibull factor with shape parameter larger than one. Part (b) follows because a network with continuous activations is bounded on the bounded interval $(\theta_1,\theta_2)$. Every fitted BNFC copula is therefore asymptotically independent, and tail dependence is only attained in the limit of a sequence of networks. This does not, however, prevent a finite network from reproducing the dependence of a tail-dependent copula at any fixed level, which is what matters for a data set of finite size.

By Proposition~\ref{M-prop:factor_generation}(b) of the main text, Proposition~\ref{M-prop:subasymptotic_convergence} applies whenever $T_{\bftheta}$ is continuous in $v$. Hence, if $T_{\bftheta}$ generates a heavy-tailed factor, the finite-level coefficients $\lambda_U^{(i,j)}(q)$ of the networks approach values close to one, as for the target, at every fixed level $q$, even though their limits as $q\to0$ are zero. This is the sense in which the BNFC model represents tail dependence, and it is how we assess it empirically in Section~\ref{app:zeta}. The results also guide the choice of the output layer of $g$ (Remark~\ref{M-rem:implementation} of the main text). By Proposition~\ref{M-thm:positive_tail_dependence}(b) of the main text, a positive output activation is appropriate when stronger dependence is expected in the upper tail than in the lower tail, whereas a linear output layer leaves both tails of the factor unrestricted.

\section{Derivation of the Conditional Copula}
\label{app:conditional_copula_derivation}

This section derives the conditional distribution used for spatial prediction in Section~\ref{M-subsec:cond_copula} of the main text. Let $\bfu=(u_1,\ldots,u_n)^\top$ be the copula data of one replicate at $\bfs_1,\ldots,\bfs_n$, and let $\bfw=(w_1,\ldots,w_n)^\top$ with $w_j=F_{1,j}^{-1}(u_j)$. For a new location $\bfs_0$, let $F_{1,0}$ and $f_{1,0}$ be the marginal distribution function and density in Equation~\eqref{M-eq:bnfc_univariate_margin} of the main text at $\bfs_0$, and let $w_0=F_{1,0}^{-1}(u_0)$. Let $c_{n+1}$ denote the copula density of $(W_0,W_1,\ldots,W_n)$, and let $f_{n+1}$ denote the corresponding joint density. The conditional copula distribution function is
\begin{equation}
C_{0\mid n}(u_0\mid\bfu)=\int_0^{u_0}\frac{c_{n+1}(r,u_1,\ldots,u_n)}{c_n(u_1,\ldots,u_n)}\,dr.
\label{eq:app_conditional_copula_definition}
\end{equation}
With the change of variable $r=F_{1,0}(t)$, $dr=f_{1,0}(t)\,dt$, and the definitions of the copula densities,
\[
c_{n+1}(r,\bfu)=\frac{f_{n+1}(t,\bfw)}{f_{1,0}(t)\prod_{j=1}^nf_{1,j}(w_j)},
\qquad
c_n(\bfu)=\frac{f_n(\bfw)}{\prod_{j=1}^nf_{1,j}(w_j)},
\]
the marginal densities cancel and
\begin{equation}
C_{0\mid n}(u_0\mid\bfu)=\int_{-\infty}^{w_0}\frac{f_{n+1}(t,\bfw)}{f_n(\bfw)}\,dt=\int_{-\infty}^{w_0}f_{0\mid n}(t\mid\bfw)\,dt,
\label{eq:app_conditional_copula_w_scale}
\end{equation}
where $f_{0\mid n}(t\mid\bfw)=f_{n+1}(t,\bfw)/f_n(\bfw)$ is the conditional density of $W_0$ given $\bfW=\bfw$. The lower limit is $F_{1,0}^{-1}(0)=-\infty$, because $F_{1,0}$ is a mixture of normal distribution functions.

Conditional on $V=v$, $(W_0,\bfW^\top)^\top$ is Gaussian with mean $\{g(v,\bfs_0),\bs{G}(v)^\top\}^\top$ and covariance matrix
\[
\begin{pmatrix}
\sigma_{00} & \bs{\sigma}_0^\top\\
\bs{\sigma}_0 & \bs{\Sigma}_{\bs{\vartheta}}
\end{pmatrix},
\]
where $\sigma_{00}=1$ and $\bs{\sigma}_0$ is the vector of correlations between $Z(\bfs_0)$ and $\bfZ$. Hence
\[
f_{n+1}(t,\bfw)=\frac{1}{\theta_2-\theta_1}\int_{\theta_1}^{\theta_2}\phi\{t;m_0(v),s_0^2\}\,\phi_{\bs{\Sigma}_{\bs{\vartheta}}}\{\bfw-\bs{G}(v)\}\,dv,
\]
where $\phi(\cdot\,;m,s^2)$ is the normal density with mean $m$ and variance $s^2$, and
\[
m_0(v)=g(v,\bfs_0)+\bs{\sigma}_0^\top\bs{\Sigma}_{\bs{\vartheta}}^{-1}\{\bfw-\bs{G}(v)\},
\qquad
s_0^2=\sigma_{00}-\bs{\sigma}_0^\top\bs{\Sigma}_{\bs{\vartheta}}^{-1}\bs{\sigma}_0
\]
are the simple kriging mean and variance given $V=v$. Dividing by $f_n(\bfw)$ in Equation~\eqref{M-eq:bnfc_marginal_pdf} of the main text gives
\[
f_{0\mid n}(t\mid\bfw)=\int_{\theta_1}^{\theta_2}\phi\{t;m_0(v),s_0^2\}\,p(v\mid\bfw)\,dv,
\qquad
p(v\mid\bfw)=\frac{\phi_{\bs{\Sigma}_{\bs{\vartheta}}}\{\bfw-\bs{G}(v)\}}{\int_{\theta_1}^{\theta_2}\phi_{\bs{\Sigma}_{\bs{\vartheta}}}\{\bfw-\bs{G}(v')\}\,dv'},
\]
where $p(v\mid\bfw)$ is the conditional density of $V$ given $\bfW=\bfw$, by Bayes' rule with the uniform prior on $(\theta_1,\theta_2)$. Substituting into \eqref{eq:app_conditional_copula_w_scale} and exchanging the order of integration, which is justified by Tonelli's theorem, gives
\begin{equation}
C_{0\mid n}(u_0\mid\bfu)=\int_{\theta_1}^{\theta_2}\Phi\left\{\frac{w_0-m_0(v)}{s_0}\right\}p(v\mid\bfw)\,dv,
\qquad w_0=F_{1,0}^{-1}(u_0),
\label{eq:app_conditional_copula_final}
\end{equation}
which is Equation~\eqref{M-eq:conditional_copula} of the main text.

The integrals in \eqref{eq:app_conditional_copula_final} are approximated as in Section~\ref{M-subsec:cond_copula} of the main text, which also averages over the posterior distribution of the parameters. The $R$ draws are split evenly between the $S$ weight samples in Equation~\eqref{M-eq:weight_average} of the main text, so that draw $r$ uses the network $g$ with weights $\bs{\varphi}^{(s(r))}$. For $r=1,\ldots,R$, let $\bs{\eta}^{(r)}=(\theta_1^{(r)},\theta_2^{(r)},\bs{\vartheta}^{(r)\top})^\top\sim q_{\hat{\bs{\lambda}}}$, let $\bfw^{(r)}=\{\hat{F}_{1,1}^{-1}(u_1),\ldots,\hat{F}_{1,n}^{-1}(u_n)\}^\top$ and $w_0^{(r)}=\hat{F}_{1,0}^{-1}(u_0)$, with the Monte Carlo estimators in Equation~\eqref{M-eq:mc_estimators} of the main text, and let $v^{(r,1)},\ldots,v^{(r,L)}\sim U(\theta_1^{(r)},\theta_2^{(r)})$ be drawn independently for each $r$. Self-normalized importance sampling of $p(v\mid\bfw)$ with the uniform proposal gives
\[
\begin{gathered}
\widehat{C}_{0\mid n}(u_0\mid\bfu)=\frac{1}{R}\sum_{r=1}^R\sum_{\ell=1}^L\omega_\ell^{(r)}\,\Phi\left\{\frac{w_0^{(r)}-m_0^{(r)}(v^{(r,\ell)})}{s_0^{(r)}}\right\},
\\
\omega_\ell^{(r)}=\frac{\phi_{\bs{\Sigma}_{\bs{\vartheta}^{(r)}}}\{\bfw^{(r)}-\bs{G}^{(r)}(v^{(r,\ell)})\}}{\sum_{\ell'=1}^L\phi_{\bs{\Sigma}_{\bs{\vartheta}^{(r)}}}\{\bfw^{(r)}-\bs{G}^{(r)}(v^{(r,\ell')})\}},
\end{gathered}
\]
where the superscript $(r)$ indicates quantities computed with the weights $\bs{\varphi}^{(s(r))}$ and the parameters $\bs{\eta}^{(r)}$; in particular, $m_0^{(r)}$ and $s_0^{2(r)}$ are the kriging mean and variance in Equation~\eqref{M-eq:kriging} of the main text with $\bs{\Sigma}_{\bs{\vartheta}^{(r)}}$ and $\bs{\sigma}_0(\bs{\vartheta}^{(r)})$, which require one Cholesky factorization of $\bs{\Sigma}_{\bs{\vartheta}^{(r)}}$ per draw. Thus $\widehat{C}_{0\mid n}$ approximates the posterior predictive distribution of $U_0$, averaged over the weight samples and over $q_{\hat{\bs{\lambda}}}$, rather than the conditional distribution at a point estimate. Sampling an index $\ell^\star$ with probabilities $\omega_1^{(r)},\ldots,\omega_L^{(r)}$, drawing $W_0^{(r)}\sim\mathcal{N}\{m_0^{(r)}(v^{(r,\ell^\star)}),s_0^{2(r)}\}$ and setting $U_0^{(r)}=\hat{F}_{1,0}(W_0^{(r)})$ gives a draw from $\widehat{C}_{0\mid n}$, which is steps (i)--(iii) of Algorithm~\ref{alg:prediction}. Unweighted draws $v^{(r,\ell)}$ from the prior $U(\theta_1^{(r)},\theta_2^{(r)})$ would instead give the unconditional distribution of the factor, which ignores the information that the observations carry about it. Averaging the conditional draws over $\ell$ would also be incorrect, because it shrinks the predictive variance.

\section{Additional simulation results}
\label{app:additional_sim}

This section gives the details of the simulation study of Section~\ref{M-sec:simulation_studies} of the main text and additional results.

\subsection{Simulation design and competing models}
\label{app:supp_design}

This section gives the details of the simulation design and of the competing models of Section~\ref{M-sec:simulation_studies} of the main text.

We consider three data-generating processes on a regular $30\times30$ grid on $[0,1]^2$. In all three, $Z(\bfs)$ is a Gaussian process with unit variance and squared exponential correlation function with range $0.1$, and each replicate is a single draw of the factor and of $Z$. The copula data are $U(\bfs)=F_{1}\{W(\bfs);\bfs\}$, where $F_1(\cdot\,;\bfs)$ is the marginal distribution function of $W(\bfs)$ under the data-generating model, evaluated by quadrature over the factor with $2{,}000$ nodes.
\begin{itemize}
\item[(i)] \emph{Gamma-factor data:} $W(\bfs)=Z(\bfs)+V_0$, where $V_0$ follows a gamma distribution with shape $6$ and scale $3$. This is the common factor model in Equation~\eqref{M-eq:classical-factor-copula} of the main text.
\item[(ii)] \emph{Sign-change data:} $W(\bfs)=Z(\bfs)+g(U,\bfs)$, where $U\sim U(0,1)$ and
\[
g(u,\bfs)=1.5\,b(\bfs)\,H^{-1}(u),\qquad b(\bfs)=\tanh\{(s_2-0.5)/0.15\},
\]
with $H$ is the gamma distribution $\mathrm{Ga}(2,1)$. The factor raises the field in the northern half of the domain, where $b(\bfs)>0$, lowers it in the southern half, where $b(\bfs)<0$, and vanishes along $s_2=0.5$. The two halves are therefore negatively dependent, with correlation $-0.82$ between the northernmost and the southernmost observed locations, and the copula has stronger dependence in the upper tail in the north and in the lower tail in the south. This is the structure that the factor networks in Figure~\ref{fig:covariance_plot} produce. By Proposition~\ref{M-thm:covariance} of the main text, the location-dependent BNFC model can represent it, whereas the common factor copula and Gaussian scale mixtures with the nonnegative correlation functions used here add positive dependence to every pair of locations and cannot represent it. The map $g(\cdot,\bfs)$ is increasing in $u$ in the north and decreasing in the south, and it is not generated by a neural network.
\item[(iii)] \emph{Regional mixture data:} $W(\bfs)=Z(\bfs)+T(U,\bfs)$, where $U\sim U(0,1)$ and
\[
T(u,\bfs)=\{1-r(\bfs)\}H_A^{-1}(u)+r(\bfs)H_B^{-1}(u),\qquad r(\bfs)=\big[1+\exp\{-(s_1-0.5)/0.1\}\big]^{-1}.
\]
Here $H_A$ is the bimodal two-component gamma mixture $0.6\,\mathrm{Ga}(2,1)+0.4\,\mathrm{Ga}(20,0.5)$, written in terms of shape and scale, and $H_B$ is the right-skewed lognormal distribution $\mathrm{LN}(1.2,0.6^2)$. The factor is therefore bimodal in the western half of the domain, unimodal and right skewed in the eastern half, and changes smoothly between the two around $s_1=0.5$. Since $T(\cdot,\bfs)$ is a convex combination of two quantile functions, it is nondecreasing in $u$, so this is a location-specific factor in the sense of Lemma~\ref{M-lemma:location_specific_factor} of the main text. It is not, however, generated by a neural network, and no single parametric family matches it, so that every model in the comparison, including BNFC, is misspecified.
\end{itemize}
For each scenario, we simulate $2{,}000$ training replicates, observed at the same $n=100$ randomly selected grid locations, and $100$ test replicates. For each test replicate, we condition on the copula data at the $100$ observed locations and predict the field at all $900$ grid locations.

We fit the stationary and location-dependent BNFC models of Section~\ref{M-sec:methodology} of the main text, denoted BNFC (stat) and BNFC (nonstat), with the weight samples of Section~\ref{M-subsec:vb_inference} of the main text. We compare them with the common factor copula model in Equation~\eqref{M-eq:classical-factor-copula} of the main text with four parametric factors: an exponential factor with rate $\delta$, a gamma factor with shape $\Delta$ and rate $\delta$, a Pareto factor with scale $\delta+1$ and shape $\Delta$, and a two-component gamma mixture $\pi\,\mathrm{Ga}(2,\text{rate }\delta_1)+(1-\pi)\,\mathrm{Ga}(20,\text{rate }\delta_2)$. The shapes of the mixture are fixed at those of $H_A$ in scenario (iii), which favors this competitor. The factor parameters $\Delta$, $\delta$, $\delta_1$ and $\delta_2$ are given half-normal priors with scale $4$, the mixture weight $\pi$ a uniform prior, and they are estimated, separately for each scenario, by Hamiltonian Monte Carlo with the No-U-Turn sampler \citep{hoffman2014no}, as implemented in Pyro \citep{bingham2019pyro}. The first three factors are scale families, $V_0=s(\delta)Q$, and the integrals in Equation~\eqref{M-eq:classical-factor-cdf} of the main text are evaluated with the $200$ quantile nodes $s(\delta)F_Q^{-1}\{(k-0.5)/200\}$, $k=1,\ldots,200$, which are shared by the marginal and joint densities, as in Equation~\eqref{M-eq:mc_estimators} of the main text. For the mixture, we use $100$ such nodes for each component, with weights $\pi/100$ and $(1-\pi)/100$, so that every factor is discretized with $200$ nodes. Because the three parameters of the mixture are strongly correlated a posteriori, its sampler is initialized at the posterior mode and uses a dense mass matrix.

As an external baseline from outside the class of factor copulas, we fit the Student-$t$ copula \citep{demarta2005t}, which is the copula of the Gaussian scale mixture $W(\bfs)=Z(\bfs)/G^{1/2}$ with $G\sim\mathrm{Ga}(\nu/2,\text{rate }\nu/2)$. Gaussian scale mixtures are a standard class of non-Gaussian spatial models with tail dependence \citep{huser2017bridging}. In the Student-$t$ copula, the common random variable acts multiplicatively rather than additively, and the copula is tail dependent for every finite $\nu$, with equal upper and lower tail dependence coefficients. To compare it with the factor copulas on equal terms, its correlation matrix is $(1-\beta)\bs{\Sigma}+\beta\mathbf{1}\mathbf{1}^\top$, where $\bs{\Sigma}$ is the Gaussian component shared by all models and $\beta\in[0,1)$ adds the constant long-range correlation that a common factor adds to $\bs{\Sigma}$. Its copula density is available in closed form, and we compute the posterior of $(\beta,\nu)$ exactly on a grid of $40\times40$ values of $(\operatorname{logit}\beta,\log\nu)$, refined to a second $40\times40$ grid around the mode, with priors $\beta\sim U(0,1)$ and $\nu\sim\mathrm{Ga}(2,\text{rate }0.1)$ \citep{juarez2010model}. Given the observations, the latent field at the unobserved locations is multivariate $t$ with $\nu+n$ degrees of freedom, so that conditional simulation is exact.

Both loading copulas use a spatially varying loading $b(\bfs)$ on a
shared factor $X$, which is Gaussian in one model and gamma in the
other. The loading controls the strength and direction of the factor
contribution $b(\bfs)X$ at each location. 
Both use the loading $b(\bfs)=\sum_{k=1}^{K}\beta_k\psi_k(\bfs)$ with $\psi_k(\bfs)=\exp\{-\|\bfs-\mathbf{c}_k\|^2/(2h^2)\}$, where $K=25$ centers $\mathbf{c}_k$ on a $5\times5$ grid over $[0,1]^2$ and bandwidth $h$ equal to the grid spacing. The priors $\beta_k\sim N(0,\sigma_\beta^2)$ and $\sigma_\beta\sim\mathrm{HN}(2)$. In the \emph{Gaussian loading copula}, $W(\bfs)=Z(\bfs)+b(\bfs)V$ with $V\sim N(0,1)$, the latent vector at $\bfs_1,\ldots,\bfs_n$ is $N_n(\mathbf{0},\bs{\Sigma}+\bfb\bfb^\top)$, so that its copula is the Gaussian copula with correlation matrix $\mathbf{R}=\mathbf{D}^{-1/2}(\bs{\Sigma}+\bfb\bfb^\top)\mathbf{D}^{-1/2}$, where $\mathbf{D}=\operatorname{diag}(1+b_1^2,\ldots,1+b_n^2)$. Its density needs no integral, and by the Woodbury identity and the matrix determinant lemma, each evaluation needs one Cholesky factor of $\bs{\Sigma}$. Since $(b,V)$ and $(-b,-V)$ give the same process, only $\bfb\bfb^\top$ is identified. Predictions are exact: given the latent data $\bfw$, $W(\bfs_0)$ is normal with mean $\mathbf{k}_0^\top(\bs{\Sigma}+\bfb\bfb^\top)^{-1}\bfw$ and variance $1+b_0^2-\mathbf{k}_0^\top(\bs{\Sigma}+\bfb\bfb^\top)^{-1}\mathbf{k}_0$, where $\mathbf{k}_0=\bs{\sigma}_0+b_0\bfb$, and the posterior of the factor is $V\mid\bfw\sim N\{\bfb^\top(\bs{\Sigma}+\bfb\bfb^\top)^{-1}\bfw,\,1-\bfb^\top(\bs{\Sigma}+\bfb\bfb^\top)^{-1}\bfb\}$. Being Gaussian, this copula has no tail dependence and is reflection symmetric. 

In the \emph{gamma loading copula}, $W(\bfs)=Z(\bfs)+b(\bfs)H_\kappa^{-1}(V)$ with $V\sim U(0,1)$ and $H_\kappa$ the $\mathrm{Ga}(\kappa,1)$ distribution, with prior $\kappa\sim\mathrm{Ga}(2,\text{rate }0.5)$; the scale of the gamma variable is absorbed by $b$. It is the special case $g(v,\bfs)=b(\bfs)H_\kappa^{-1}(v)$ of BNFC (nonstat), and its copula density is evaluated as for the common factor models, with the $M=100$ quantile nodes $q_m=H_\kappa^{-1}\{(m-0.5)/M\}$, which depend on $\kappa$ and are differentiated through the gamma quantile function. Because the gamma factor is skewed, the sign of $b(\bfs)$ determines whether the upper or the lower tail is heavier at $\bfs$, and pairs of locations with loadings of opposite sign link the upper tail at one location with the lower tail at the other. Both loading copulas are fitted by Hamiltonian Monte Carlo with the No-U-Turn sampler, and their centered factor contributions are $b(\bfs)V$ and $b(\bfs)\{H_\kappa^{-1}(V)-\kappa\}$.

All models use the same Gaussian component, which in the simulations has the squared exponential correlation with range $\ell$ estimated using the same Bayesian inference. Since the copula data are obtained with the true probability integral transforms, all comparisons are made on the copula scale, where the models differ only in their copulas and no marginal model enters.

\subsection{Performance measures}
\label{app:supp_measures}

This section gives the details of the performance measures of Section~\ref{M-sec:simulation_studies} of the main text. All measures are computed on the copula scale and averaged over the $100$ test replicates.

Our main measure is the held-out copula log score, the log posterior predictive density of the copula data $\bfu$ of a test replicate,
\[
\log\Big\{\frac1R\sum_{r=1}^R c_n(\bfu\mid\bs{\eta}^{(r)})\Big\},
\]
where $c_n$ is the copula density in Equation~\eqref{M-eq:bnfc_copula_density} of the main text at the $n$ locations at which the score is evaluated and $\bs{\eta}^{(1)},\ldots,\bs{\eta}^{(R)}$ are $R=20$ posterior draws of the model parameters. For BNFC, the draws are split evenly between the weight samples of Section~\ref{M-subsec:vb_inference} of the main text, so that the score also averages over the posterior of the network weights. For the factor copulas, the factor integrals in Equation~\eqref{M-eq:bnfc_copula_density} of the main text are evaluated on the same deterministic grid of $1{,}000$ factor nodes for every model, the $1{,}000$ midpoints of $U(\theta_1,\theta_2)$ for BNFC and the corresponding quantile nodes for the parametric factors; the density of the Student-$t$ copula is available in closed form. The log score is a strictly proper scoring rule for the joint distribution \citep{gneiting2007strictly} and assesses the whole dependence structure, including its tails, rather than a summary of it. We evaluate it at the $100$ observed locations, ``Log score (observed)'' in Table~\ref{M-tab:simulation_comparison} of the main text, and at all $900$ grid locations, ``Log score (all locations)''. The latter also assesses the copula at the $800$ locations that were not used for fitting on the test replicates. Higher values are better.

Predictions are generated with Algorithm~\ref{alg:prediction}, conditioning on the copula data at the observed locations, with $100$ posterior predictive draws $\mathbf{U}^{(1)},\ldots,\mathbf{U}^{(100)}$ of the copula field per test replicate and $L=200$ factor draws per predictive draw. For each test replicate with true copula field $\bfu$, we compute at all $900$ grid locations the continuous ranked probability score (CRPS), $\hat{\mathbb{E}}|U(\bfs)-u(\bfs)|-\tfrac12\hat{\mathbb{E}}|U(\bfs)-U'(\bfs)|$, averaged over the locations, the energy score of the predicted copula field, $\hat{\mathbb{E}}\|\mathbf{U}-\bfu\|-\tfrac12\hat{\mathbb{E}}\|\mathbf{U}-\mathbf{U}'\|$, where $\|\cdot\|$ is the Euclidean norm on $\mathbb{R}^{900}$ and which therefore also assesses the joint dependence of the predictions, and the mean absolute error (MAE) of the predictive mean \citep{gneiting2007strictly}. Here $\hat{\mathbb{E}}$ denotes the average over the predictive draws, and $\mathbf{U}$ and $\mathbf{U}'$ are independent draws. Lower values of these three measures are better.

At the conditioning locations, every model predicts the copula data almost exactly, so the calibration of the predictive distributions is assessed at the unobserved grid locations only, the $800$ locations not observed, or the $875$ locations not used for conditioning in Section~\ref{app:supp_sparse}. We compute the empirical coverage of the central prediction intervals with nominal levels $10\%,20\%,\ldots,90\%$ and $95\%$, the average coverage error (ACE), which is the mean absolute difference between empirical and nominal coverage over these ten levels, the mean width of the $95\%$ intervals, and the histograms of the probability integral transform (PIT) of the true copula values under the predictive draws, with randomized ranks \citep{gneiting2007probabilistic}. A calibrated model has empirical coverage equal to the nominal level and a flat PIT histogram, and among calibrated models the one with narrower intervals is preferable. These measures are reported in Section~\ref{app:supp_calibration}. The coverage of the $95\%$ intervals in Table~\ref{tab:supp_summary_n25} is computed, like the CRPS, at all grid locations, and therefore differs slightly from that in Table~\ref{tab:supp_calibration_n25}.

Finally, we compare the posterior of the factor given the observations with its true posterior. By Remark~\ref{M-rem:identifiability} of the main text, the conditional distribution of the centered factor contribution $\tilde g(V,\bfs)$ given the copula data of a replicate is comparable across models and with the data-generating model. For the factor copulas, we take the draws $\tilde g(v^{(\ell^\star)},\bfs)$ from step~(ii) of Algorithm~\ref{alg:prediction}, which is $V_0-\mathbb{E}V_0$ at every location for the common factor models. Under the data-generating model, the observations are the latent field itself, so the true posterior of the centered factor contribution is available exactly by quadrature over the factor, with $2{,}000$ nodes and weights proportional to $\phi_{\bs{\Sigma}}\{\bfw-\bs{T}(u)\}$ at the conditioning locations. Figure~\ref{fig:factor_posterior} compares the model posterior with the true posterior along a transect through the domain for the first test replicate of each scenario. The Student-$t$ copula has no additive factor and is not included in this comparison.

The fit times in Table~\ref{M-tab:simulation_comparison} of the main text are wall-clock times on one GPU. For BNFC they include the variational fit and the stochastic gradient MCMC for the network weights, for the parametric factor copulas the Hamiltonian Monte Carlo, and for the Student-$t$ copula the evaluation of the posterior on the grid.

Because the test replicates are shared by all models, we compare two models by the average paired differences of their scores over the test replicates, whose standard errors are much smaller than those of the averages in Table~\ref{M-tab:simulation_comparison} of the main text. In scenarios (ii) and (iii), the mean log scores of the Pareto factor model are dominated by a few test replicates to which it assigns almost zero density, and its median log scores are also clearly below those of BNFC (nonstat).

\subsection{Results for all common factor copulas}
\label{app:supp_all_models}

\begin{figure}[p]
\centering
\subfloat[Gamma-factor data: BNFC (stat); best common factor: gamma.]{\includegraphics[width=0.9\textwidth]{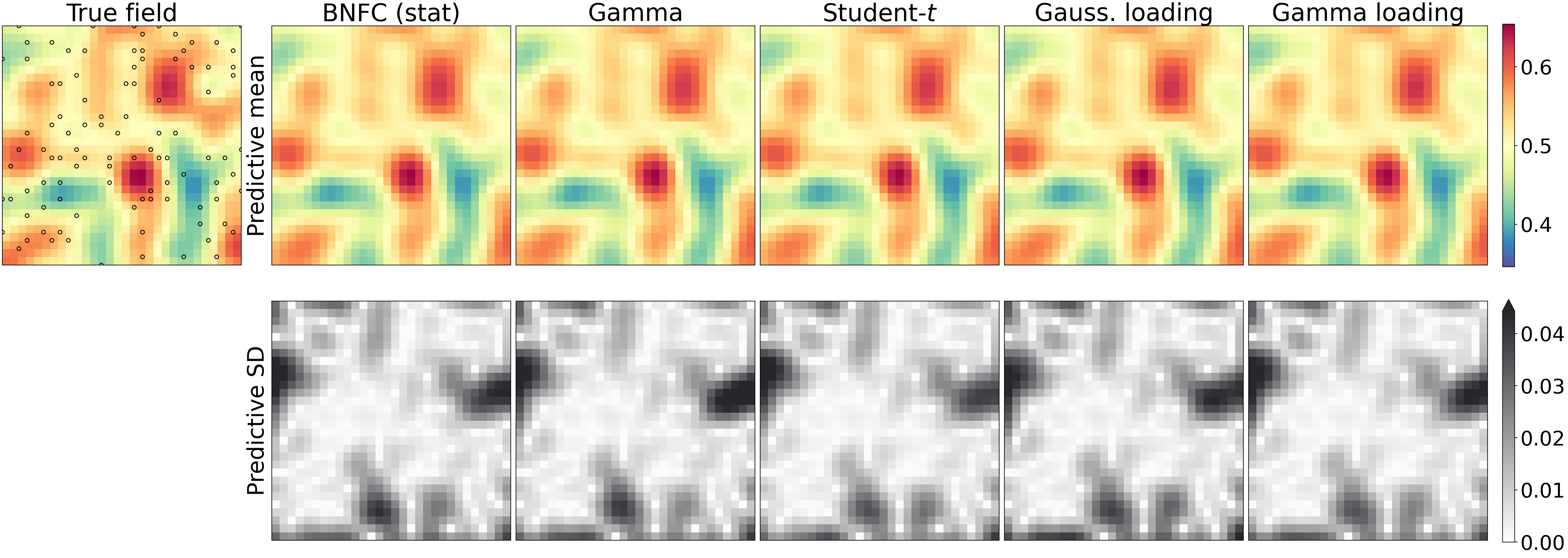}\label{fig:heatmaps_gamma}}

\subfloat[Sign-change data: BNFC (nonstat); best common factor: exponential.]{\includegraphics[width=0.9\textwidth]{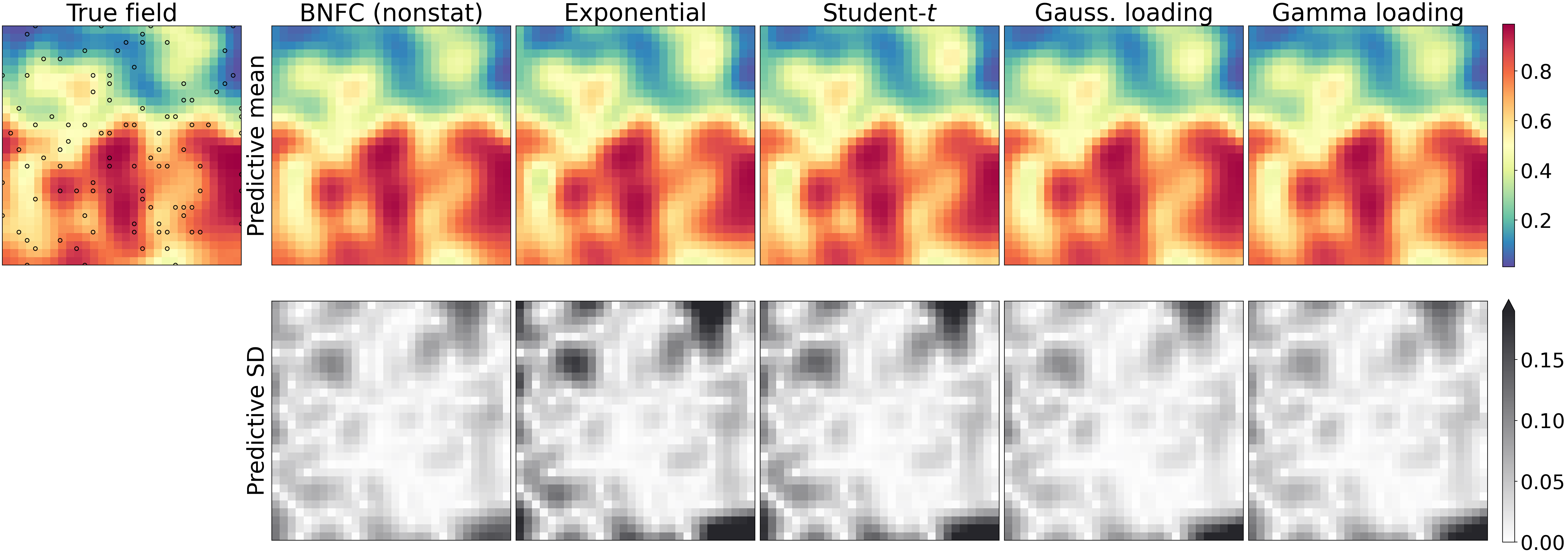}\label{fig:heatmaps_signchange}}

\subfloat[Regional mixture data: BNFC (nonstat); best common factor: gamma mixture.]{\includegraphics[width=0.9\textwidth]{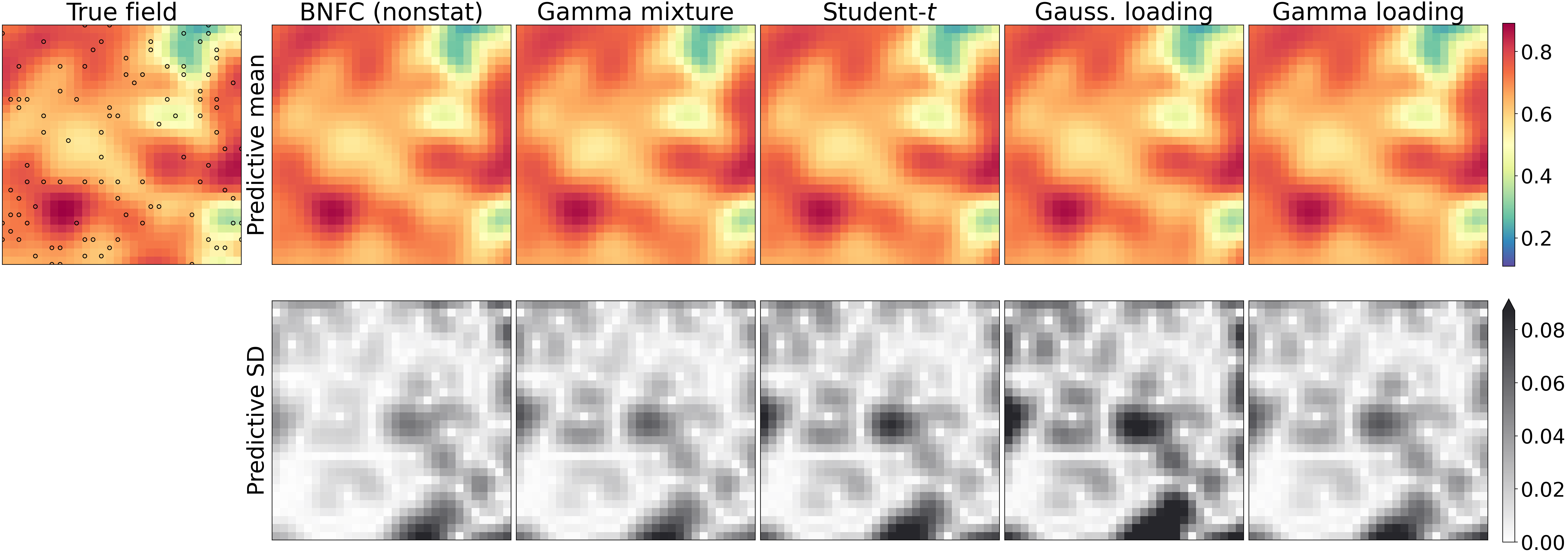}\label{fig:heatmaps_mixgamma}}
\caption{First test replicate of each scenario on the copula scale, given the $100$ observed locations, for the models of Table~\ref{M-tab:simulation_comparison} of the main text: BNFC, the best common factor copula, the Student-$t$ copula and the Gaussian and gamma loading copulas. In each panel, top row: true copula field, with the $100$ observed locations marked by circles, and the predictive mean of each model, on a common color scale centered at the median $0.5$; bottom row: the predictive standard deviation of each model, on a second common scale on which smaller values are lighter.}
\label{fig:heatmaps}
\end{figure}

Table~\ref{M-tab:simulation_comparison} of the main text reports, for each scenario, only the best of the four common factor copulas. Table~\ref{tab:supp_sim_all} gives the results of all four, together with BNFC, the Student-$t$ copula and the Gaussian and gamma loading copulas, and it adds the PICP and the MPIW of every model. The choice of the factor family matters: in scenario (i), the exponential and, in particular, the Pareto factor models are clearly worse in CRPS than the data-generating gamma factor model, and in scenarios (ii) and (iii) the mean log scores of the Pareto factor model are dominated by a few test replicates to which it assigns almost zero density. The prediction intervals show the same pattern. Those of the Pareto factor model are much too wide in scenarios (i) and (iii) and too narrow in scenario (ii). The other common factor copulas are close to the nominal level, but in scenario (ii) their intervals are much wider than those of BNFC (nonstat). The Student-$t$ copula undercovers slightly in all three scenarios. The loading copulas lie between the common factor copulas and BNFC. In scenario (i), the gamma loading copula, which contains the data-generating model, performs about as well as the best common factor copula and BNFC, while the Gaussian loading copula has a lower log score. In scenario (ii), a loading that can change sign recovers much of the advantage of BNFC (nonstat) over the common factor copulas, the gamma loading copula more than the Gaussian one, which also undercovers. BNFC (nonstat) remains best on all the scoring rules. In scenario (iii), the gamma loading copula is comparable with the common factor copulas and the Gaussian loading copula is clearly worse, and neither reaches BNFC (nonstat). Figure~\ref{fig:heatmaps} shows the predictions of all models of Table~\ref{M-tab:simulation_comparison} of the main text for the first test replicate of each scenario.

\begin{sidewaystable}
\centering
\caption{As Table~\ref{M-tab:simulation_comparison} of the main text, for all models: BNFC, the four common factor copulas, the Student-$t$ copula and the Gaussian and gamma loading copulas, including the PICP and the MPIW of the nominal $90\%$ prediction intervals at the $800$ unobserved locations (closest to $0.90$ and lowest are better). The best value of each measure is in bold.}
\label{tab:supp_sim_all}
\renewcommand{\arraystretch}{1.15}
\resizebox{\textwidth}{!}{%
\begin{tabular}{lccccccccc}
\toprule
 & \multicolumn{2}{c}{BNFC} & \multicolumn{4}{c}{Common factor copula} & \multicolumn{1}{c}{Scale mixture} & \multicolumn{2}{c}{Varying loading} \\
\cmidrule(lr){2-3}\cmidrule(lr){4-7}\cmidrule(lr){8-8}\cmidrule(lr){9-10}
Measure & stat & nonstat & Exponential & Gamma & Pareto & Gamma mix. & Student-$t$ & {Gaussian} & {Gamma} \\
\midrule
\multicolumn{10}{l}{\textbf{(i) Gamma-factor data}} \\[2pt]
\quad CRPS & $0.0045$\,{\scriptsize$(0.0002)$} & $0.0045$\,{\scriptsize$(0.0002)$} & $0.0049$\,{\scriptsize$(0.0002)$} & $0.0045$\,{\scriptsize$(0.0002)$} & $0.0107$\,{\scriptsize$(0.0004)$} & $0.0045$\,{\scriptsize$(0.0002)$} & $0.0045$\,{\scriptsize$(0.0002)$} & {$0.0045$\,{\scriptsize$(0.0002)$}} & {$\mathbf{0.0045}$\,{\scriptsize$(0.0002)$}} \\
\quad Energy score & $0.232$\,{\scriptsize$(0.009)$} & $0.233$\,{\scriptsize$(0.009)$} & $0.254$\,{\scriptsize$(0.009)$} & $\mathbf{0.232}$\,{\scriptsize$(0.009)$} & $0.563$\,{\scriptsize$(0.017)$} & $0.234$\,{\scriptsize$(0.009)$} & $0.232$\,{\scriptsize$(0.009)$} & {$0.233$\,{\scriptsize$(0.009)$}} & {$0.232$\,{\scriptsize$(0.009)$}} \\
\quad MAE & $\mathbf{0.0063}$\,{\scriptsize$(0.0002)$} & $0.0063$\,{\scriptsize$(0.0002)$} & $0.0064$\,{\scriptsize$(0.0002)$} & $0.0063$\,{\scriptsize$(0.0002)$} & $0.0085$\,{\scriptsize$(0.0003)$} & $0.0063$\,{\scriptsize$(0.0002)$} & $0.0063$\,{\scriptsize$(0.0002)$} & {$0.0063$\,{\scriptsize$(0.0002)$}} & {$0.0063$\,{\scriptsize$(0.0002)$}} \\
\quad Log score (observed) & $266.5$\,{\scriptsize$(4.9)$} & $265.1$\,{\scriptsize$(4.6)$} & $249.9$\,{\scriptsize$(5.3)$} & $\mathbf{266.8}$\,{\scriptsize$(5.0)$} & $151.4$\,{\scriptsize$(6.5)$} & $262.1$\,{\scriptsize$(4.9)$} & $265.1$\,{\scriptsize$(4.9)$} & {$260.0$\,{\scriptsize$(4.3)$}} & {$266.3$\,{\scriptsize$(4.9)$}} \\
\quad Log score (all locations) & $4899$\,{\scriptsize$(44)$} & $4892$\,{\scriptsize$(41)$} & $4754$\,{\scriptsize$(48)$} & $\mathbf{4908}$\,{\scriptsize$(45)$} & $3883$\,{\scriptsize$(61)$} & $4855$\,{\scriptsize$(44)$} & $4893$\,{\scriptsize$(45)$} & {$4839$\,{\scriptsize$(38)$}} & {$4902$\,{\scriptsize$(44)$}} \\
{\quad PICP $90\%$} & {$0.916$\,{\scriptsize$(0.004)$}} & {$0.930$\,{\scriptsize$(0.004)$}} & {$0.958$\,{\scriptsize$(0.004)$}} & {$0.914$\,{\scriptsize$(0.004)$}} & {$1.000$\,{\scriptsize$(0.000)$}} & {$0.908$\,{\scriptsize$(0.006)$}} & {$0.876$\,{\scriptsize$(0.005)$}} & {$\mathbf{0.902}$\,{\scriptsize$(0.008)$}} & {$0.893$\,{\scriptsize$(0.004)$}} \\
{\quad MPIW $90\%$} & {$0.030$\,{\scriptsize$(0.001)$}} & {$0.032$\,{\scriptsize$(0.001)$}} & {$0.044$\,{\scriptsize$(0.002)$}} & {$0.030$\,{\scriptsize$(0.001)$}} & {$0.147$\,{\scriptsize$(0.005)$}} & {$0.030$\,{\scriptsize$(0.001)$}} & {$\mathbf{0.028}$\,{\scriptsize$(0.001)$}} & {$0.031$\,{\scriptsize$(0.001)$}} & {$0.029$\,{\scriptsize$(0.001)$}} \\
\quad Fit time (s) & {$126$} & {$726$} & $140$ & $143$ & $136$ & $4548$ & $19$ & {$325$} & {$3601$} \\
\midrule
\multicolumn{10}{l}{\textbf{(ii) Sign-change data}} \\[2pt]
\quad CRPS & $0.0215$\,{\scriptsize$(0.0004)$} & $\mathbf{0.0163}$\,{\scriptsize$(0.0005)$} & $0.0219$\,{\scriptsize$(0.0004)$} & $0.0217$\,{\scriptsize$(0.0004)$} & $0.0334$\,{\scriptsize$(0.0019)$} & $0.0216$\,{\scriptsize$(0.0004)$} & $0.0218$\,{\scriptsize$(0.0004)$} & {$0.0177$\,{\scriptsize$(0.0004)$}} & {$0.0166$\,{\scriptsize$(0.0005)$}} \\
\quad Energy score & $1.087$\,{\scriptsize$(0.014)$} & $\mathbf{0.831}$\,{\scriptsize$(0.023)$} & $1.106$\,{\scriptsize$(0.014)$} & $1.097$\,{\scriptsize$(0.015)$} & $1.677$\,{\scriptsize$(0.091)$} & $1.091$\,{\scriptsize$(0.015)$} & $1.099$\,{\scriptsize$(0.016)$} & {$0.904$\,{\scriptsize$(0.020)$}} & {$0.857$\,{\scriptsize$(0.021)$}} \\
\quad MAE & $0.0297$\,{\scriptsize$(0.0005)$} & $\mathbf{0.0229}$\,{\scriptsize$(0.0007)$} & $0.0299$\,{\scriptsize$(0.0005)$} & $0.0298$\,{\scriptsize$(0.0005)$} & $0.0482$\,{\scriptsize$(0.0029)$} & $0.0298$\,{\scriptsize$(0.0005)$} & $0.0303$\,{\scriptsize$(0.0005)$} & {$0.0247$\,{\scriptsize$(0.0006)$}} & {$0.0233$\,{\scriptsize$(0.0006)$}} \\
\quad Log score (observed) & $106.7$\,{\scriptsize$(2.2)$} & $\mathbf{144.2}$\,{\scriptsize$(5.7)$} & $105.6$\,{\scriptsize$(2.5)$} & $106.1$\,{\scriptsize$(2.2)$} & $-3164.2$\,{\scriptsize$(1420.7)$} & $106.1$\,{\scriptsize$(2.2)$} & $111.7$\,{\scriptsize$(3.8)$} & {$133.8$\,{\scriptsize$(4.6)$}} & {$143.2$\,{\scriptsize$(5.7)$}} \\
\quad Log score (all locations) & $2741$\,{\scriptsize$(176)$} & $\mathbf{3764}$\,{\scriptsize$(46)$} & $2767$\,{\scriptsize$(121)$} & $2721$\,{\scriptsize$(154)$} & $-589234$\,{\scriptsize$(187842)$} & $2661$\,{\scriptsize$(181)$} & $3072$\,{\scriptsize$(38)$} & {$3336$\,{\scriptsize$(56)$}} & {$3688$\,{\scriptsize$(35)$}} \\
{\quad PICP $90\%$} & {$0.880$\,{\scriptsize$(0.006)$}} & {$0.892$\,{\scriptsize$(0.004)$}} & {$\mathbf{0.893}$\,{\scriptsize$(0.005)$}} & {$0.880$\,{\scriptsize$(0.006)$}} & {$0.802$\,{\scriptsize$(0.008)$}} & {$0.877$\,{\scriptsize$(0.006)$}} & {$0.873$\,{\scriptsize$(0.003)$}} & {$0.841$\,{\scriptsize$(0.007)$}} & {$0.871$\,{\scriptsize$(0.004)$}} \\
{\quad MPIW $90\%$} & {$0.160$\,{\scriptsize$(0.003)$}} & {$\mathbf{0.104}$\,{\scriptsize$(0.003)$}} & {$0.172$\,{\scriptsize$(0.004)$}} & {$0.163$\,{\scriptsize$(0.004)$}} & {$0.197$\,{\scriptsize$(0.008)$}} & {$0.160$\,{\scriptsize$(0.004)$}} & {$0.157$\,{\scriptsize$(0.002)$}} & {$0.109$\,{\scriptsize$(0.003)$}} & {$0.107$\,{\scriptsize$(0.003)$}} \\
\quad Fit time (s) & $93$ & {$960$} & $524$ & $558$ & $690$ & $1562$ & $20$ & {$41$} & {$4355$} \\
\midrule
\multicolumn{10}{l}{\textbf{(iii) Regional mixture data}} \\[2pt]
\quad CRPS & $0.0109$\,{\scriptsize$(0.0004)$} & $\mathbf{0.0105}$\,{\scriptsize$(0.0004)$} & $0.0111$\,{\scriptsize$(0.0005)$} & $0.0113$\,{\scriptsize$(0.0005)$} & $0.0139$\,{\scriptsize$(0.0005)$} & $0.0109$\,{\scriptsize$(0.0005)$} & $0.0118$\,{\scriptsize$(0.0005)$} & {$0.0119$\,{\scriptsize$(0.0005)$}} & {$0.0110$\,{\scriptsize$(0.0005)$}} \\
\quad Energy score & $0.544$\,{\scriptsize$(0.023)$} & $\mathbf{0.536}$\,{\scriptsize$(0.023)$} & $0.556$\,{\scriptsize$(0.023)$} & $0.560$\,{\scriptsize$(0.023)$} & $0.704$\,{\scriptsize$(0.022)$} & $0.545$\,{\scriptsize$(0.023)$} & $0.575$\,{\scriptsize$(0.025)$} & {$0.586$\,{\scriptsize$(0.024)$}} & {$0.550$\,{\scriptsize$(0.023)$}} \\
\quad MAE & $0.0151$\,{\scriptsize$(0.0006)$} & $\mathbf{0.0147}$\,{\scriptsize$(0.0006)$} & $0.0153$\,{\scriptsize$(0.0006)$} & $0.0155$\,{\scriptsize$(0.0006)$} & $0.0172$\,{\scriptsize$(0.0006)$} & $0.0152$\,{\scriptsize$(0.0006)$} & $0.0162$\,{\scriptsize$(0.0007)$} & {$0.0160$\,{\scriptsize$(0.0007)$}} & {$0.0152$\,{\scriptsize$(0.0006)$}} \\
\quad Log score (observed) & $188.0$\,{\scriptsize$(6.9)$} & $\mathbf{197.4}$\,{\scriptsize$(7.0)$} & $184.3$\,{\scriptsize$(7.1)$} & $184.9$\,{\scriptsize$(7.7)$} & $-42.9$\,{\scriptsize$(182.8)$} & $188.9$\,{\scriptsize$(7.5)$} & $181.1$\,{\scriptsize$(7.8)$} & {$172.7$\,{\scriptsize$(7.0)$}} & {$187.7$\,{\scriptsize$(6.9)$}} \\
\quad Log score (all locations) & $4097$\,{\scriptsize$(65)$} & $\mathbf{4248}$\,{\scriptsize$(63)$} & $3976$\,{\scriptsize$(74)$} & $4039$\,{\scriptsize$(74)$} & $-237389$\,{\scriptsize$(240995)$} & $4093$\,{\scriptsize$(67)$} & $3989$\,{\scriptsize$(77)$} & {$3870$\,{\scriptsize$(72)$}} & {$4034$\,{\scriptsize$(58)$}} \\
{\quad PICP $90\%$} & {$\mathbf{0.902}$\,{\scriptsize$(0.005)$}} & {$0.908$\,{\scriptsize$(0.005)$}} & {$0.910$\,{\scriptsize$(0.005)$}} & {$0.885$\,{\scriptsize$(0.007)$}} & {$0.977$\,{\scriptsize$(0.003)$}} & {$0.889$\,{\scriptsize$(0.005)$}} & {$0.875$\,{\scriptsize$(0.006)$}} & {$0.889$\,{\scriptsize$(0.011)$}} & {$0.874$\,{\scriptsize$(0.006)$}} \\
{\quad MPIW $90\%$} & {$0.073$\,{\scriptsize$(0.003)$}} & {$\mathbf{0.069}$\,{\scriptsize$(0.003)$}} & {$0.082$\,{\scriptsize$(0.003)$}} & {$0.073$\,{\scriptsize$(0.003)$}} & {$0.146$\,{\scriptsize$(0.005)$}} & {$0.070$\,{\scriptsize$(0.003)$}} & {$0.076$\,{\scriptsize$(0.003)$}} & {$0.082$\,{\scriptsize$(0.003)$}} & {$0.071$\,{\scriptsize$(0.003)$}} \\
\quad Fit time (s) & {$66$} & {$681$} & $148$ & $142$ & $186$ & $1465$ & $20$ & {$32$} & {$3327$} \\
\bottomrule
\end{tabular}%
}
\end{sidewaystable}

\subsection{Calibration of the predictive distributions}
\label{app:supp_calibration}

\begin{figure}[tbp]
\centering
\subfloat[$100$ conditioning locations.]{\includegraphics[width=\textwidth]{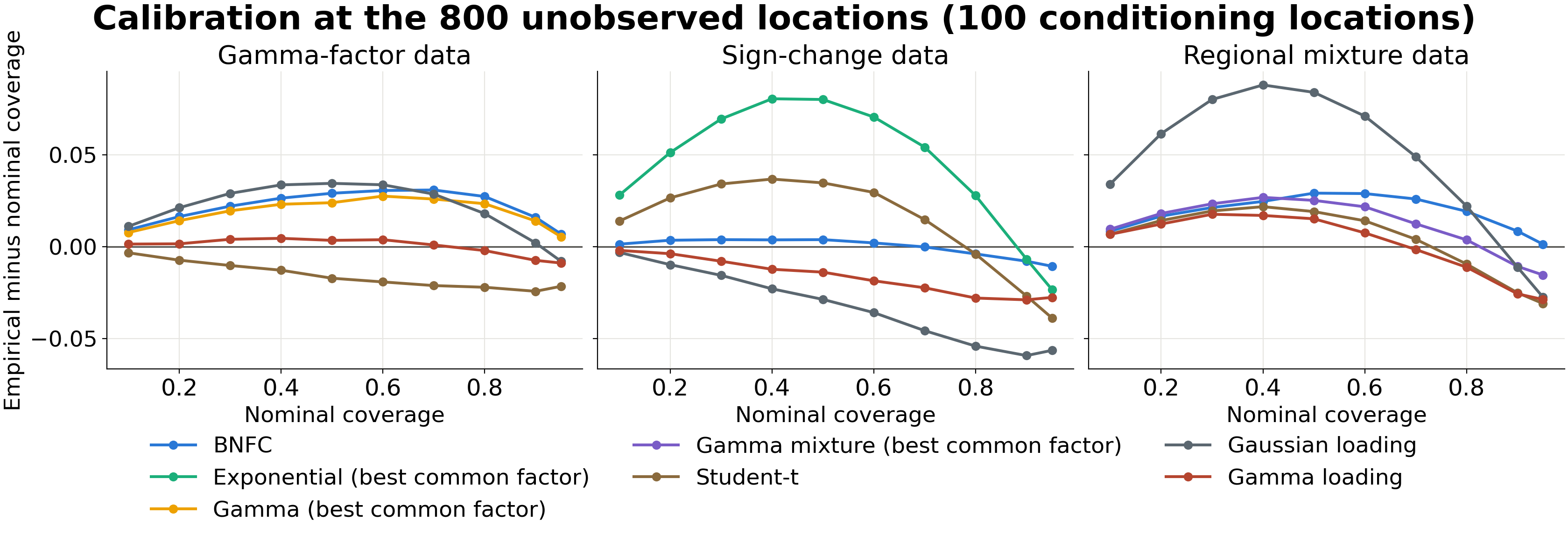}}

\subfloat[$25$ conditioning locations.]{\includegraphics[width=\textwidth]{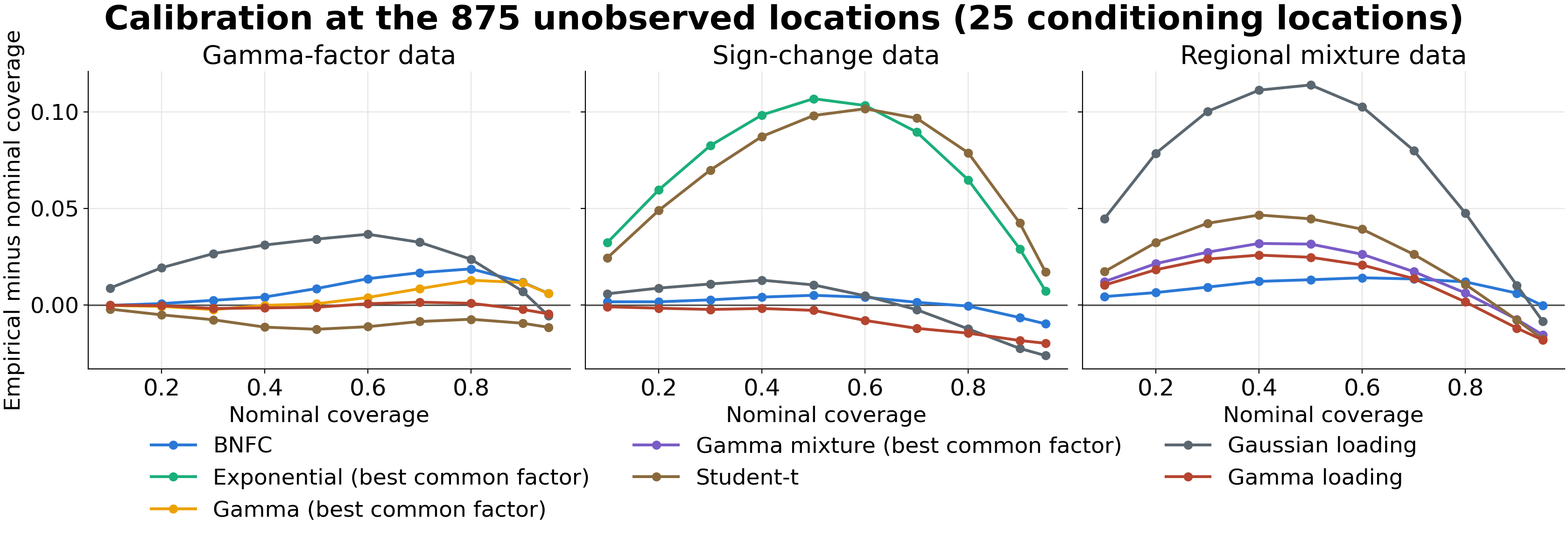}}
\caption{Empirical minus nominal coverage of the central prediction intervals of the copula field at the unobserved locations, at nominal levels $10\%,\ldots,90\%$ and $95\%$, averaged over the $100$ test replicates. Values above zero indicate intervals that are too wide, values below zero intervals that are too narrow. BNFC is BNFC (stat) for the gamma-factor data and BNFC (nonstat) otherwise. As in Table~\ref{M-tab:simulation_comparison} of the main text, the common factor copulas are represented by the best of them in each scenario; all models are in Tables~\ref{tab:calibration} and~\ref{tab:supp_calibration_n25}.}
\label{fig:supp_calibration}
\end{figure}

\begin{figure}[tbp]
\centering
\subfloat[$100$ conditioning locations.]{\includegraphics[width=0.88\textwidth]{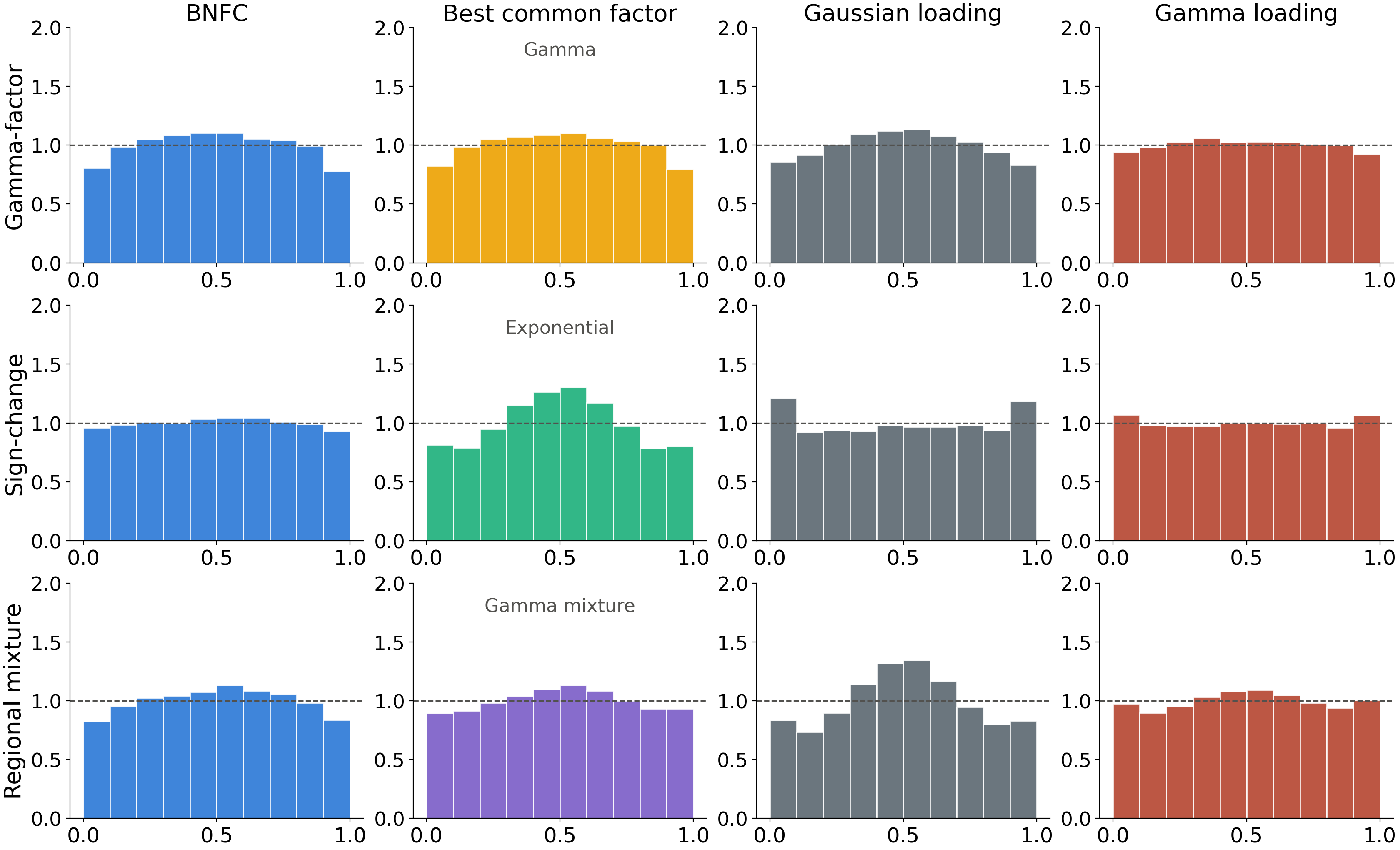}}

\subfloat[$25$ conditioning locations.]{\includegraphics[width=0.88\textwidth]{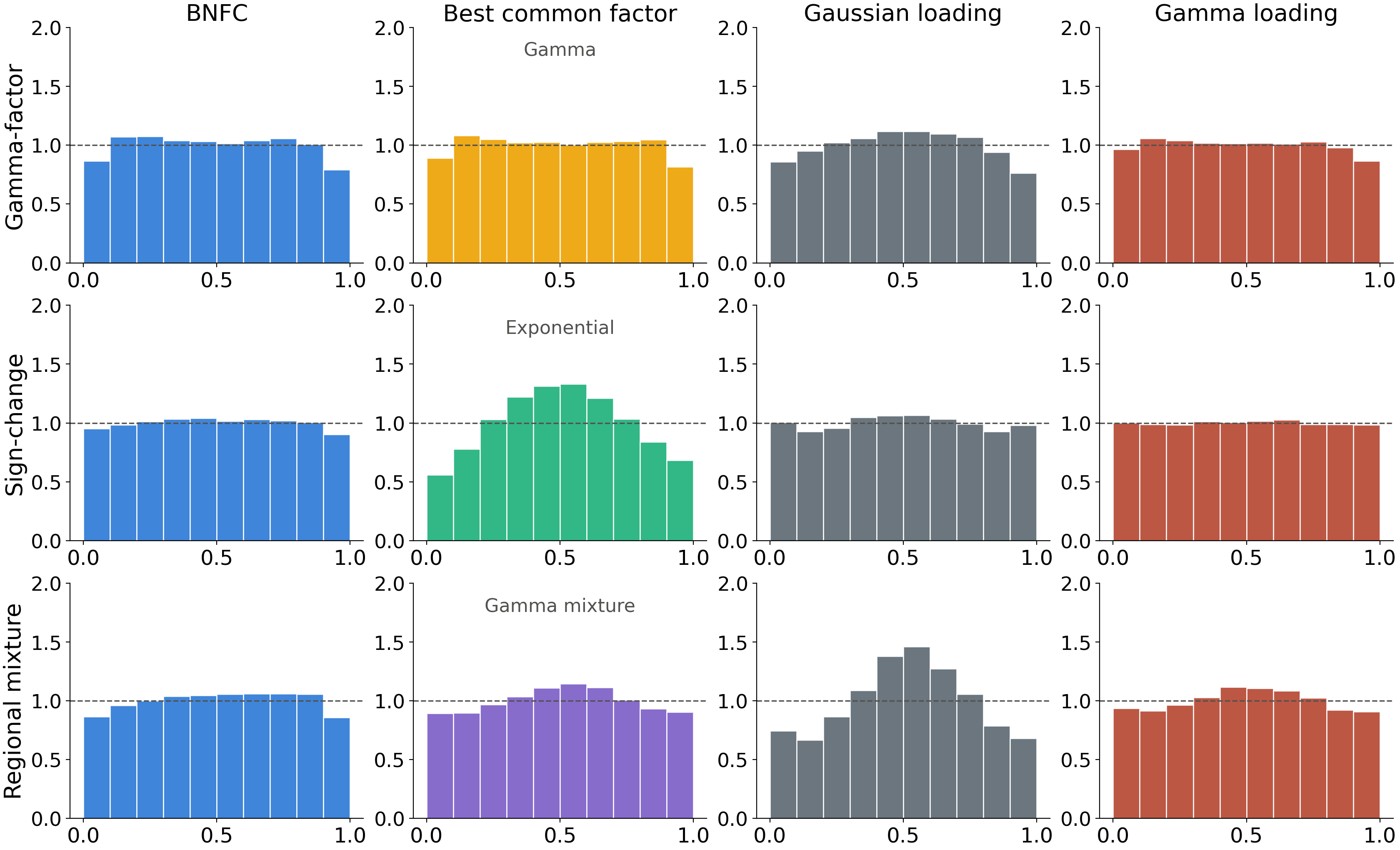}}
\caption{Histograms of the probability integral transform of the true copula values at the unobserved locations under the predictive draws, pooled over the $100$ test replicates, for the three simulation scenarios (rows): BNFC, which is BNFC (stat) for the gamma-factor data and BNFC (nonstat) otherwise, the best common factor copula of each scenario (named in the panel), and the Gaussian and gamma loading copulas. The dashed line is the density of the uniform distribution.}
\label{fig:supp_pit}
\end{figure}

The calibration measures are defined in Section~\ref{app:supp_measures}. Figure~\ref{fig:supp_calibration} shows the difference between the empirical and the nominal coverage of the central prediction intervals at the unobserved locations, and Tables~\ref{tab:calibration} and~\ref{tab:supp_calibration_n25} give the coverage at three levels, the ACE, the width of the $95\%$ intervals and, with $100$ conditioning locations, the held-out log score. With the stochastic gradient MCMC for the network weights, the $95\%$ intervals of BNFC (nonstat) are close to the nominal level in every scenario. For the sign-change data, BNFC (nonstat) is the best-calibrated model.
BNFC (stat), the common factor copulas, and the Student-$t$ copula have
more restrictive dependence structures and produce substantially wider
prediction intervals with empirical coverage below the nominal $95\%$
level. The loading copulas yield intervals comparable in width to those
of BNFC (nonstat), with lower empirical coverage. For the regional mixture data, the gamma loading copula has the smallest
ACE, and its $95\%$ prediction intervals have empirical coverage below
the nominal level. BNFC (nonstat) yields the narrowest intervals among
the models compared, with empirical coverage close to $95\%$. For the gamma-factor data, most models are well calibrated. The gamma
loading copula, which includes the data-generating model as a special
case, achieves the best calibration. BNFC (stat) closely approximates
the data-generating model. The predictive distributions are slightly
overdispersed for BNFC (nonstat) and more markedly overdispersed for
the exponential and Pareto factor models. The results with $25$ conditioning locations
(Table~\ref{tab:supp_calibration_n25}) show similar calibration patterns.
BNFC (nonstat) remains among the better-calibrated models for both the
sign-change and regional mixture data. Figure~\ref{fig:supp_pit} shows histograms of the probability integral
transform (PIT) of the true copula values, computed using predictive
draws from BNFC, the best common factor copula, and the two loading
copulas. For a calibrated predictive distribution, the PIT histogram
is flat. With both $100$ and $25$ conditioning locations, the PIT histograms
for BNFC and the gamma loading copula are generally close to flat,
particularly for the sign-change data. The clearest departures from
uniformity are concentrations of PIT values near $0.5$ for the
exponential common factor copula in the sign-change scenario and
for the Gaussian loading copula in the regional mixture scenario.

\begin{table}[tbp]
\centering
\caption{Calibration of the predictive distributions of the copula field at the $800$ unobserved locations, given the $100$ observed locations: empirical coverage of the central $50\%$, $80\%$ and $95\%$ prediction intervals, ACE over the nominal levels $10\%,\ldots,90\%$ and $95\%$, mean width of the $95\%$ intervals, and the held-out log score at all grid locations. For each measure, the value closest to the nominal level, the smallest ACE and the largest log score are in bold.}
\label{tab:calibration}
\renewcommand{\arraystretch}{1.15}
\resizebox{\textwidth}{!}{%
\begin{tabular}{lccccccccc}
\toprule
 & \multicolumn{2}{c}{BNFC} & \multicolumn{4}{c}{Common factor copula} & \multicolumn{1}{c}{Scale mixture} & \multicolumn{2}{c}{Varying loading} \\
\cmidrule(lr){2-3}\cmidrule(lr){4-7}\cmidrule(lr){8-8}\cmidrule(lr){9-10}
Measure & stat & nonstat & Exp. & Gamma & Pareto & Gamma mix. & Student-$t$ & Gaussian & Gamma \\
\midrule
\multicolumn{10}{l}{\textbf{(i) Gamma-factor data}} \\[2pt]
\quad Coverage 50\% & $0.529$ & $0.553$ & $0.659$ & $0.524$ & $0.982$ & $0.534$ & $0.483$ & $0.534$ & $\mathbf{0.503}$ \\
\quad Coverage 80\% & $0.827$ & $0.847$ & $0.901$ & $0.823$ & $0.999$ & $0.823$ & $0.778$ & $0.818$ & $\mathbf{0.798}$ \\
\quad Coverage 95\% & $0.957$ & $0.967$ & $0.981$ & $0.955$ & $1.000$ & $\mathbf{0.949}$ & $0.928$ & $0.942$ & $0.941$ \\
\quad ACE & $0.021$ & $0.038$ & $0.105$ & $0.018$ & $0.341$ & $0.022$ & $0.016$ & $0.022$ & $\mathbf{0.004}$ \\
\quad Width 95\% & $0.036$ & $0.038$ & $0.053$ & $0.036$ & $0.171$ & $0.036$ & $0.033$ & $0.036$ & $0.034$ \\
\quad Log score (all) & $4899$ & $4892$ & $4754$ & $\mathbf{4908}$ & $3883$ & $4855$ & $4893$ & $4839$ & $4902$ \\
\midrule
\multicolumn{10}{l}{\textbf{(ii) Sign-change data}} \\[2pt]
\quad Coverage 50\% & $0.554$ & $\mathbf{0.504}$ & $0.580$ & $0.554$ & $0.459$ & $0.548$ & $0.535$ & $0.471$ & $0.486$ \\
\quad Coverage 80\% & $0.808$ & $0.796$ & $0.828$ & $0.809$ & $0.715$ & $\mathbf{0.803}$ & $0.796$ & $0.746$ & $0.772$ \\
\quad Coverage 95\% & $0.916$ & $\mathbf{0.939}$ & $0.927$ & $0.916$ & $0.851$ & $0.913$ & $0.911$ & $0.894$ & $0.922$ \\
\quad ACE & $0.034$ & $\mathbf{0.004}$ & $0.049$ & $0.036$ & $0.051$ & $0.032$ & $0.026$ & $0.033$ & $0.017$ \\
\quad Width 95\% & $0.186$ & $0.123$ & $0.200$ & $0.189$ & $0.231$ & $0.186$ & $0.183$ & $0.127$ & $0.124$ \\
\quad Log score (all) & $2741$ & $\mathbf{3764}$ & $2767$ & $2721$ & $-589234$ & $2661$ & $3072$ & $3336$ & $3688$ \\
\midrule
\multicolumn{10}{l}{\textbf{(iii) Regional mixture data}} \\[2pt]
\quad Coverage 50\% & $0.547$ & $0.529$ & $0.564$ & $0.533$ & $0.749$ & $0.525$ & $0.519$ & $0.584$ & $\mathbf{0.515}$ \\
\quad Coverage 80\% & $0.821$ & $0.819$ & $0.833$ & $\mathbf{0.801}$ & $0.944$ & $0.804$ & $0.791$ & $0.822$ & $0.789$ \\
\quad Coverage 95\% & $0.944$ & $0.951$ & $\mathbf{0.949}$ & $0.927$ & $0.990$ & $0.935$ & $0.919$ & $0.923$ & $0.921$ \\
\quad ACE & $0.027$ & $0.018$ & $0.038$ & $0.023$ & $0.166$ & $0.017$ & $0.016$ & $0.053$ & $\mathbf{0.014}$ \\
\quad Width 95\% & $0.086$ & $0.082$ & $0.096$ & $0.086$ & $0.171$ & $0.083$ & $0.090$ & $0.096$ & $0.084$ \\
\quad Log score (all) & $4097$ & $\mathbf{4248}$ & $3976$ & $4039$ & $-237389$ & $4093$ & $3989$ & $3870$ & $4034$ \\
\bottomrule
\end{tabular}
}
\end{table}

\begin{table}[tbp]
\centering
\caption{As Table~\ref{tab:calibration}, without the log score, for predictions conditioned on $25$ of the $100$ observed locations, at the $875$ unobserved locations.}
\label{tab:supp_calibration_n25}
\renewcommand{\arraystretch}{1.15}
\resizebox{\textwidth}{!}{%
\begin{tabular}{lccccccccc}
\toprule
 & \multicolumn{2}{c}{BNFC} & \multicolumn{4}{c}{Common factor copula} & \multicolumn{1}{c}{Scale mixture} & \multicolumn{2}{c}{Varying loading} \\
\cmidrule(lr){2-3}\cmidrule(lr){4-7}\cmidrule(lr){8-8}\cmidrule(lr){9-10}
Measure & stat & nonstat & Exp. & Gamma & Pareto & Gamma mix. & Student-$t$ & Gaussian & Gamma \\
\midrule
\multicolumn{10}{l}{\textbf{(i) Gamma-factor data}} \\[2pt]
\quad Coverage 50\% & $0.508$ & $0.538$ & $0.640$ & $\mathbf{0.501}$ & $0.948$ & $0.515$ & $0.487$ & $0.534$ & $0.499$ \\
\quad Coverage 80\% & $0.819$ & $0.844$ & $0.897$ & $0.813$ & $0.992$ & $0.811$ & $0.793$ & $0.824$ & $\mathbf{0.801}$ \\
\quad Coverage 95\% & $0.956$ & $0.967$ & $0.979$ & $0.956$ & $0.999$ & $\mathbf{0.947}$ & $0.938$ & $0.944$ & $0.945$ \\
\quad ACE & $0.008$ & $0.030$ & $0.094$ & $0.005$ & $0.318$ & $0.009$ & $0.009$ & $0.022$ & $\mathbf{0.002}$ \\
\quad Width 95\% & $0.119$ & $0.126$ & $0.174$ & $0.118$ & $0.539$ & $0.119$ & $0.113$ & $0.122$ & $0.114$ \\
\midrule
\multicolumn{10}{l}{\textbf{(ii) Sign-change data}} \\[2pt]
\quad Coverage 50\% & $0.575$ & $0.505$ & $0.607$ & $0.582$ & $0.515$ & $0.569$ & $0.598$ & $0.510$ & $\mathbf{0.497}$ \\
\quad Coverage 80\% & $0.843$ & $\mathbf{0.799}$ & $0.865$ & $0.853$ & $0.775$ & $0.840$ & $0.879$ & $0.788$ & $0.785$ \\
\quad Coverage 95\% & $0.949$ & $0.940$ & $0.957$ & $\mathbf{0.949}$ & $0.895$ & $0.944$ & $0.967$ & $0.924$ & $0.930$ \\
\quad ACE & $0.045$ & $\mathbf{0.004}$ & $0.067$ & $0.051$ & $0.021$ & $0.042$ & $0.066$ & $0.012$ & $0.008$ \\
\quad Width 95\% & $0.620$ & $0.431$ & $0.647$ & $0.625$ & $0.572$ & $0.616$ & $0.644$ & $0.436$ & $0.427$ \\
\midrule
\multicolumn{10}{l}{\textbf{(iii) Regional mixture data}} \\[2pt]
\quad Coverage 50\% & $0.551$ & $\mathbf{0.513}$ & $0.562$ & $0.546$ & $0.763$ & $0.531$ & $0.545$ & $0.614$ & $0.525$ \\
\quad Coverage 80\% & $0.828$ & $0.812$ & $0.831$ & $0.810$ & $0.946$ & $0.806$ & $0.811$ & $0.848$ & $\mathbf{0.802}$ \\
\quad Coverage 95\% & $0.947$ & $\mathbf{0.950}$ & $0.949$ & $0.933$ & $0.988$ & $0.934$ & $0.932$ & $0.942$ & $0.932$ \\
\quad ACE & $0.031$ & $\mathbf{0.009}$ & $0.038$ & $0.028$ & $0.175$ & $0.020$ & $0.028$ & $0.070$ & $0.017$ \\
\quad Width 95\% & $0.278$ & $0.262$ & $0.306$ & $0.278$ & $0.506$ & $0.268$ & $0.280$ & $0.313$ & $0.272$ \\
\bottomrule
\end{tabular}
}
\end{table}

\clearpage

\subsection{Sparse conditioning}
\label{app:supp_sparse}

With $100$ observed locations, the observations determine the value of the factor in each replicate almost exactly, so that differences in spatial prediction between the models can be small even when their log scores differ considerably. To assess the predictions in a setting in which the data are less informative, we repeat the prediction of Section~\ref{M-sec:simulation_studies} of the main text for the same test replicates conditioning on only $25$ of the $100$ observed locations, chosen at random and shared by all models and replicates, without refitting the models.

\begin{figure}[tbp]
\centering
\includegraphics[width=\textwidth]{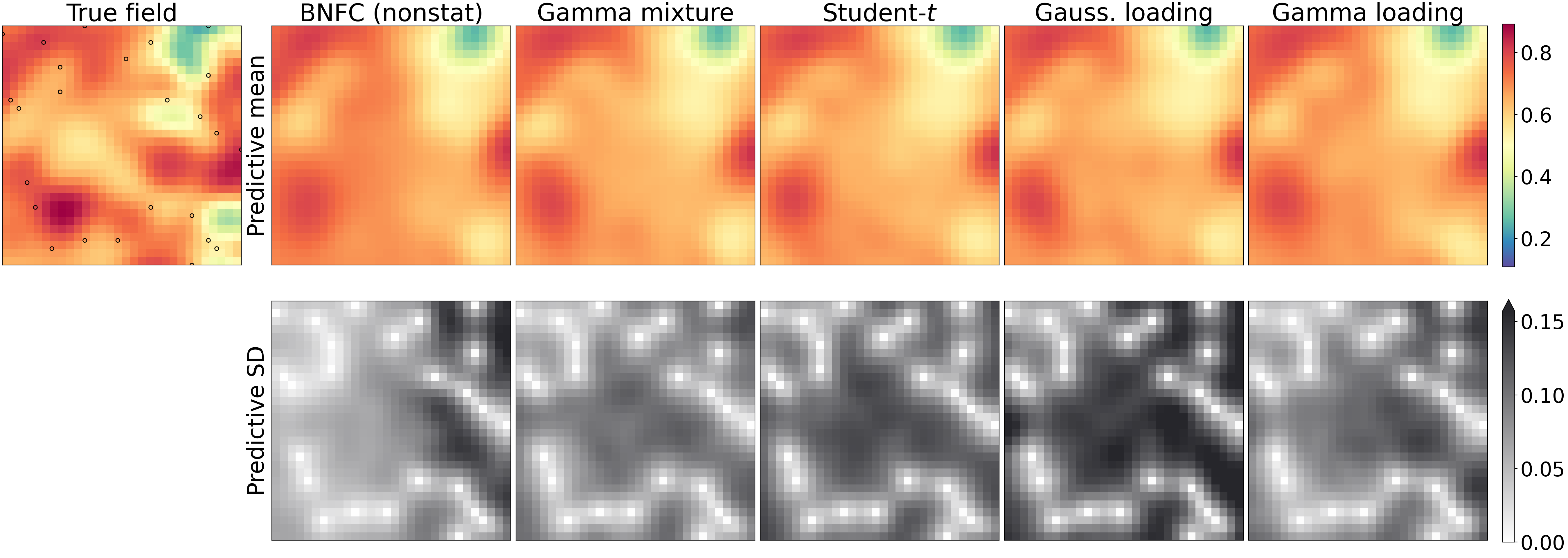}
\caption{Regional mixture data, first test replicate, on the copula scale, with predictions conditioned on $25$ observed locations, for the models of Table~\ref{M-tab:simulation_comparison} of the main text: BNFC (nonstat), the gamma mixture (the best common factor copula), the Student-$t$ copula and the Gaussian and gamma loading copulas. Top row: true copula field, with the $25$ conditioning locations marked by circles, and the predictive mean of each model, on a common color scale centered at the median $0.5$; bottom row: the predictive standard deviation of each model, on a second common scale on which smaller values are lighter.}
\label{fig:heatmaps_n25}
\end{figure}

Table~\ref{tab:supp_summary_n25} summarizes the results on the copula scale. It has the same models and measures as Table~\ref{tab:supp_sim_all}, except for the log scores and the fit times, which do not depend on the conditioning locations and are therefore those of Table~\ref{tab:supp_sim_all}; the PICP and the MPIW are computed at the $875$ locations that are not conditioned on. The advantage of BNFC (nonstat) persists. For the sign-change data, BNFC (stat), the common factor copulas and the Student-$t$ copula have a much higher CRPS. The loading copulas, whose loadings can change sign, come much closer, but they remain behind BNFC (nonstat) by more than two paired standard errors. For the regional mixture data, BNFC (nonstat) again has the lowest CRPS; BNFC (stat), the gamma loading copula and the gamma mixture are closest to it, the exponential and gamma factor models and the Student-$t$ copula are somewhat further behind, the Gaussian loading copula is further behind still, and the Pareto factor model is clearly worse. For the gamma-factor data, BNFC (stat), BNFC (nonstat), the loading copulas and the Student-$t$ copula perform essentially as well as the data-generating gamma factor model, whereas the exponential and, in particular, the Pareto factor models are clearly worse. The prediction intervals are wider than with $100$ conditioning locations, and their calibration shows the same pattern as in Table~\ref{tab:supp_sim_all}. The Pareto factor model overcovers, with much too wide intervals, for the gamma-factor and regional mixture data and undercovers for the sign-change data, and the exponential factor model has too wide intervals for the gamma-factor data. For the sign-change data, the intervals of BNFC (nonstat) and the loading copulas are much narrower than those of the other models; the loading copulas undercover slightly, and the Student-$t$ copula overcovers. For the regional mixture data, BNFC (nonstat) has the narrowest intervals, with coverage close to the nominal level. With fewer conditioning locations, the observations provide less
information about the spatial field, so kriging predictions rely more
on the model's spatial dependence structure. In BNFC (nonstat), the same factor value has a different effect in different parts of the domain, whereas in every common factor model, including the gamma mixture, it shifts the whole latent field by the same amount. Figure~\ref{fig:heatmaps_n25} shows this for one test replicate of the regional mixture data.

\begin{sidewaystable}
\centering
\caption{As Table~\ref{tab:supp_sim_all}, for predictions conditioned on $25$ of the $100$ observed locations: CRPS, energy score and MAE of the predictive draws of the copula field, and the PICP and the MPIW of the nominal $90\%$ prediction intervals at the $875$ locations that are not conditioned on (closest to $0.90$ and lowest are better), averaged over the $100$ test replicates, with standard errors in parentheses. The log scores and fit times do not depend on the conditioning locations and are those of Table~\ref{tab:supp_sim_all}. The best value of each measure is in bold.}
\label{tab:supp_summary_n25}
\renewcommand{\arraystretch}{1.15}
\resizebox{\textwidth}{!}{%
\begin{tabular}{lccccccccc}
\toprule
 & \multicolumn{2}{c}{BNFC} & \multicolumn{4}{c}{Common factor copula} & \multicolumn{1}{c}{Scale mixture} & \multicolumn{2}{c}{Varying loading} \\
\cmidrule(lr){2-3}\cmidrule(lr){4-7}\cmidrule(lr){8-8}\cmidrule(lr){9-10}
Measure & stat & nonstat & Exponential & Gamma & Pareto & Gamma mix. & Student-$t$ & {Gaussian} & {Gamma} \\
\midrule
\multicolumn{10}{l}{\textbf{(i) Gamma-factor data}} \\[2pt]
\quad CRPS & $\mathbf{0.0162}$\,{\scriptsize$(0.0006)$} & $0.0163$\,{\scriptsize$(0.0006)$} & $0.0178$\,{\scriptsize$(0.0006)$} & $0.0162$\,{\scriptsize$(0.0006)$} & $0.0420$\,{\scriptsize$(0.0013)$} & $0.0164$\,{\scriptsize$(0.0006)$} & $0.0162$\,{\scriptsize$(0.0006)$} & $0.0164$\,{\scriptsize$(0.0006)$} & $0.0163$\,{\scriptsize$(0.0006)$} \\
\quad Energy score & ${0.647}$\,{\scriptsize$(0.023)$} & $0.650$\,{\scriptsize$(0.023)$} & $0.713$\,{\scriptsize$(0.024)$} & $\mathbf{0.647}$\,{\scriptsize$(0.023)$} & $1.744$\,{\scriptsize$(0.052)$} & $0.657$\,{\scriptsize$(0.023)$} & $0.649$\,{\scriptsize$(0.023)$} & $0.655$\,{\scriptsize$(0.023)$} & $0.650$\,{\scriptsize$(0.024)$} \\
\quad MAE & $0.0229$\,{\scriptsize$(0.0008)$} & $0.0229$\,{\scriptsize$(0.0008)$} & $0.0237$\,{\scriptsize$(0.0008)$} & $\mathbf{0.0229}$\,{\scriptsize$(0.0008)$} & $0.0448$\,{\scriptsize$(0.0022)$} & $0.0231$\,{\scriptsize$(0.0008)$} & $\mathbf{0.0229}$\,{\scriptsize$(0.0008)$} & $0.0229$\,{\scriptsize$(0.0008)$} & $0.0230$\,{\scriptsize$(0.0008)$} \\
\quad PICP $90\%$ & $0.912$\,{\scriptsize$(0.004)$} & $0.930$\,{\scriptsize$(0.004)$} & $0.956$\,{\scriptsize$(0.005)$} & $0.912$\,{\scriptsize$(0.004)$} & $0.997$\,{\scriptsize$(0.002)$} & $\mathbf{0.901}$\,{\scriptsize$(0.007)$} & $0.890$\,{\scriptsize$(0.006)$} & $0.907$\,{\scriptsize$(0.009)$} & $0.898$\,{\scriptsize$(0.004)$} \\
\quad MPIW $90\%$ & $0.100$\,{\scriptsize$(0.003)$} & $0.106$\,{\scriptsize$(0.003)$} & $0.148$\,{\scriptsize$(0.007)$} & $0.099$\,{\scriptsize$(0.003)$} & $0.476$\,{\scriptsize$(0.013)$} & $0.100$\,{\scriptsize$(0.003)$} & $\mathbf{0.096}$\,{\scriptsize$(0.003)$} & $0.104$\,{\scriptsize$(0.003)$} & $0.097$\,{\scriptsize$(0.003)$} \\
\midrule
\multicolumn{10}{l}{\textbf{(ii) Sign-change data}} \\[2pt]
\quad CRPS & $0.0856$\,{\scriptsize$(0.0009)$} & $\mathbf{0.0643}$\,{\scriptsize$(0.0016)$} & $0.0859$\,{\scriptsize$(0.0008)$} & $0.0852$\,{\scriptsize$(0.0009)$} & $0.0919$\,{\scriptsize$(0.0021)$} & $0.0852$\,{\scriptsize$(0.0009)$} & $0.0844$\,{\scriptsize$(0.0009)$} & $0.0657$\,{\scriptsize$(0.0016)$} & $0.0649$\,{\scriptsize$(0.0016)$} \\
\quad Energy score & $3.591$\,{\scriptsize$(0.051)$} & $\mathbf{2.704}$\,{\scriptsize$(0.056)$} & $3.607$\,{\scriptsize$(0.049)$} & $3.577$\,{\scriptsize$(0.051)$} & $3.865$\,{\scriptsize$(0.110)$} & $3.574$\,{\scriptsize$(0.050)$} & $3.589$\,{\scriptsize$(0.056)$} & $2.758$\,{\scriptsize$(0.051)$} & $2.732$\,{\scriptsize$(0.055)$} \\
\quad MAE & $0.1259$\,{\scriptsize$(0.0019)$} & $\mathbf{0.0926}$\,{\scriptsize$(0.0023)$} & $0.1260$\,{\scriptsize$(0.0020)$} & $0.1254$\,{\scriptsize$(0.0019)$} & $0.1317$\,{\scriptsize$(0.0028)$} & $0.1254$\,{\scriptsize$(0.0019)$} & $0.1278$\,{\scriptsize$(0.0022)$} & $0.0939$\,{\scriptsize$(0.0022)$} & $0.0934$\,{\scriptsize$(0.0023)$} \\
\quad PICP $90\%$ & $0.914$\,{\scriptsize$(0.008)$} & $\mathbf{0.893}$\,{\scriptsize$(0.004)$} & $0.929$\,{\scriptsize$(0.007)$} & $0.918$\,{\scriptsize$(0.008)$} & $0.852$\,{\scriptsize$(0.012)$} & $0.911$\,{\scriptsize$(0.008)$} & $0.942$\,{\scriptsize$(0.003)$} & $0.877$\,{\scriptsize$(0.007)$} & $0.882$\,{\scriptsize$(0.004)$} \\
\quad MPIW $90\%$ & $0.556$\,{\scriptsize$(0.005)$} & $0.377$\,{\scriptsize$(0.008)$} & $0.584$\,{\scriptsize$(0.006)$} & $0.561$\,{\scriptsize$(0.005)$} & $0.511$\,{\scriptsize$(0.006)$} & $0.552$\,{\scriptsize$(0.005)$} & $0.579$\,{\scriptsize$(0.003)$} & $0.381$\,{\scriptsize$(0.008)$} & $\mathbf{0.375}$\,{\scriptsize$(0.008)$} \\
\midrule
\multicolumn{10}{l}{\textbf{(iii) Regional mixture data}} \\[2pt]
\quad CRPS & $0.0386$\,{\scriptsize$(0.0016)$} & $\mathbf{0.0380}$\,{\scriptsize$(0.0016)$} & $0.0392$\,{\scriptsize$(0.0016)$} & $0.0393$\,{\scriptsize$(0.0016)$} & $0.0489$\,{\scriptsize$(0.0019)$} & $0.0388$\,{\scriptsize$(0.0016)$} & $0.0393$\,{\scriptsize$(0.0016)$} & $0.0404$\,{\scriptsize$(0.0016)$} & $0.0387$\,{\scriptsize$(0.0016)$} \\
\quad Energy score & $1.577$\,{\scriptsize$(0.064)$} & $\mathbf{1.570}$\,{\scriptsize$(0.064)$} & $1.596$\,{\scriptsize$(0.064)$} & $1.596$\,{\scriptsize$(0.064)$} & $1.998$\,{\scriptsize$(0.074)$} & $1.582$\,{\scriptsize$(0.065)$} & $1.590$\,{\scriptsize$(0.064)$} & $1.641$\,{\scriptsize$(0.062)$} & $1.583$\,{\scriptsize$(0.064)$} \\
\quad MAE & $0.0545$\,{\scriptsize$(0.0023)$} & $0.0545$\,{\scriptsize$(0.0023)$} & $0.0550$\,{\scriptsize$(0.0023)$} & $0.0546$\,{\scriptsize$(0.0023)$} & $0.0635$\,{\scriptsize$(0.0027)$} & $0.0547$\,{\scriptsize$(0.0023)$} & $0.0546$\,{\scriptsize$(0.0023)$} & $0.0550$\,{\scriptsize$(0.0023)$} & $\mathbf{0.0544}$\,{\scriptsize$(0.0023)$} \\
\quad PICP $90\%$ & $0.908$\,{\scriptsize$(0.005)$} & $\mathbf{0.906}$\,{\scriptsize$(0.005)$} & $0.909$\,{\scriptsize$(0.006)$} & $0.892$\,{\scriptsize$(0.007)$} & $0.976$\,{\scriptsize$(0.004)$} & $0.892$\,{\scriptsize$(0.005)$} & $0.892$\,{\scriptsize$(0.007)$} & $0.910$\,{\scriptsize$(0.010)$} & $0.888$\,{\scriptsize$(0.006)$} \\
\quad MPIW $90\%$ & $0.239$\,{\scriptsize$(0.009)$} & $\mathbf{0.226}$\,{\scriptsize$(0.008)$} & $0.264$\,{\scriptsize$(0.011)$} & $0.238$\,{\scriptsize$(0.008)$} & $0.448$\,{\scriptsize$(0.016)$} & $0.230$\,{\scriptsize$(0.008)$} & $0.239$\,{\scriptsize$(0.008)$} & $0.269$\,{\scriptsize$(0.008)$} & $0.234$\,{\scriptsize$(0.009)$} \\
\bottomrule
\end{tabular}
}
\end{sidewaystable}

\subsection{Comparison with a Bayesian transport map}
\label{app:supp_transport_map}

Section~\ref{M-sec:simulation_studies} of the main text compares BNFC with the Bayesian transport map of \citet{katzfuss2024scalable}, a nonparametric model for the joint distribution of replicated non-Gaussian spatial fields. The map expresses the joint density of the normal scores
$\mathbf{x}=\bigl(\Phi^{-1}(u_1),\ldots,\Phi^{-1}(u_n)\bigr)^\top$
as a product of conditional densities under a maximin ordering of the
locations. A Gaussian process regression models each conditional density
using the normal scores at up to $30$ nearest preceding locations as
predictors. The priors shrink the map toward that of a Gaussian field
with smoothness parameter $1.5$. We fit it with the \texttt{batram} package in Python to the same $2{,}000$ training replicates at the $100$ observed locations, by $300$ steps of the Adam optimizer with learning rate $0.01$ and a cosine schedule; over the last $30$ steps, the loss changed by less than $0.014$ per replicate in every scenario. Let $f_{\mathbf{X}}$ denote the joint predictive density of the normal
scores under the fitted transport map. For each test replicate $\bfu$,
the log predictive density on the copula scale is
\[
\log f_{\mathbf{U}}(\bfu)
=
\log f_{\mathbf{X}}(\mathbf{x})
-\sum_{j=1}^{n}\log\phi(x_j),
\]
where $\phi$ is the standard normal density and the sum accounts for
the change of variables from $\mathbf{x}$ to $\bfu$. We evaluate this
score on the same held-out replicates at the same $100$ observed
locations as in Section~\ref{app:supp_measures}. The map is defined only at the locations with training data, so that it gives no score over the full grid and no predictions at unobserved locations. Table~\ref{tab:supp_transport_map} gives the results, together with those for the ERA5 precipitation data of Section~\ref{M-sec:application} of the main text. Because the memory requirement of the map grows quadratically with the number of training replicates, it is fitted there to the $2{,}000$ most recent training days, from July 2017 to December 2022, by $100$ steps of the Adam optimizer with otherwise the same settings as in the simulations, whereas the other models use all training days.

\begin{table}[tbp]
\centering
\caption{Held-out copula log score at the $100$ observed locations, averaged over the $100$ test replicates of the simulations and the $200$ test days of the ERA5 data, with standard errors in parentheses, of BNFC (nonstat), the Bayesian transport map, the Student-$t$ copula, the gamma loading copula and the best common factor copula of Tables~\ref{M-tab:simulation_comparison} and~\ref{M-tab:application_comparison} of the main text, and the average paired difference between the transport map and BNFC (nonstat). The highest log score in each column is in bold. For the ERA5 data, the transport map is fitted to the $2{,}000$ most recent training days. Fit times of the transport map are on $24$ CPU threads.}
\label{tab:supp_transport_map}
\renewcommand{\arraystretch}{1.15}
\resizebox{\textwidth}{!}{%
\begin{tabular}{lcccc}
\toprule
 & (i) Gamma factor & (ii) Sign change & (iii) Regional mixture & ERA5 \\
\midrule
BNFC (nonstat) & $265.1$\,{\scriptsize$(4.6)$} & $\mathbf{144.2}$\,{\scriptsize$(5.7)$} & $\mathbf{197.4}$\,{\scriptsize$(7.0)$} & $122.2$\,{\scriptsize$(3.6)$} \\
Transport map & $258.0$\,{\scriptsize$(4.3)$} & $135.7$\,{\scriptsize$(5.5)$} & $175.2$\,{\scriptsize$(7.1)$} & $\mathbf{123.4}$\,{\scriptsize$(4.0)$} \\
Student-$t$ copula & $265.1$\,{\scriptsize$(4.9)$} & $111.7$\,{\scriptsize$(3.8)$} & $181.1$\,{\scriptsize$(7.8)$} & $113.5$\,{\scriptsize$(3.6)$} \\
Gamma loading copula & $266.3$\,{\scriptsize$(4.9)$} & $143.2$\,{\scriptsize$(5.7)$} & $187.7$\,{\scriptsize$(6.9)$} & $111.7$\,{\scriptsize$(3.7)$} \\
Best common factor & $\mathbf{266.8}$\,{\scriptsize$(5.0)$} & $105.6$\,{\scriptsize$(2.5)$} & $188.9$\,{\scriptsize$(7.5)$} & $104.4$\,{\scriptsize$(3.7)$} \\
\midrule
Transport map $-$ BNFC (nonstat) & $-7.1$\,{\scriptsize$(0.8)$} & $-8.5$\,{\scriptsize$(0.7)$} & $-22.3$\,{\scriptsize$(0.8)$} & $1.2$\,{\scriptsize$(5.3)$} \\
Fit time of the transport map (min) & $97$ & $97$ & $105$ & $33$ \\
\bottomrule
\end{tabular}%
}
\end{table}

\subsection{Posterior of the factor}\label{app:supp_factor_posterior}

\begin{figure}[tbp]
\centering
\includegraphics[width=\textwidth]{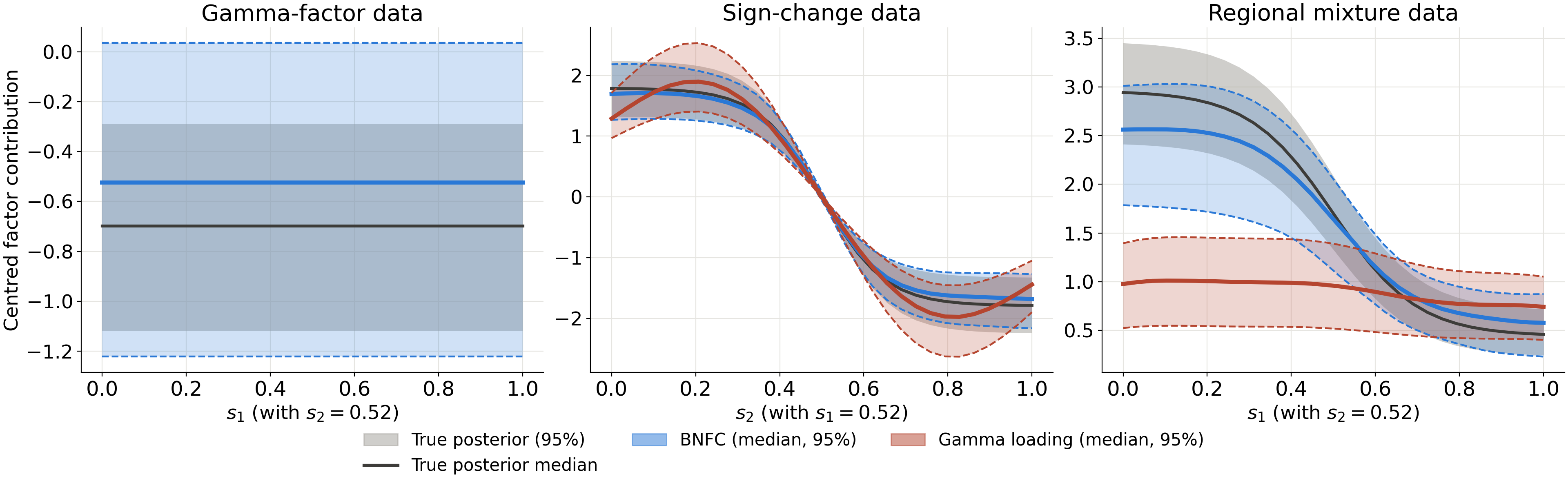}
\caption{Posterior of the centered factor contribution along a transect through the domain for the first test replicate of each scenario, given the $100$ observed locations: BNFC (stat) in (i), and BNFC (nonstat) and the gamma loading copula in (ii) and (iii). Gray: true posterior ($95\%$ interval and median); colors: posterior median (solid) and $95\%$ interval (shaded, dashed edges) of each model. The transect is along $s_1$ for the gamma-factor and regional mixture data and along $s_2$, across the sign change, for the sign-change data.}
\label{fig:factor_posterior}
\end{figure}

Section~\ref{M-sec:simulation_studies} of the main text summarizes the comparison of the posterior of the centered factor contribution given the $100$ observed locations with its true posterior under the data-generating model (Section~\ref{app:supp_measures}). Figure~\ref{fig:factor_posterior} shows both along a transect for the first test replicate of each scenario. For the sign-change and regional mixture data, the posterior of BNFC (nonstat) follows the true posterior closely, across the sign change and across the change of factor family, whereas the gamma loading copula departs from it in parts of the domain, and the factor of a common factor copula, which shifts the whole field by the same amount, cannot follow it at all. For the gamma-factor data, the posterior of BNFC (stat) is constant along the transect, as is the true posterior, and covers it. The calibration of the resulting predictions is assessed in Section~\ref{app:supp_calibration}.

\subsection{Tuning of BNFC}
\label{app:supp_tuning}

This section describes how the optimization settings of BNFC in Section~\ref{M-sec:simulation_studies} of the main text are chosen. The settings are tuned for each scenario without using the test replicates. Each candidate is fitted to training replicates $1$ to $1{,}800$, and its held-out copula log score, computed as in Section~\ref{app:supp_measures}, is evaluated on training replicates $1{,}801$ to $2{,}000$, at the $100$ observed locations and at all $900$ grid locations. BNFC (stat) is tuned on the gamma-factor data and BNFC (nonstat) on the sign-change and regional mixture data; the other BNFC model in each scenario uses the default settings of Section~\ref{M-sec:empirical} of the main text, that is, Adam with learning rate $10^{-3}$, halved every $20$ epochs, for at most $100$ epochs, early stopping after $25$ epochs without an increase of the ELBO per replicate of more than $0.01$, $M_1=100$, $M_2=50$ and mini-batches of $20$ replicates. The candidate with the highest validation log score over all locations is refitted to all $2{,}000$ training replicates.

The candidates vary the learning rate, the number of epochs and the patience of early stopping, the number $M_1$ of factor draws, the mini-batch size, the step size of the stochastic gradient MCMC and the weight decay, and, for BNFC (nonstat), the width, depth and output activation of the network and Fourier features of the location. Table~\ref{tab:supp_tuning} lists them with their validation log scores. All candidates other than the default settings use at most $200$ epochs, a learning rate halved every $40$ epochs and early stopping after $60$ (stat) or $40$ (nonstat) epochs without an increase of the ELBO per replicate of more than $0.001$; the smaller mini-batches of some candidates are needed to fit the larger networks and numbers of factor draws into the memory of the GPU. The selected settings use, in addition, $M_1=300$ and learning rate $3\times10^{-3}$ for the gamma-factor data, learning rate $3\times10^{-3}$ for the sign-change data, and $M_1=200$ with mini-batches of $10$ replicates for the regional mixture data. In every scenario, their validation log score over all locations is clearly higher than that of the default settings.

In every scenario, the longer training improves on the default settings, and a larger learning rate or number $M_1$ of factor draws improves further. Fourier features of the location fit the observed locations as well as the other candidates but interpolate poorly to the unobserved ones. The selected setting of BNFC (stat) is considerably closer to the data-generating gamma factor model over all locations than the default settings. The longer training of the tuned models increases their fit times (Table~\ref{M-tab:simulation_comparison} of the main text).

\begin{table}[tbp]
\centering
\caption{Validation log scores of the candidate settings of BNFC: held-out copula log score at the $100$ observed and at all $900$ grid locations of training replicates $1{,}801$ to $2{,}000$, for fits to replicates $1$ to $1{,}800$, with standard errors in parentheses. ``Long'' denotes the longer training described above; unless stated otherwise, $M_1=100$, $M_2=50$, learning rate $10^{-3}$ and mini-batches of $20$ replicates. The selected setting, with the highest log score over all locations, is in bold. For scenario (i), the data-generating gamma factor model is included for reference.}
\label{tab:supp_tuning}
\renewcommand{\arraystretch}{1.15}
\resizebox{\textwidth}{!}{%
\begin{tabular}{lcc}
\toprule
Candidate & Log score (observed) & Log score (all locations) \\
\midrule
\multicolumn{3}{l}{\textbf{(i) Gamma-factor data, BNFC (stat)}} \\[2pt]
\quad Default settings & $272.0$\,{\scriptsize$(3.9)$} & $4953.8$\,{\scriptsize$(35.6)$} \\
\quad Long: 200 epochs, halving every 40, patience 60, tolerance $0.001$ & $272.7$\,{\scriptsize$(4.1)$} & $4962.1$\,{\scriptsize$(36.9)$} \\
\quad Long, learning rate $3\times10^{-3}$ & $273.9$\,{\scriptsize$(4.3)$} & $4975.2$\,{\scriptsize$(39.0)$} \\
\quad Long, $M_1=300$ & $275.0$\,{\scriptsize$(4.3)$} & $4980.0$\,{\scriptsize$(39.1)$} \\
\quad Long, SG-MCMC step $3\times10^{-4}$ & $272.1$\,{\scriptsize$(4.0)$} & $4952.5$\,{\scriptsize$(36.1)$} \\
\quad Long, SG-MCMC step $3\times10^{-3}$ & $273.3$\,{\scriptsize$(4.2)$} & $4968.5$\,{\scriptsize$(37.5)$} \\
\quad Long, weight decay $10^{-5}$ & $272.5$\,{\scriptsize$(4.0)$} & $4962.4$\,{\scriptsize$(36.5)$} \\
\quad Long, mini-batches of $50$ & $270.4$\,{\scriptsize$(3.7)$} & $4945.0$\,{\scriptsize$(33.9)$} \\
\quad Long, $M_1=300$, learning rate $3\times10^{-3}$ & $275.8$\,{\scriptsize$(4.5)$} & $\mathbf{4987.2}$\,{\scriptsize$(40.7)$} \\
\quad \emph{Data-generating gamma factor model} & $277.5$\,{\scriptsize$(4.9)$} & $5009.5$\,{\scriptsize$(45.0)$} \\
\midrule
\multicolumn{3}{l}{\textbf{(ii) Sign-change data, BNFC (nonstat)}} \\[2pt]
\quad Default settings & $147.2$\,{\scriptsize$(3.7)$} & $3777.2$\,{\scriptsize$(31.2)$} \\
\quad Long: 200 epochs, halving every 40, patience 40, tolerance $0.001$ & $147.7$\,{\scriptsize$(3.8)$} & $3797.9$\,{\scriptsize$(32.5)$} \\
\quad Long, linear output & $147.8$\,{\scriptsize$(3.8)$} & $3799.2$\,{\scriptsize$(32.5)$} \\
\quad Long, $4$ hidden layers of width $128$, mini-batches of $5$ & $148.1$\,{\scriptsize$(3.9)$} & $3806.5$\,{\scriptsize$(33.7)$} \\
\quad Long, $16$ Fourier features of $\bfs$ (scale $2$), mini-batches of $10$ & $148.0$\,{\scriptsize$(3.9)$} & $3179.4$\,{\scriptsize$(15.6)$} \\
\quad Long, learning rate $3\times10^{-3}$ & $148.1$\,{\scriptsize$(3.9)$} & $\mathbf{3811.1}$\,{\scriptsize$(34.4)$} \\
\quad Long, $M_1=200$, mini-batches of $10$ & $147.7$\,{\scriptsize$(3.8)$} & $3790.0$\,{\scriptsize$(32.0)$} \\
\quad Long, $4$ hidden layers of width $128$, linear output, mini-batches of $5$ & $148.2$\,{\scriptsize$(3.9)$} & $3811.0$\,{\scriptsize$(34.6)$} \\
\midrule
\multicolumn{3}{l}{\textbf{(iii) Regional mixture data, BNFC (nonstat)}} \\[2pt]
\quad Default settings & $191.1$\,{\scriptsize$(3.9)$} & $4180.1$\,{\scriptsize$(35.4)$} \\
\quad Long: 200 epochs, halving every 40, patience 40, tolerance $0.001$ & $191.3$\,{\scriptsize$(4.0)$} & $4186.2$\,{\scriptsize$(36.1)$} \\
\quad Long, linear output & $191.3$\,{\scriptsize$(4.0)$} & $4187.0$\,{\scriptsize$(36.3)$} \\
\quad Long, $4$ hidden layers of width $128$, mini-batches of $5$ & $191.4$\,{\scriptsize$(4.0)$} & $4192.1$\,{\scriptsize$(36.6)$} \\
\quad Long, $16$ Fourier features of $\bfs$ (scale $2$), mini-batches of $10$ & $190.9$\,{\scriptsize$(3.9)$} & $1948.9$\,{\scriptsize$(281.7)$} \\
\quad Long, learning rate $3\times10^{-3}$ & $191.4$\,{\scriptsize$(4.0)$} & $4189.8$\,{\scriptsize$(36.6)$} \\
\quad Long, $M_1=200$, mini-batches of $10$ & $192.0$\,{\scriptsize$(4.0)$} & $\mathbf{4201.7}$\,{\scriptsize$(36.4)$} \\
\quad Long, $4$ hidden layers of width $128$, linear output, mini-batches of $5$ & $191.4$\,{\scriptsize$(4.0)$} & $4190.4$\,{\scriptsize$(36.7)$} \\
\bottomrule
\end{tabular}%
}
\end{table}

\section{Additional results for the precipitation application}
\label{app:application_tail}

The ERA5 daily total precipitation of Section~\ref{M-sec:application} of the main text, which is available on a regular $32\times32$ grid, is averaged over two consecutive days and detrended, and the coordinates are rescaled to $[0,1]^2$. Because $6.0\%$ of the values are zero, the copula data are the randomized rank scores
\[
\hat{u}_{ij}=\frac{\#\{k:y_{kj}<y_{ij}\}+r_i\,(\#\{k:y_{kj}=y_{ij}\}+1)}{N+1},\qquad r_i\sim U(0,1),
\]
where $k$ runs over the training days and $r_i$ is shared by all locations of day $i$; the same transform is applied to the test days. The marginal network $T_{\hat{\bfpsi}}$ of Section~\ref{M-subsec:marginals} of the main text, trained on the training days at all $1{,}024$ grid locations, is used only to map predictions to the data scale.

For the ERA5 data, we fit BNFC (stat), BNFC (nonstat), the exponential,
gamma, and Pareto common factor copulas, the Student-$t$ copula, and the
Gaussian and gamma loading copulas from
Section~\ref{M-sec:simulation_studies} of the main text. We use $K=25$
loading bumps for both loading copulas and also fit the Gaussian
loading copula with $K=49$ bumps on a $7\times7$ grid. The architecture of BNFC (nonstat) is chosen by the held-out log score on the days of 2022 after fitting candidate networks to the days from 1994 to 2021: four hidden layers of width $128$, a linear output and an exponential correlation function. With this architecture, BNFC (nonstat) is refitted to all training days, with estimated range $\hat\ell=0.61$ and nugget $\hat\tau^2=1\times10^{-5}$, and $20$ weight samples. The Student-$t$ copula has priors $\nu\sim\mathrm{Ga}(2,\text{rate }0.1)$ and $\beta\sim U(0,1)$, with posterior means $\hat\nu=17.5$ and $\hat\beta=0.35$. For each test day, we condition on the $100$ observed locations and predict all $1{,}024$ grid locations.

Table~\ref{tab:supp_app_all} gives the results of all models fitted to the ERA5 precipitation data of Section~\ref{M-sec:application} of the main text, including the gamma and Pareto common factor copulas and the Gaussian loading copula with $K=49$ loading bumps on a $7\times7$ grid. The common factor copulas are clearly behind the Student-$t$ copula and the loading copulas, and the Pareto factor model is misspecified, with a few test days of almost zero density. With $K=49$, the Gaussian loading copula is slightly worse than with $K=25$ over the full grid and in CRPS and energy score, so that the results do not depend on a finer loading. The gamma and Gaussian loading copulas yield similar predictive scores
and interval coverage. The Hamiltonian Monte Carlo chain for the gamma
loading copula mixes slowly, with a minimum effective sample size of
$4$ across parameters based on $300$ draws retained after $300$ warm-up
iterations. Fitting the gamma loading copula takes $13.4$ hours. The slow mixing is due to a weak identification of the shape $\kappa$. The centered factor contribution is $b(\bfs)\sqrt{\kappa}\,S_\kappa$, where $S_\kappa=(H_\kappa^{-1}(V)-\kappa)/\sqrt{\kappa}$ is the standardized $\mathrm{Ga}(\kappa,1)$ variable, with skewness $2/\sqrt{\kappa}$. The data therefore determine the scale $b(\bfs)\sqrt{\kappa}$ well, but $\kappa$ only through the skewness of $S_\kappa$, which changes little once $\kappa$ is large. The posterior of $(\bs{\beta},\kappa)$ then concentrates near a long, curved ridge on which $b(\bfs)\sqrt{\kappa}$ is nearly constant, and the sampler, whose step size and mass matrix are adapted to the narrow direction across the ridge, moves only slowly along it. In our fit, the mode search that initializes the chain stopped at a small value of $\kappa$, and the chain drifted towards larger values throughout. Sampling the standardized coefficients $\bs{\beta}\sqrt{\kappa}$ instead removes the curvature of the ridge, but even with this parameterization a mode search with $3{,}000$ steps was still increasing $\kappa$ at its end, so that $\kappa$ remains weakly identified. This is a limitation of fitting the gamma loading copula by Hamiltonian Monte Carlo. We do not expect it to change the comparison: the predictions depend on $\kappa$ only through the skewness of the factor, which is small for all values of $\kappa$ visited by the chain, and the gamma loading copula is almost indistinguishable from the Gaussian loading copula in Table~\ref{tab:supp_app_all}.

\begin{table}[tbp]
\centering
\caption{As Table~\ref{M-tab:application_comparison} of the main text, for all models, with the same measures. Gaussian loading ($K$): Gaussian loading copula with $K$ loading bumps. The best value of each measure is in bold; for the PICP, the value closest to $0.90$.}
\label{tab:supp_app_all}
\renewcommand{\arraystretch}{1.15}
\resizebox{\textwidth}{!}{%
\begin{tabular}{lccccccccc}
\toprule
 & \multicolumn{2}{c}{BNFC} & \multicolumn{3}{c}{Common factor copula} & Scale mixture & \multicolumn{3}{c}{Varying loading} \\
\cmidrule(lr){2-3}\cmidrule(lr){4-6}\cmidrule(lr){7-7}\cmidrule(lr){8-10}
Measure & stat & nonstat & Exponential & Gamma & Pareto & Student-$t$ & Gaussian ($25$) & Gaussian ($49$) & Gamma \\
\midrule
\quad CRPS & $0.0373$\,{\scriptsize$(0.0009)$} & $\mathbf{0.0326}$\,{\scriptsize$(0.0007)$} & $0.0378$\,{\scriptsize$(0.0009)$} & $0.0377$\,{\scriptsize$(0.0009)$} & $0.0420$\,{\scriptsize$(0.0012)$} & $0.0374$\,{\scriptsize$(0.0009)$} & $0.0373$\,{\scriptsize$(0.0009)$} & $0.0377$\,{\scriptsize$(0.0008)$} & $0.0373$\,{\scriptsize$(0.0009)$} \\
\quad Energy score & $1.717$\,{\scriptsize$(0.038)$} & $\mathbf{1.573}$\,{\scriptsize$(0.034)$} & $1.793$\,{\scriptsize$(0.041)$} & $1.789$\,{\scriptsize$(0.042)$} & $2.029$\,{\scriptsize$(0.058)$} & $1.777$\,{\scriptsize$(0.040)$} & $1.726$\,{\scriptsize$(0.037)$} & $1.772$\,{\scriptsize$(0.034)$} & $1.726$\,{\scriptsize$(0.037)$} \\
\quad MAE & $0.0502$\,{\scriptsize$(0.0012)$} & $\mathbf{0.0443}$\,{\scriptsize$(0.0010)$} & $0.0506$\,{\scriptsize$(0.0012)$} & $0.0504$\,{\scriptsize$(0.0012)$} & $0.0558$\,{\scriptsize$(0.0015)$} & $0.0507$\,{\scriptsize$(0.0012)$} & $0.0503$\,{\scriptsize$(0.0012)$} & $0.0508$\,{\scriptsize$(0.0011)$} & $0.0502$\,{\scriptsize$(0.0012)$} \\
\quad Log score (observed) & $112.9$\,{\scriptsize$(3.5)$} & $\mathbf{122.2}$\,{\scriptsize$(3.6)$} & $104.4$\,{\scriptsize$(3.7)$} & $104.7$\,{\scriptsize$(3.8)$} & $-774.0$\,{\scriptsize$(241.1)$} & $113.5$\,{\scriptsize$(3.6)$} & $111.6$\,{\scriptsize$(3.7)$} & $111.7$\,{\scriptsize$(3.7)$} & $111.7$\,{\scriptsize$(3.7)$} \\
\quad Log score (all locations) & $1684$\,{\scriptsize$(37)$} & $\mathbf{1995}$\,{\scriptsize$(32)$} & $1619$\,{\scriptsize$(46)$} & $1600$\,{\scriptsize$(48)$} & $-13973$\,{\scriptsize$(3674)$} & $1817$\,{\scriptsize$(36)$} & $1685$\,{\scriptsize$(38)$} & $1678$\,{\scriptsize$(38)$} & $1690$\,{\scriptsize$(38)$} \\
\quad {PICP $90\%$} & {$0.850$\,{\scriptsize$(0.006)$}} & {$\mathbf{0.887}$\,{\scriptsize$(0.005)$}} & {$0.826$\,{\scriptsize$(0.008)$}} & {$0.818$\,{\scriptsize$(0.008)$}} & {$0.743$\,{\scriptsize$(0.010)$}} & {$0.829$\,{\scriptsize$(0.003)$}} & {$0.858$\,{\scriptsize$(0.006)$}} & {$0.855$\,{\scriptsize$(0.007)$}} & {$0.858$\,{\scriptsize$(0.007)$}} \\
\quad {MPIW $90\%$} & {$0.194$\,{\scriptsize$(0.003)$}} & {$0.204$\,{\scriptsize$(0.003)$}} & {$0.191$\,{\scriptsize$(0.003)$}} & {$0.184$\,{\scriptsize$(0.003)$}} & {$\mathbf{0.173}$\,{\scriptsize$(0.002)$}} & {$0.190$\,{\scriptsize$(0.004)$}} & {$0.205$\,{\scriptsize$(0.003)$}} & {$0.203$\,{\scriptsize$(0.003)$}} & {$0.204$\,{\scriptsize$(0.003)$}} \\
\quad Fit time (s) & $274$ & $4745$ &$2314$ & $2436$ & $1160$ & $74$ & $136$ & $197$ & $48114$ \\
\bottomrule
\end{tabular}%
}
\end{table}

\paragraph{Comparison with a Bayesian transport map.} 
At the observed locations, the nonparametric Bayesian transport map of \citet{katzfuss2024scalable}, fitted as in Section~\ref{app:supp_transport_map} to the $2{,}000$ most recent training days, has a slightly higher held-out log score than BNFC (nonstat), even though the difference is not statistically significant, unlike in the simulations (Table~\ref{tab:supp_transport_map}). It is important to note that the Bayesian transport map is defined only at the locations with training data, so that it gives no score over the full grid and no predictions at unobserved locations, and it has no interpretable factor.

\subsection{Tail dependence and relation to the relief}

\begin{figure}[tbp]
\centering
\includegraphics[width=\textwidth]{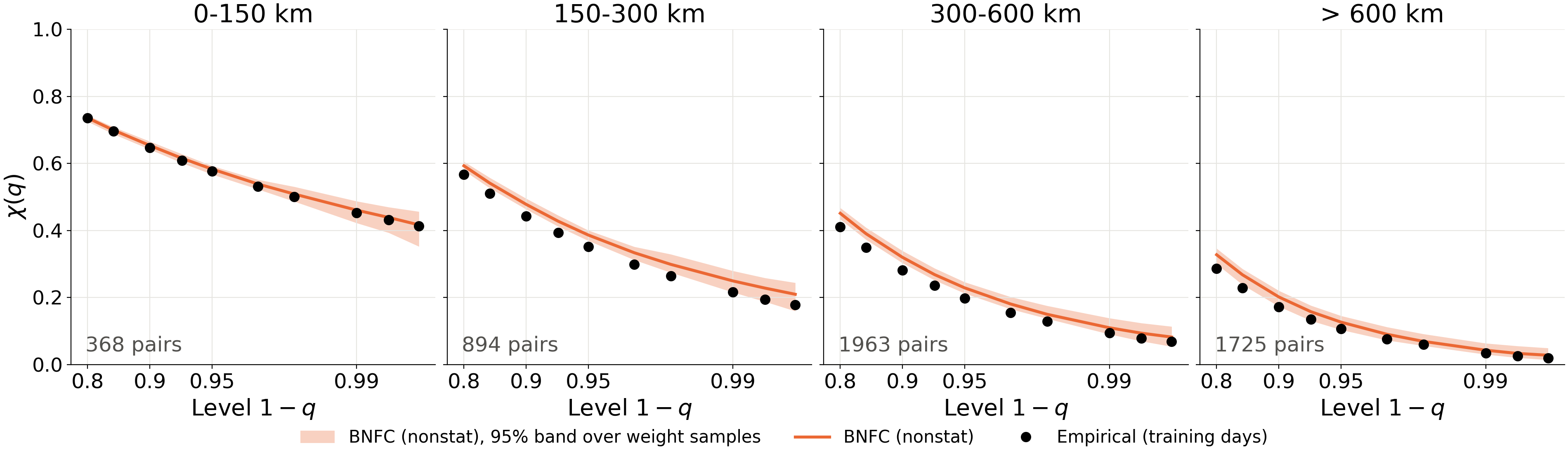}
\caption{ERA5 precipitation: upper tail dependence $\chi(q)=P(U_k>1-q\mid U_j>1-q)$ against the level $1-q$, averaged over the pairs of the $100$ observed locations in four distance classes, for the training days (dots) and for $40{,}000$ days simulated from BNFC (nonstat) (line), with the $95\%$ range of the averages over the $20$ weight samples (band).}
\label{fig:application_dependence}
\end{figure}

Figure~\ref{fig:application_dependence} shows the upper tail dependence of BNFC (nonstat), fitted to the ERA5 precipitation data of Section~\ref{M-sec:application} of the main text, through $\chi(q)=P(U_k>1-q\mid U_j>1-q)$, averaged over the pairs of observed locations in four distance classes, for levels $1-q$ between $0.8$ and $0.995$, together with the empirical values of the training days. The model values are computed from $40{,}000$ simulated days, $2{,}000$ from each weight sample.

The model captures the extremal dependence of precipitation closely. In the data, the co-occurrence of extremes decreases both with the level and with distance, and the model reproduces both decreases in all four distance classes. Up to $150$~km, the model values lie on the empirical ones at every level; beyond $150$~km, they are slightly larger at moderate levels and converge to the empirical values at the highest levels, where both lie within the band over the weight samples. Beyond $600$~km, $\chi(q)$ falls below $0.05$ at the highest levels in both the model and the data, so that the model makes the extremes at distant locations nearly independent, as the data do, which a common factor copula or the Student-$t$ copula, whose tail dependence does not vanish with distance, cannot achieve. By Proposition~\ref{M-thm:asymptotic_independence} of the main text, a finite network gives asymptotic independence, so that $\chi(q)$ of the model eventually decreases to zero, but within the range of levels that the data can inform, the model matches the joint extremes at every distance. The bands over the weight samples are narrow, so that these features are stable across the retained weight samples.

The first map in Figure~\ref{M-fig:application_factor_maps} of the
main text shows that the factor explains a smaller share of latent
variance at higher elevations, with the smallest shares in the
mountains \citep[ETOPO1;][]{amante2009etopo1}. Over the plains,
frontal systems spanning hundreds of kilometers account for most
precipitation \citep{catto2012relating}. These systems cause
precipitation to rise and fall across large areas, consistent
with a larger contribution from the shared factor. In the Alps
and the Carpathians, orographic lifting and convection make
precipitation more sensitive to local conditions
\citep{roe2005orographic}. The Gaussian component represents this
local variation.

The second map shows a change in factor skewness across the Carpathian
arc and the Eastern Alps. The North European Plain and the Pannonian
Basin are both lowlands with similar factor shares. Skewness is negative
over the plain and predominantly positive over the basin. The plain
is exposed to Atlantic and Baltic air masses, and wet conditions bring
precipitation across much of it. Occasional dry pockets during these
periods may help explain the negative skewness. In the mountain-enclosed
Pannonian Basin, occasional cyclones of Mediterranean origin account
for much of the heavy precipitation
\citep{messmer2015climatology,volosciuk2016rising}. These episodes
of unusually heavy precipitation are consistent with predominantly
positive skewness.

The model uses coordinates as its only spatial inputs, so these
patterns are learned from the precipitation data. Confirming the
physical interpretations would require a separate meteorological
analysis.

\begin{figure}[tbp]
\centering
\includegraphics[width=\textwidth]{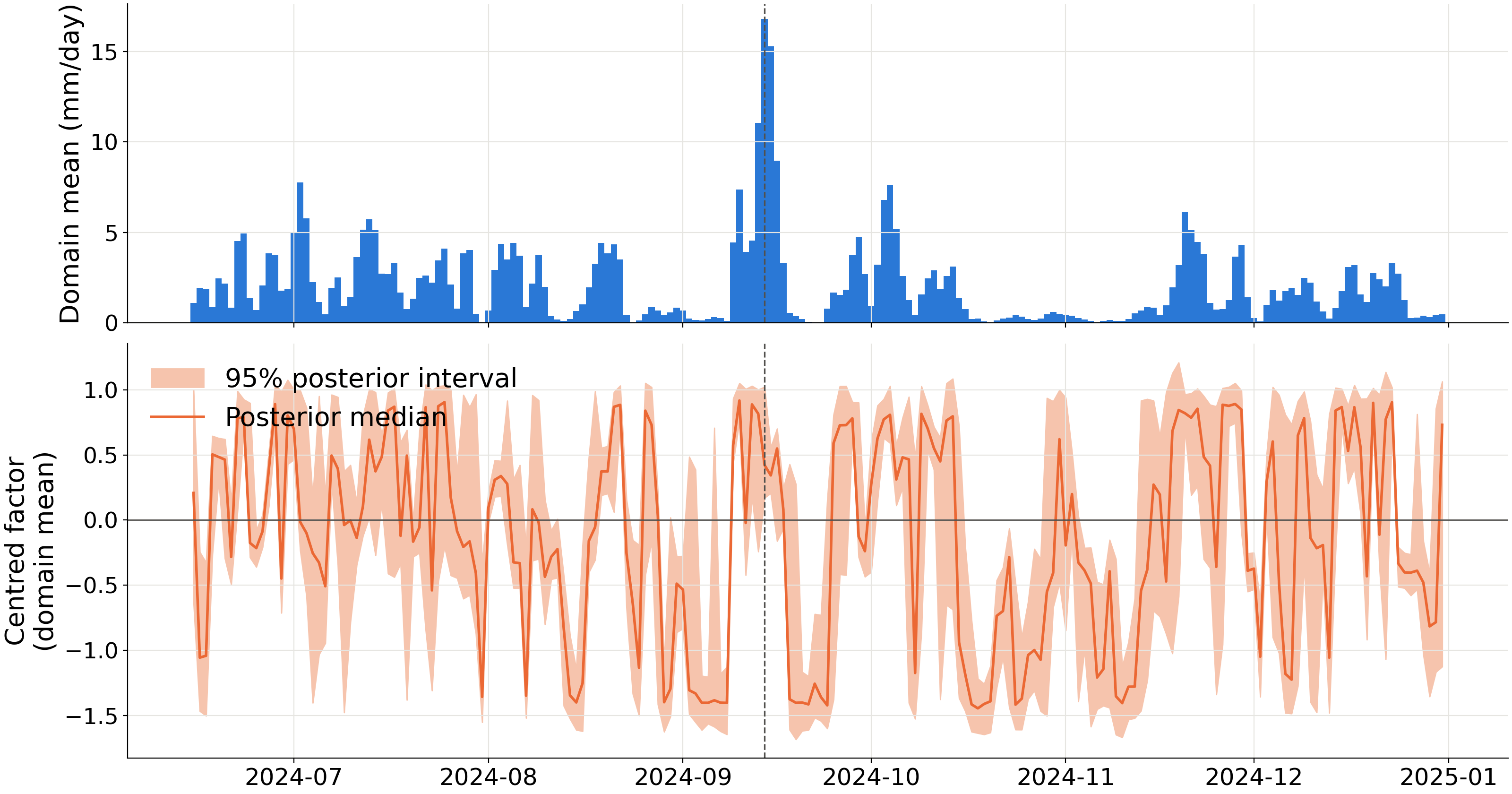}
\caption{ERA5 precipitation, test period from 15 June to 31 December 2024. Top: domain-average precipitation of the ERA5 fields. Bottom: posterior median and $95\%$ interval of the domain average of the centered factor contribution of BNFC (nonstat), given the $100$ observed locations of each day. The dashed line marks 14 September 2024.}
\label{fig:application_factor_time}
\end{figure}

Figure~\ref{fig:application_factor_time} shows the posterior of the domain average of the centered factor contribution for each test day, given the $100$ observed locations, together with the domain-average precipitation. The factor tracks the large-scale wet and dry spells: it is negative during the dry spells of early September and from late October to mid-November, and it reaches its largest value of the test period on 10 September, at the onset of the extreme event of mid-September 2024. On 14 and 15 September, when precipitation peaks, it is lower but still positive, with $95\%$ intervals that exclude zero; on these days, the very high values are described by the Gaussian component, whose long range spreads them over large parts of the domain. The factor thus describes the domain-wide state rather than the domain-average amount, and on days of transition its value remains uncertain given $100$ observations. What the posterior of the network weights adds to these conclusions, compared with a maximum likelihood fit with plug-in prediction, is discussed below.

\subsection{Posterior of the factor and its contribution to uncertainty}

The conclusions of Section~\ref{M-sec:application} of the main text rest on the posterior of the model rather than on a single fitted network, and several of them could not be drawn from a maximum likelihood fit with plug-in prediction, as in \citet{krupskii2018factor} (Remark~\ref{M-rem:identifiability} of the main text). We summarize the posterior by the $20$ weight samples of the stochastic gradient MCMC, each combined with the variational posterior of the factor parameters. First, the spatial patterns of Figure~\ref{M-fig:application_factor_maps} of the main text are supported by the posterior and are not features of one fit: the sign of the skewness of the factor contribution is determined with high posterior probability at most grid locations, the skewness changes sign over the domain in every weight sample, and the posterior band of the factor share $\varrho(\bfs)$ is narrow compared with its range over the domain. Second, the posterior quantifies how much of the uncertainty about the factor on a given day is due to the model rather than to the data of that day. By the law of total variance, the posterior variance of the factor contribution on a test day splits into the expected conditional variance given the network, the uncertainty that a plug-in analysis would report, and the variance of the conditional mean over the weight samples. The second part is a noticeable fraction of the total on average and a large one on some days and at some locations, so that a plug-in analysis would understate the uncertainty about the large-scale state where it matters most. Third, the posterior gives direct probability statements about the large-scale state, which is positive with posterior probability close to one at the onset and the peak of the extreme event of September 2024. The same posterior propagates to the predictions of Figure~\ref{M-fig:application_map} of the main text, whose intervals average over the weight samples, and to the coverage of the $95\%$ prediction intervals in Table~\ref{tab:supp_app_all}, which is the closest to the nominal level of all models. Figure~\ref{fig:application_dependence} gives the posterior bands of the tail dependence.

\section{Scalability with the Vecchia approximation}
\label{app:vecchia}

The cost of the copula density in Equation~\eqref{M-eq:bnfc_copula_density} of the main text is dominated by the Gaussian joint density $\phi_{\bs{\Sigma}_{\bs{\vartheta}}}\{\bfw-\bs{G}(v)\}$, which is evaluated at every factor node $v^{(m)}$ and every replicate. With a Cholesky factor of $\bs{\Sigma}_{\bs{\vartheta}}$, each evaluation costs $O(n^2)$, after an $O(n^3)$ factorization that must be repeated whenever the range and nugget are updated, so that the exact density limits the number of locations that can be used for fitting. Because $\bfW$ is Gaussian given $V=v$, the Vecchia approximation \citep{vecchia1988estimation,katzfuss2021general} applies directly to the residual $\bs{r}=\bfw-\bs{G}(v)$, with the same sparse factor at every factor node.

We order the locations by the maximin ordering \citep{guinness2018permutation} and, for the $i$th location in this ordering, take as conditioning set $c(i)$ its $m$ nearest neighbors among the preceding locations. The Vecchia approximation of $\bs{\Sigma}_{\bs{\vartheta}}^{-1}$ is $\bs{U}_v\bs{U}_v^\top$, where $\bs{U}_v$ is the sparse upper-triangular matrix whose $i$th column is nonzero only on $\{i\}\cup c(i)$, with
\[
\bs{U}_{v,\{i\}\cup c(i),\,i}=\frac{\bs{\Sigma}_{\{i\}\cup c(i)}^{-1}\mathbf{e}_1}{\big(\mathbf{e}_1^\top\bs{\Sigma}_{\{i\}\cup c(i)}^{-1}\mathbf{e}_1\big)^{1/2}},
\]
where $\bs{\Sigma}_{\{i\}\cup c(i)}$ is the $(m+1)\times(m+1)$ submatrix of $\bs{\Sigma}_{\bs{\vartheta}}$ and $\mathbf{e}_1$ is the first unit vector \citep{schafer2021sparse}. Then
\[
\log\phi_{\bs{\Sigma}_{\bs{\vartheta}}}(\mathbf{r})\approx-\frac12\|\bs{U}_v^\top\mathbf{r}\|^2+\sum_{i=1}^n\log U_{v,ii}-\frac n2\log(2\pi).
\]
Building $\bs{U}_v$ costs $O(nm^3)$ and each evaluation $O(nm)$, instead of $O(n^3)$ and $O(n^2)$. The submatrices are computed directly from the correlation function, so that $\bs{U}_v$ is differentiable in the range and nugget, which are estimated jointly with the ELBO as before, and the product $\bs{U}_v^\top\mathbf{r}$ is computed as a sparse matrix product for all factor nodes and replicates at once. With $m=n-1$, the approximation is exact. For $m<n-1$, the approximate covariance matrix $(\bs{U}_v\bs{U}_v^\top)^{-1}$ is a proper covariance matrix, but its diagonal is not exactly one for general conditioning sets. We do not renormalize it: the marginal distributions $F_{1,j}$ in Equation~\eqref{M-eq:bnfc_univariate_margin} of the main text are computed with the exact unit variance, and the Vecchia approximation is used only for the joint density, so that the small departure of the approximate diagonal from one is part of the approximation error, which vanishes as $m$ approaches $n-1$. Renormalizing to the correlation matrix $\bs{R}_v=\bs{D}^{-1}(\bs{U}_v\bs{U}_v^\top)^{-1}\bs{D}^{-1}$, with $\bs{D}=\operatorname{diag}(d_1,\ldots,d_n)^{1/2}$ and $d_1,\ldots,d_n$ the approximate variances, the diagonal entries of $(\bs{U}_v\bs{U}_v^\top)^{-1}$, would make the margins exactly consistent, but it would remove the computational advantage of the approximation. The quadratic form and the log-determinant still need only the sparse factor, $\mathbf{r}^\top\bs{R}_v^{-1}\mathbf{r}=\|\bs{U}_v^\top\bs{D}\mathbf{r}\|^2$ and $\log|\bs{R}_v|=-2\sum_{i}\log(\bs{U}_v)_{ii}-\sum_{i}\log d_i$, but the variances form the diagonal of the inverse of the sparse approximate precision matrix, and this inverse is dense: the approximate variance at a location depends on the approximate covariances among its neighbors, and through them on the whole matrix. Computing the variances exactly requires triangular solves with the full $n\times n$ factor, which cost $O(n^3)$ operations and $O(n^2)$ memory with dense linear algebra, and still $O(n^2m)$ operations with sparse solves. Since $\bs{D}$ depends on the range and nugget, these solves would be repeated at every gradient step, with gradients propagated through them, so that the cost of a step would again grow faster than linearly in $n$. The accuracy gained is small. Table~\ref{tab:supp_vecchia_renorm} reports, for the $1{,}024$ locations of the ERA5 grid and the exponential correlation with the estimated range and nugget of BNFC (nonstat) in Section~\ref{M-sec:application} of the main text, the approximate variances and the error of the approximate Gaussian log-density against the exact one. With $m=30$, the approximate variances lie within $10^{-4}$ of one, and the error is the same with and without renormalization, so that renormalizing does not improve the approximation. This is expected: $\bs{D}\bs{U}_v$ has the same sparsity pattern as $\bs{U}_v$, and among all factors with this pattern, $\bs{U}_v$ minimizes the Kullback--Leibler divergence of the approximation from the exact Gaussian distribution \citep{schafer2021sparse}, so that renormalization cannot reduce this divergence. The factor network, the Monte Carlo estimators in Equation~\eqref{M-eq:mc_estimators} of the main text, the implicit gradients and the variational procedure are unchanged.

\begin{table}[tbp]
\centering
\caption{Vecchia approximation of the Gaussian component on the $1{,}024$ locations of the ERA5 grid, for $m$ nearest previous neighbors, with the exponential correlation and the estimated range and nugget of BNFC (nonstat) in Section~\ref{M-sec:application} of the main text: range and mean absolute deviation from one of the approximate variances $d_i$, the diagonal entries of $(\bs{U}_v\bs{U}_v^\top)^{-1}$, and the mean difference between the approximate and the exact Gaussian log-density per location, over $1{,}000$ draws from the exact Gaussian distribution, without and with renormalization to $\bs{D}^{-1}(\bs{U}_v\bs{U}_v^\top)^{-1}\bs{D}^{-1}$, with standard errors in parentheses. The mean difference estimates minus the Kullback--Leibler divergence of the approximation from the exact distribution, divided by $n$.}
\label{tab:supp_vecchia_renorm}
\renewcommand{\arraystretch}{1.15}
\small
\begin{tabular}{@{}rccccc@{}}
\toprule
 & \multicolumn{3}{c}{Approximate variances $d_i$} & \multicolumn{2}{c}{Log-density error per location} \\
\cmidrule(r){2-4}\cmidrule(l){5-6}
$m$ & Min & Max & Mean $|d_i-1|$ & Not renormalized & Renormalized \\
\midrule
$10$ & $0.9949$ & $1.0007$ & $0.00063$ & $-0.00314$\,{\scriptsize$(0.00008)$} & $-0.00314$\,{\scriptsize$(0.00008)$} \\
$20$ & $0.9996$ & $1.0000$ & $0.00007$ & $-0.00021$\,{\scriptsize$(0.00002)$} & $-0.00021$\,{\scriptsize$(0.00002)$} \\
$30$ & $0.9999$ & $1.0000$ & $0.00001$ & $-0.00003$\,{\scriptsize$(0.00001)$} & $-0.00003$\,{\scriptsize$(0.00001)$} \\
\bottomrule
\end{tabular}
\end{table}

Table~\ref{tab:supp_vecchia_timing} reports two benchmarks on one NVIDIA GeForce RTX 4090 GPU. The first evaluates the copula log-density of the fitted BNFC (nonstat) model of Section~\ref{M-sec:application} of the main text with $500$ factor nodes, exactly and with the Vecchia approximation with $m=30$, on regular grids of $n$ locations, for $5$ replicates simulated from the fitted model, since the network can be evaluated at any location. The cost of the exact density grows faster than linearly in $n$, because of the Cholesky factorization and the dense triangular solves, whereas that of the Vecchia approximation grows about as fast as $n$, so that the approximation is faster at the larger grids and the gap widens with $n$. The maximin ordering and the neighbor search are computed once per set of locations. The second benchmark measures the time to evaluate the Monte Carlo
ELBO estimator in Equation~\eqref{M-eq:elbo_mc} of the main text
(forward pass) and compute its gradients with respect to the
variational parameters and network weights (backward pass).
Each mini-batch contains $4$ replicates. As in the complete-data fit
described below, we use $M_2=5$ draws from the variational distribution
and $M_1=100$ factor draws per variational draw. Its time and memory grow about linearly with $n$, and up to $n=1{,}024$ they are almost the same for the exact density and the Vecchia approximation. At these sizes, the cost is dominated by the parts of the step that are linear in $n$, the network and the marginal estimators, so that the fit is linear in $n$ once the Vecchia approximation removes the cubic cost of the Gaussian part at larger $n$.

\begin{table}[tbp]
\centering
\caption{Computing times on one GPU. Left: evaluation of the copula log-density of the fitted BNFC (nonstat) model, in milliseconds per replicate, exactly and with the Vecchia approximation with $m=30$, on regular grids of $n$ locations; the last column is the one-off time for the ordering and the neighbor search. Right: time and peak memory of one gradient step of the ELBO with mini-batches of $4$ replicates, $M_1=100$ and $M_2=5$, with the Vecchia approximation with $m=30$; with the exact density, the times are within $5\%$ and the memory is the same to the precision shown.}
\label{tab:supp_vecchia_timing}
\renewcommand{\arraystretch}{1.15}
\small
\begin{tabular}{@{}rccc@{\qquad}rcc@{}}
\toprule
\multicolumn{4}{c}{Copula log-density} & \multicolumn{3}{c}{ELBO gradient step} \\
\cmidrule(r){1-4}\cmidrule(l){5-7}
$n$ & Exact (ms) & Vecchia (ms) & Ordering (s) & $n$ & Time (ms) & Memory (GB) \\
\midrule
$576$ & $2.8$ & $2.2$ & $0.01$ & $100$ & $8.6$ & $0.47$ \\
$1{,}024$ & $4.7$ & $3.3$ & $0.03$ & $256$ & $15.9$ & $1.13$ \\
$2{,}304$ & $12.2$ & $7.9$ & $0.11$ & $576$ & $30.4$ & $2.46$ \\
$4{,}096$ & $30.4$ & $17.2$ & $0.32$ & $1{,}024$ & $51.1$ & $4.32$ \\
$9{,}216$ & $126.7$ & $40.7$ & $1.53$ & & & \\
\bottomrule
\end{tabular}
\end{table}

We fit BNFC (nonstat) to the complete ERA5 data, with all $1{,}024$ grid locations observed on each of the $10{,}591$ training days, with $m=30$ neighbors. To keep the memory requirements small, the fit uses mini-batches of $4$ replicates and $M_1=100$ and $M_2=5$ Monte Carlo samples, and it was run for $100$ epochs. The estimated range and nugget of the Gaussian component are $\hat\ell=0.062$ and $\hat\tau^2=0.015$. The fit took $0.84$ hours on one NVIDIA GeForce RTX 4090 GPU: the $100$ training epochs took $2{,}713$ seconds, about $136$ seconds per epoch, and the stochastic gradient MCMC for the $20$ weight samples took $318$ seconds; the peak GPU memory was $4.6$~GB. 

For prediction, we condition on the data at all $1{,}024$ grid locations and predict the field on a regular $80\times80$ grid over the same domain, with $6{,}400$ locations, using the posterior predictive sampler of Section~\ref{M-subsec:cond_copula} of the main text with the Vecchia approximated Gaussian conditional distribution. The marginal quantile function $T_{\hat{\bfpsi}}$ is trained on the training days at all $1{,}024$ grid locations and evaluated at the $6{,}400$ prediction locations. With $100$ predictive draws, five from each of the $20$ weight samples, and $200$ factor draws per predictive draw, the prediction takes about $1.6$ seconds per day on one GPU, after a one-off $0.4$ seconds for the conditional covariance. 

\begin{figure}[tbp]
\centering
\includegraphics[width=\textwidth]{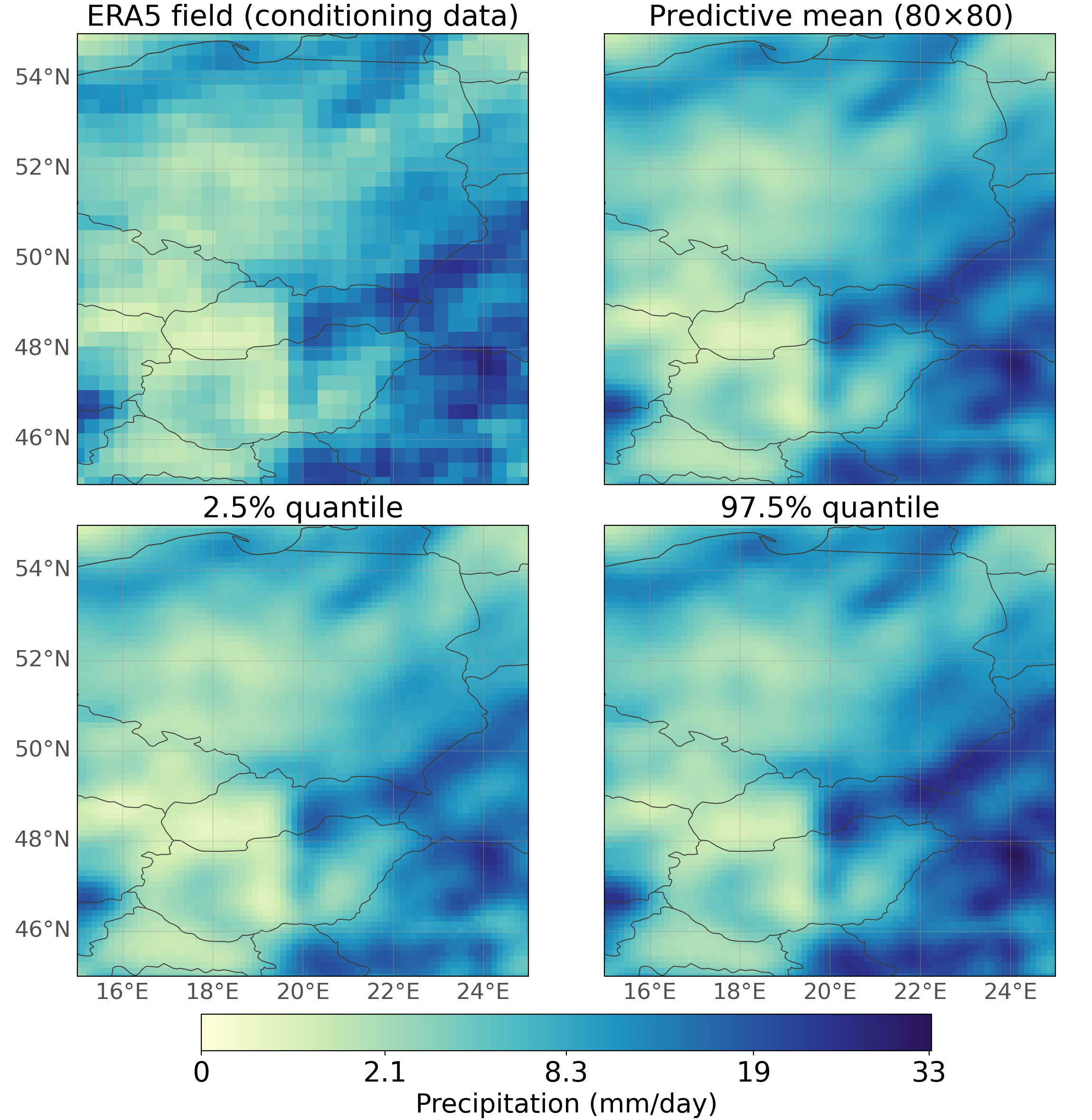}
\caption{ERA5 precipitation on 2 July 2024 (mm/day): predictions of BNFC (nonstat) fitted to all $1{,}024$ grid locations with the Vecchia approximation ($m=30$), on an $80\times80$ grid, given the data at all $1{,}024$ grid locations. Top row: ERA5 field on the $32\times32$ grid, which is the conditioning data, and predictive mean; bottom row: $2.5\%$ and $97.5\%$ predictive quantiles. All panels share one square-root color scale.}
\label{fig:supp_vecchia_map}
\end{figure}

Figure~\ref{fig:supp_vecchia_map} shows the predictive mean and the
$2.5\%$ and $97.5\%$ predictive quantiles of precipitation in mm/day
for 2 July 2024, when the heaviest rainfall was concentrated in the
east and southeast of the domain. 

\section{Implementation details of the estimation procedure}
\label{app:implementation}

{This section collects implementation details of the estimation procedure of Section~\ref{M-sec:estimation} of the main text: the marginal quantile network (Section~\ref{app:T_psi}), the computation of the latent values and their gradients (Section~\ref{app:supp_latent}), the computational cost (Section~\ref{app:supp_cost}), the stochastic gradient MCMC (Section~\ref{app:supp_psgld}).

\subsection{The marginal quantile network \texorpdfstring{$T_{\bfpsi}$}{T}}\label{app:T_psi}

This section describes the monotone neural network $T_{\bfpsi}$ used in Section~\ref{M-subsec:marginals} of the main text to approximate the marginal quantile function $G^{-1}(u;\bfs)$ of the data. We describe the location-dependent version, $T_{\bfpsi}:(0,1)\times D\to\mathbb{R}$. The stationary version is obtained by removing the location input, so that the coefficients below are constants.

The input is first mapped to the real line by the logit transformation, $z=\operatorname{logit}(u)=\log\{u/(1-u)\}$, so that the transformation acts on an unconstrained domain. The location is processed by a multilayer perceptron $\bs{h}=h_{\bfpsi_1}(\bfs)$, where $h_{\bfpsi_1}:\mathbb{R}^2\to\mathbb{R}^{d_1}$ has two hidden layers of width $d_1=128$ with GELU activations. The network outputs six sets of location-dependent coefficients,
\[
a(\bfs) = \mathbf{W}_a\bs{h}+\bs{b}_a,
\qquad
\varsigma(\bfs) = \exp(\mathbf{W}_\varsigma\bs{h}+\bs{b}_\varsigma),
\]
\[
\alpha_k(\bfs) = 0.05\operatorname{softplus}(\mathbf{W}_{\ell,k}\bs{h}+b_{\ell,k}),
\qquad
\beta_k(\bfs) = \operatorname{softplus}(\mathbf{W}_{\beta,k}\bs{h}+b_{\beta,k})+\epsilon,
\]
\[
\gamma_k(\bfs) = 0.05\operatorname{softplus}(\mathbf{W}_{\gamma,k}\bs{h}+b_{\gamma,k}),
\qquad
\delta_k(\bfs) = \operatorname{softplus}(\mathbf{W}_{\delta,k}\bs{h}+b_{\delta,k})+\epsilon,
\qquad k = 1, \dots, K,
\]
where $\epsilon=10^{-4}$ ensures strictly positive slopes. Let $c_k=-6+12(k-1)/(K-1)$, $k=1,\ldots,K$, be fixed knots on $[-6,6]$, with $K=32$. The marginal quantile function is approximated by
\begin{equation}\label{eq:neural_inverse}
\begin{aligned}
T_{\bfpsi}(u,\bfs)
={}&a(\bfs)+\varsigma(\bfs)z+\sum_{k=1}^{K}\alpha_k(\bfs)\operatorname{softplus}\{\beta_k(\bfs)(z-c_k)\}\\
&-\sum_{k=1}^{K}\gamma_k(\bfs)\operatorname{softplus}\{-\delta_k(\bfs)(z-c_k)\},
\qquad z=\operatorname{logit}(u),
\end{aligned}
\end{equation}
where $\bfpsi$ collects the weights and biases of $h_{\bfpsi_1}$ and of the six output layers. The first two terms form a location-dependent affine transformation of $z$, and the two sums provide a flexible nonlinear correction through shifted softplus basis functions. The terms of the first sum are convex in $z$ and those of the second sum are concave, so that $T_{\bfpsi}$ can steepen as well as flatten in either tail. Both are needed. With the first sum only, $T_{\bfpsi}(\cdot,\bfs)$ is a convex function of $z$, which cannot represent a quantile function whose tails are lighter than those of the logistic distribution, such as the margins in our simulations. The weights of the output layers are initialized at zero, and the biases so that $T_{\bfpsi}(u,\bfs)=\bar y+\hat\sigma_y\sqrt{3}\,z/\pi$ at initialization, where $\bar y$ and $\hat\sigma_y$ are the mean and standard deviation of the data, that is, $T_{\bfpsi}$ starts at the logistic quantile function with the same mean and variance as the data. The network is trained by Adam for $100$ epochs with learning rate $10^{-3}$, cosine decay and mini-batches of $256$ pairs.

Since the coefficients do not depend on $u$, and since $\alpha_k(\bfs),\gamma_k(\bfs)\ge0$, $\beta_k(\bfs),\delta_k(\bfs)>0$ and $\varsigma(\bfs)>0$,
\begin{equation}\label{eq:neural_inverse_derivative}
\begin{aligned}
\frac{\partial T_{\bfpsi}(u,\bfs)}{\partial u}
=\frac{1}{u(1-u)}\Bigg[&\varsigma(\bfs)+\sum_{k=1}^{K}\alpha_k(\bfs)\beta_k(\bfs)\,\sigma\{\beta_k(\bfs)(z-c_k)\}\\
&+\sum_{k=1}^{K}\gamma_k(\bfs)\delta_k(\bfs)\,\sigma\{-\delta_k(\bfs)(z-c_k)\}\Bigg]>0,
\end{aligned}
\end{equation}
where $\sigma(x)=1/(1+e^{-x})$ is the logistic function. Hence $T_{\bfpsi}(\cdot,\bfs)$ is strictly increasing and continuous at every location, so it defines a valid quantile function, and it is invertible. Monotonicity relies on the coefficients depending on $\bfs$ only. If they also depended on $u$, the derivative would contain additional terms of arbitrary sign.


The probability integral transform $u=G(y;\bfs)$ is obtained by solving $T_{\bfpsi}(u,\bfs)=y$. Since \eqref{eq:neural_inverse} is strictly increasing in $z$, we solve for $z$ by eight iterations of Newton's method, started at $z=y$, and set $u=1/(1+e^{-z})$. By the inverse function theorem, the marginal density of the data is
\[
G'(y;\bfs)=\left\{\frac{\partial T_{\bfpsi}(u,\bfs)}{\partial u}\right\}^{-1},\qquad u=G(y;\bfs),
\qquad\text{so that}\qquad
\log G'(y;\bfs)=-\log\frac{\partial T_{\bfpsi}(u,\bfs)}{\partial u},
\]
which is available in closed form from \eqref{eq:neural_inverse_derivative}. It is not needed in the two-step procedure of the main text, but it would allow the marginal and copula parameters to be estimated jointly.


The network $T_{\bfpsi}$ approximates the quantile function of the data, $G^{-1}(\cdot\,;\bfs)$. It does not approximate the latent quantile function $F_{1,j}^{-1}$ of the copula, which depends on $(\bftheta,\bs{\varphi})$ and changes during the fit. That quantile function is computed by bisection on the Monte Carlo approximation in Equation~\eqref{M-eq:mc_estimators} of the main text, with the gradients in Equation~\eqref{M-eq:implicit_gradient}, as described in Section~\ref{app:supp_latent}. The two maps are composed in spatial prediction. A latent draw $W_0$ is mapped to the copula scale by $\hat{F}_{1,0}$ and then to the data scale by $T_{\bfpsi}(\cdot,\bfs_0)$, see Section~\ref{M-subsec:cond_copula} of the main text.

\subsection{Latent values and their gradients}\label{app:supp_latent}

This section gives the details of the Monte Carlo approximation in Equation~\eqref{M-eq:mc_estimators} of the main text and of the computation of the latent values $\hat{w}_{ij}=\hat{F}_{1,j}^{-1}(u_{ij})$ in Section~\ref{M-subsec:likelihood} of the main text.


\emph{Common random numbers} means that, within one evaluation of the likelihood, the same $M_1$ draws are used for the marginal distribution functions, the marginal densities and the joint density. Fresh draws are generated at every evaluation, that is, at every step of the optimization and for every draw of the factor parameters in Section~\ref{M-subsec:vb_inference} of the main text; the draws are not fixed over iterations. For a given set of draws, $\hat{F}_{1,j}$, $\hat{f}_{1,j}$ and $\hat{f}_n$ are exactly the marginal distribution functions, marginal densities and joint density of model~\eqref{M-eq:bnfc_finite_model} of the main text in which $V$ is replaced by a random variable that is uniformly distributed on the finite set $\{v^{(1)},\ldots,v^{(M_1)}\}$. The resulting approximation $\hat{c}_n$ of the copula density is therefore itself the density of a valid factor copula, whose factor takes the $M_1$ values $g(v^{(m)},\cdot)$ with equal probability. The factor contribution is $g(V,\cdot)$, where $V$ is continuously
distributed. We approximate the integrals over $V$ using $M_1$ Monte
Carlo draws. The resulting estimate $\hat{c}_n$ converges to the
model copula density $c_n$ as $M_1\to\infty$. If different draws were used for $\hat{F}_{1,j}$ and $\hat{f}_n$, the margins of $\hat{f}_n$ would not match $\hat{F}_{1,j}$, $\hat{c}_n$ would not be a copula density, and the Monte Carlo error would be larger. The averages in \eqref{M-eq:mc_estimators} of the main text are unbiased for fixed arguments, but $\log\hat{c}_n$ is evaluated at $\hat{w}_{ij}=\hat{F}_{1,j}^{-1}(u_{ij})$ and involves the logarithm of an average, so that it is a consistent but biased estimator of $\log c_n$, with bias of order $1/M_1$. 



Since $\hat{F}_{1,j}$ is a mixture of normal distribution functions, it is continuous and strictly increasing, and $\hat{w}_{ij}=\hat{F}_{1,j}^{-1}(u_{ij})$ is the unique root of $\hat{F}_{1,j}(w)=u_{ij}$. Moreover,
\[
\hat{F}_{1,j}\Big\{\min_m g(v^{(m)},\bfs_j)-10\Big\}\le\Phi(-10)
\quad\text{and}\quad
\hat{F}_{1,j}\Big\{\max_m g(v^{(m)},\bfs_j)+10\Big\}\ge\Phi(10),
\]
so the interval $[\min_m g(v^{(m)},\bfs_j)-10,\ \max_m g(v^{(m)},\bfs_j)+10]$ contains the root for every $u_{ij}\in[\Phi(-10),\Phi(10)]$. We compute $\hat{w}_{ij}$ by $60$ bisection steps on this interval. The copula data are clipped to $[10^{-6},1-10^{-6}]$ before the inversion, which is inside $[\Phi(-10),\Phi(10)]$ and avoids probabilities that round to zero or one in single precision, where $\Phi(10)$ is one. Because the interval moves with the network output, it contains the root wherever $g$ takes its values. Since $g$ does not depend on $w$, each bisection step only requires evaluating $\hat{F}_{1,j}$ at the cached values $g(v^{(m)},\bfs_j)$.


On line~5 of Algorithm~S2, we use bisection to solve
$\hat{F}_{1,j}(\hat{w}_{ij})=u_{ij}$ for each variational draw,
mini-batch replicate, and location. Line~6 evaluates the joint and
marginal densities at these latent values to compute the ELBO
estimator in Equation~\eqref{M-eq:elbo_mc} and take an Adam step.
A change in the parameters affects both the densities and the latent
values at which we evaluate them.

Let $\bs{\upsilon}$ denote a parameter of $(\bftheta,\bs{\varphi})$.
Holding $u_{ij}$ and the Monte Carlo draws fixed, implicit
differentiation gives
\[
\frac{\partial\hat{w}_{ij}}{\partial\bs{\upsilon}}
=
-\frac{
\left.\partial\hat{F}_{1,j}(w;\bs{\upsilon})/
\partial\bs{\upsilon}\right|_{w=\hat{w}_{ij}}
}{
\hat{f}_{1,j}(\hat{w}_{ij};\bs{\upsilon})
},
\]
as in Equation~\eqref{M-eq:implicit_gradient} of the main text.
We compute $\hat{w}_{ij}$ by bisection without tracking gradients
and use
\[
\tilde{w}_{ij}
=\hat{w}_{ij}
-\frac{
\hat{F}_{1,j}(\hat{w}_{ij};\bs{\upsilon})
-\operatorname{sg}\{\hat{F}_{1,j}(\hat{w}_{ij};\bs{\upsilon})\}
}{
\operatorname{sg}\{\hat{f}_{1,j}(\hat{w}_{ij};\bs{\upsilon})\}
},
\]
where $\operatorname{sg}$ returns its input and has derivative zero.
Thus $\tilde{w}_{ij}$ equals the bisection root in the forward pass
and supplies the implicit derivative in the backward pass. This
derivative contributes to the updates of the network weights and,
through the reparameterized factor parameters, of the variational
parameters. It also contributes to the weight gradient during
weight sampling.

Adding a constant $a$ to $g$ also shifts $\hat{w}_{ij}$ by $a$.
The two shifts cancel in the density residuals, leaving the copula
log-likelihood invariant (Remark~\ref{M-rem:identifiability} of the
main text). Recomputing the roots gives this cancellation in the
forward pass. The implicit derivative gives the corresponding
cancellation in the likelihood gradient. 


\subsection{Computational cost}\label{app:supp_cost}

The cost of a gradient step of the estimator in Equation~\eqref{M-eq:elbo_mc} of the main text does not depend on the number $N$ of replicates, so that the cost of an epoch is linear in $N$ and long records can be used. At each gradient step, the combined cost of evaluating the network,
computing the marginal estimates in Equation~\eqref{M-eq:mc_estimators}
of the main text, and finding the latent values by bisection is
$O(|\mathcal{B}|M_2M_1n)$. Here $\mathcal{B}$ is the set of training
replicate indices in the current mini-batch, and $|\mathcal{B}|$ is its
size. For fixed mini-batch and Monte Carlo sample sizes, this cost
grows linearly with the number of locations $n$. The only part that grows faster is the Gaussian density $\phi_{\bs{\Sigma}_{\bs{\vartheta}}}$, which needs an $O(n^3)$ Cholesky factorization for each draw of $\bs{\vartheta}$ and $O(n^2)$ operations and memory for each of the $|\mathcal{B}|M_2M_1$ residual vectors $\tilde{\bfw}_i^{(r)}-\bs{G}\{v^{(m,r)}\}$, one for each replicate of the mini-batch, outer draw $r$ and inner draw $m$. Because $\bfW$ is Gaussian given $V=v$, we replace it by the Vecchia approximation \citep{vecchia1988estimation,katzfuss2021general}, in which each location is conditioned only on $m$ nearest previous neighbors. A single sparse factor, which is differentiable in $\bs{\vartheta}$, then serves all factor nodes and replicates, the cost becomes $O(nm^3)$ for the factor and $O(nm)$ per residual, and the memory becomes $O(nm)$ instead of $O(n^2)$. The factor network, the common random numbers, the implicit gradients and the variational procedure are unchanged, so that, for fixed $m$, Monte Carlo sizes, mini-batch size and network, the cost of an epoch is linear in both $N$ and $n$. The number of epochs and the Monte Carlo sizes needed for a given accuracy may change with $n$ and must be assessed separately. For $n=10^5$ locations, for example, the dense covariance matrix would occupy $80$~GB in double precision, whereas the Vecchia factor with $m=30$ occupies about $25$~MB, and the remaining memory, which is linear in $n$, can be controlled through $|\mathcal{B}|$ and $M_1$. The construction of the Vecchia factor and computing times are given in Section~\ref{app:vecchia}.

\subsection{Preconditioned stochastic gradient Langevin dynamics}\label{app:supp_psgld}


The mean-field family matches the supports of the parameters and is easy to reparameterize, but it omits posterior dependence between the components of $\bs{\eta}$ and tends to understate posterior spread \citep{blei2017variational}; we assess its effect through the calibration of the predictions in Section~\ref{M-sec:empirical} of the main text. The two-step procedure conditions on the copula data and therefore omits the uncertainty about the margins. In the simulations, the margins are known; in the application, the rank-based scores are computed from more than $10{,}000$ days at each location, so that their sampling error is small compared with the uncertainty about the copula.

The optimization produces a point estimate $\hat{\bs{\varphi}}$ of the weights and a fitted variational distribution $q_{\hat{\bs{\lambda}}}(\bs{\eta})$ for $\bs{\eta}$. Variational Bayes is used for $\bs{\eta}$ because it has four components with simple supports; for the thousands of network weights, mean-field variational approximations are known to underestimate the posterior spread, and they would add variance to an objective that is already a nested Monte Carlo estimator. The point estimate $\hat{\bs{\varphi}}$ ignores the uncertainty about the shape of the factor distribution, which is typically the dominant source of uncertainty about the copula, and it can make predictive intervals too narrow. After the fit, we therefore explore the weights by stochastic gradient Markov chain Monte Carlo (SG-MCMC), with $q_{\hat{\bs{\lambda}}}(\bs{\eta})$ held fixed. The exact marginal posterior of the weights is $p(\bs{\varphi}\mid\mathcal{U})\propto p(\bs{\varphi})\,p(\mathcal{U}\mid\bs{\varphi})$, which requires the evidence $p(\mathcal{U}\mid\bs{\varphi})$. Replacing $\log p(\mathcal{U}\mid\bs{\varphi})$ by the ELBO gives the approximate weight target
\[
\tilde{\pi}(\bs{\varphi}\mid\mathcal{U})\propto\exp\{\mathcal{L}(\hat{\bs{\lambda}},\bs{\varphi})\}\,\mathcal{N}\{\bs{\varphi};\mathbf{0},(N\omega)^{-1}I\}
\;\propto\;p(\bs{\varphi}\mid\mathcal{U})\exp\{-D(\bs{\varphi})\}
\]
in Equation~\eqref{M-eq:weight_posterior} of the main text, where the Gaussian prior $p(\bs{\varphi})$ corresponds to the weight decay $\omega=10^{-4}$ used in the optimization, and $D(\bs{\varphi})=D_{\mathrm{KL}}\{q_{\hat{\bs{\lambda}}}(\bs{\eta})\,\|\,\pi(\bs{\eta}\mid\mathcal{U},\bs{\varphi})\}$ by Equation~\eqref{M-eq:elbo_decomposition} of the main text. The target $\tilde{\pi}$ therefore differs from the exact marginal posterior by the factor $\exp\{-D(\bs{\varphi})\}$, which favors weights for which the fixed-weight posterior of $\bs{\eta}$ is close to $q_{\hat{\bs{\lambda}}}$. In addition, the same $q_{\hat{\bs{\lambda}}}$ is combined with every weight sample, so that the procedure omits the posterior dependence between the weights and $\bs{\eta}$, beyond the independence assumed within the mean-field family. We use preconditioned stochastic gradient Langevin dynamics \citep{welling2011bayesian,li2016preconditioned}, whose update is given below; the preconditioner continues to adapt during sampling. The chain starts at $\hat{\bs{\varphi}}$, a first epoch estimates the preconditioner without moving the weights, a second is discarded as burn-in, and we then retain $S$ thinned samples $\bs{\varphi}^{(1)},\ldots,\bs{\varphi}^{(S)}$. The predictive mixture in Equation~\eqref{M-eq:weight_average} of the main text approximates the idealized mixture $\iint p(\cdot\mid\bs{\eta},\bs{\varphi})\,q_{\hat{\bs{\lambda}}}(\bs{\eta})\,\tilde{\pi}(\bs{\varphi}\mid\mathcal{U})\,d\bs{\eta}\,d\bs{\varphi}$. This is much cheaper than a deep ensemble \citep{lakshminarayanan2017simple}, which requires several complete fits. We assess the resulting predictive distributions by their calibration in Section~\ref{M-sec:empirical} of the main text. All results for BNFC in the main text are based on Equation~\eqref{M-eq:weight_average} of the main text.



We now give the update of the stochastic gradient MCMC for the network weights in Section~\ref{M-subsec:vb_inference} of the main text, which targets the approximate weight target $\tilde{\pi}(\bs{\varphi}\mid\mathcal{U})$ in Equation~\eqref{M-eq:weight_posterior} of the main text. We use preconditioned stochastic gradient Langevin dynamics \citep{welling2011bayesian,li2016preconditioned}, which updates the weights by
\[
\bs{\varphi}\leftarrow\bs{\varphi}-\frac{\epsilon}{2}\,\mathbf{P}\odot\hat{\bfg}+(\epsilon\mathbf{P})^{1/2}\odot\bs{\xi},\qquad\bs{\xi}\sim\mathcal{N}(\mathbf{0},I),
\]
where $\epsilon>0$ is the step size, $\hat{\bfg}$ is the gradient of $-\hat{\mathcal{L}}+N\omega\|\bs{\varphi}\|^2/2$ on a mini-batch, with $\hat{\mathcal{L}}$ in Equation~\eqref{M-eq:elbo_mc} of the main text, $\odot$ denotes elementwise multiplication, and the preconditioner is $\mathbf{P}=1/(N\omega+\mathbf{a}^{1/2})$, elementwise, with $\mathbf{a}$ an exponentially weighted average of the squared gradients with decay $0.99$. The floor $N\omega$, the prior precision, ensures that weights which the data do not inform, for example those of inactive units, move on the scale of their prior rather than receiving noise of unbounded variance. The preconditioner $\mathbf{a}$ is updated at every step, during the first epoch, in which the weights are not moved, and during burn-in and sampling, so that the adaptation continues while the samples are retained. The step size, the number $S$ of retained samples and the thinning are given in Section~\ref{M-sec:empirical} of the main text.

\section{Algorithms}\label{app:supp_algorithm}

\subsection{Algorithm for prediction}\label{app:supp_alg_prediction}

We draw from the approximate posterior predictive distribution of $Y(\bfs_0)$ in Equation~\eqref{M-eq:weight_average} of the main text with Algorithm~\ref{alg:prediction}. Each of the $R$ predictive draws repeats steps (i)--(iii) with its own draw of the parameters and uses the network with weights $\bs{\varphi}^{(s(r))}$, where the $R$ draws are split evenly between the $S$ weight samples. In step (ii), $L$ candidate factor values are weighted by their conditional density given the observations, and one of them is selected for the predictive draw; repeating this selection over the $R$ draws produces a Monte Carlo sample from the conditional distribution \eqref{M-eq:conditional_copula} of the main text, averaged over the parameter uncertainty. Because the correlation parameters are drawn from $q_{\hat{\bs{\lambda}}}$ together with $\bftheta$, the predictive distribution also reflects the uncertainty about the Gaussian component. Predictive means and standard errors are computed from the draws, and several locations can be predicted jointly by replacing \eqref{M-eq:kriging} of the main text with the multivariate kriging mean and covariance.

\begin{algorithm}[tbp]
\caption{Posterior predictive sampling at an unobserved location $\bfs_0$}
\label{alg:prediction}
\begin{algorithmic}[1]
\Require copula data $\bfu$ of one replicate at $\bfs_1,\ldots,\bfs_n$; $q_{\hat{\bs{\lambda}}}$, weight samples $\bs{\varphi}^{(1)},\ldots,\bs{\varphi}^{(S)}$; numbers $R$ of predictive draws and $L$ of candidate factor values
\For{$r=1,\ldots,R$}
  \State (i) use $g$ with the weight sample $\bs{\varphi}^{(s(r))}$ assigned to draw $r$; draw $\bs{\eta}^{(r)}=(\bftheta^{(r)\top},\bs{\vartheta}^{(r)\top})^\top\sim q_{\hat{\bs{\lambda}}}$, compute $\bs{\Sigma}_{\bs{\vartheta}^{(r)}}$ and $\bs{\sigma}_0(\bs{\vartheta}^{(r)})$, and compute $\bfw^{(r)}=\{\hat{F}_{1,1}^{-1}(u_1),\ldots,\hat{F}_{1,n}^{-1}(u_n)\}^\top$ under $\bs{\eta}^{(r)}$
  \State (ii) draw $v^{(1)},\ldots,v^{(L)}\sim U(\theta_1^{(r)},\theta_2^{(r)})$, compute the self-normalized weights $\omega_\ell\propto\phi_{\bs{\Sigma}_{\bs{\vartheta}^{(r)}}}\{\bfw^{(r)}-\bs{G}(v^{(\ell)})\}$, and select one candidate $v^{(\ell^\star)}$ with probabilities $\omega_1,\ldots,\omega_L$
  \State (iii) draw $W_0^{(r)}\sim\mathcal{N}\{m_0^{(r)}(v^{(\ell^\star)}),\sigma_{K,0}^{2(r)}\}$, with $m_0^{(r)}$ and $\sigma_{K,0}^{2(r)}$ in \eqref{M-eq:kriging} of the main text evaluated at $\bs{\vartheta}^{(r)}$; set $U_0^{(r)}=\hat{F}_{1,0}(W_0^{(r)})$ and $Y_0^{(r)}=T_{\hat{\bfpsi}}(U_0^{(r)},\bfs_0)$
\EndFor
\Ensure predictive draws $W_0^{(r)}$, $U_0^{(r)}$ and $Y_0^{(r)}$, $r=1,\ldots,R$, on the latent, copula and data scales
\end{algorithmic}
\end{algorithm}

\subsection{Algorithm for estimation and weight sampling}\label{app:supp_alg_estimation}

Algorithm~\ref{alg:estimation} summarizes the estimation procedure of Section~\ref{M-subsec:vb_inference} of the main text. 



\begin{algorithm}[tbp]
\caption{Estimation and weight sampling for BNFC}
\label{alg:estimation}
\begin{algorithmic}[1]
\Require copula data $\mathcal{U}$; Monte Carlo sizes $M_1$, $M_2$; mini-batch size; learning-rate schedule and patience; SG-MCMC step size $\epsilon$, number of samples $S$ and thinning
\State Initialize $\bs{\lambda}$ so that the factor interval is close to $(1,2)$, and the weights $\bs{\varphi}$ at random
\Repeat \Comment{variational EM by stochastic gradient ascent}
  \For{each mini-batch $\mathcal{B}$}
    \State draw $\bs{\eta}^{(r)}\sim q_{\bs{\lambda}}(\bs{\eta})$ by \eqref{M-eq:variational_reparam} of the main text, $r=1,\ldots,M_2$, and fresh $\varepsilon^{(m,r)}\sim U(0,1)$, $m=1,\ldots,M_1$
    \State compute $\tilde{w}_{ij}^{(r)}=\hat{F}_{1,j}^{-1}(u_{ij})$ by bisection, with the gradients \eqref{M-eq:implicit_gradient} of the main text
    \State compute $\hat{\mathcal{L}}(\bs{\lambda},\bs{\varphi})$ in \eqref{M-eq:elbo_mc} of the main text and take an Adam step in $(\bs{\lambda},\bs{\varphi})$
  \EndFor
\Until{the epoch average of $\hat{\mathcal{L}}/N$ has shown no
improvement for the specified number of consecutive epochs}
\State keep $(\hat{\bs{\lambda}},\hat{\bs{\varphi}})$ of the best epoch; hold $q_{\hat{\bs{\lambda}}}(\bs{\eta})$ fixed from now on
\State set $\bs{\varphi}\leftarrow\hat{\bs{\varphi}}$; run one epoch that only estimates the preconditioner, then one burn-in epoch of preconditioned SGLD targeting \eqref{M-eq:weight_posterior} of the main text
\State continue preconditioned SGLD and retain every $k$th state until $S$ samples $\bs{\varphi}^{(1)},\ldots,\bs{\varphi}^{(S)}$ are kept, where $k$ is the thinning
\Ensure $q_{\hat{\bs{\lambda}}}(\bs{\eta})$ and $\bs{\varphi}^{(1)},\ldots,\bs{\varphi}^{(S)}$, used in \eqref{M-eq:weight_average} of the main text
\end{algorithmic}
\end{algorithm}

\clearpage

\bibliographystyle{dcu}
\bibliography{ref}